\documentclass{aa}
\usepackage{times,epsfig,amssymb,amsmath}

\newcommand{\chandra}{{\it Chandra}}
\newcommand{\xmm}{{\it XMM-Newton}}

\newcommand{\meti}{{\it Method 1}}
\newcommand{\metii}{{\it Method 2}}

\title{Mass profiles and $c-M_{\rm DM}$ relation \\
in X-ray luminous galaxy clusters}
\titlerunning{Mass profiles and $c-M_{\rm DM}$ relation in X-ray luminous galaxy clusters}

\author{S. Ettori\inst{1,2} \and F. Gastaldello\inst{3,5,6} \and A. Leccardi\inst{3,4}
  \and S. Molendi\inst{3} \and M. Rossetti\inst{3} \and D. Buote\inst{5}
  \and M. Meneghetti\inst{1,2}
} 
\authorrunning{S. Ettori et al.}

\institute{
 INAF, Osservatorio Astronomico di Bologna, via Ranzani 1, I-40127 Bologna, Italy
 \and INFN, Sezione di Bologna, viale Berti Pichat 6/2, I-40127 Bologna, Italy
 \and INAF, IASF, via Bassini 15, I-20133 Milano, Italy
 \and Universit\`a degli Studi di Milano, Dip. di Fisica, via Celoria 16, I-20133 Milano, Italy
 \and Department of Physics and Astronomy, University of California, Irvine, CA 92697-4575
 \and Occhialini Fellow
}

\offprints{S. Ettori}

\mail{stefano.ettori@oabo.inaf.it}

\date{Received 24 June 2010 / Accepted 14 September 2010}

\begin{document}

\abstract
% context heading (optional)
{Galaxy clusters represent valuable cosmological probes using tests that
mainly rely on measurements of cluster masses and baryon fractions. 
X-ray observations represent one of the main tools for uncovering these quantities.}
% aims heading (mandatory)
{We aim to constrain the cosmological parameters
$\Omega_{\rm m}$ and $\sigma_8$ using the observed distribution
of the both values of the concentrations and dark mass within $R_{200}$ and of the gas mass
fraction within $R_{500}$.}
% methods heading (mandatory)
{We applied two different techniques to recover the profiles the gas and dark mass,
described according to the Navarro, Frenk \& White (1997, ApJ, 490, 493) functional form, 
of a sample of 44 X-ray luminous galaxy clusters
observed with \xmm\ in the redshift range $0.1-0.3$.
We made use of the spatially resolved spectroscopic data and of the PSF--deconvolved
surface brightness and assumed that hydrostatic equilibrium holds between 
the intracluster medium and the gravitational potential.
We evaluated several systematic uncertainties that affect our reconstruction of the X-ray masses.}
% results heading (mandatory)
{We measured the concentration $c_{200}$, the dark mass $M_{200}$ and the gas
mass fraction in all the objects of our sample, providing the largest dataset 
of mass parameters for galaxy clusters in the redshift range $0.1-0.3$.
We confirm that a tight correlation between $c_{200}$ and $M_{200}$
is present and in good agreement with the predictions from numerical simulations
and previous observations.
When we consider a subsample of relaxed clusters that host a low entropy core, 
we measure a flatter $c-M$ relation with a total scatter that is lower by 40 per cent.
We conclude, however, that the slope of the $c-M$ relation
cannot be reliably determined from the fitting over a narrow mass range
as the one considered in the present work.
From the distribution of the estimates of $c_{200}$ and $M_{200}$, with
associated statistical (15--25\%) and systematic (5--15\%) errors,
we used the predicted values from semi-analytic prescriptions calibrated through 
N-body numerical runs and obtain
$\sigma_8 \, \Omega_{\rm m}^{0.60 \pm 0.03} = 0.45 \pm 0.01$ 
(at $2 \sigma$ level, statistical only)
for the subsample of the clusters where the mass reconstruction has been obtained
more robustly and $\sigma_8 \, \Omega_{\rm m}^{0.56 \pm 0.04} = 0.39 \pm 0.02$
for the subsample of the 11 more relaxed LEC objects.
With the further constraint from the gas mass fraction distribution in our
sample, we break the degeneracy in the $\sigma_8-\Omega_{\rm m}$ plane
and obtain the best-fit values $\sigma_8 \approx 1.0 \pm 0.2$ 
($0.83 \pm 0.1$ when the subsample of the more relaxed objects is considered)
and $\Omega_{\rm m} = 0.26 \pm 0.02$.}
% conclusions heading (optional), leave it empty if necessary
{We demonstrate that the analysis of the distribution of the 
$c_{200}-M_{200}-f_{\rm gas}$ values represents a 
mature and competitive technique in the present era of precision cosmology,
even though it needs more detailed analysis of the output of larger sets of 
cosmological numerical simulations to provide definitive and robust results.}


\maketitle

% ==================================================================

\section{Introduction}

The distribution of the total and baryonic mass in galaxy clusters
is a fundamental ingredient to validate the scenario of structure
formation in a Cold Dark Matter (CDM) Universe. Within this scenario,
the massive virialized objects are powerful cosmological tools
able to constrain the fundamental parameters of a given CDM model. 
The $N-$body simulations of structure formation in CDM models indicate
that dark matter halos aggregate with a typical mass density profile 
characterized by only 2 parameters,
the concentration $c$ and the scale radius $r_{\rm s}$ (e.g. Navarro et al. 1997,
hereafter NFW).
The product of these two quantities fixes the radius within which the mean cluster 
density is 200 times the critical value at the cluster's redshift 
[i.e. $R_{200} = c_{200} \times r_{\rm s}$ and the cluster's volume
$V = 4/3 \pi R_{200}^3$ is equal to $M_{200} / (200 \rho_{c,z})$,
where $M_{200}$ is the cluster gravitating mass within $R_{200}$].
With this prescription, the structural properties of DM halos from galaxies to
galaxy clusters are dependent on the halo mass, with systems at higher masses 
less concentrated. Moreover, the concentration depends upon the
assembly redshift (e.g. Bullock et al. 2001, Wechsler et al. 2002, 
Zhao et al. 2003, Li et al. 2007),
which happens to be later in cosmologies with lower matter density,
$\Omega_{\rm m}$, and lower normalization of the linear power spectrum
on scale of $8 h^{-1}$ Mpc, $\sigma_8$, implying less concentrated DM halos
of given mass.
The concentration -- mass relation, and its evolution in redshift, is therefore 
a strong prediction obtained from CDM simulations of structure formation
and is quite sensitive to the assumed cosmological parameters
(NFW; Bullock et al. 2001; Eke, Navarro \& Steinmetz 2001; 
Dolag et al. 2004; Neto et al. 2007; Macci\`o et al. 2008). 
In this context, NFW, Bullock et al. 2001 (with revision after
Macci\`o et al. 2008) and Eke et al. 2001 have provided simple
and powerful models that match the predictions from numerical
simulations and allow comparison with the observational measurements.

Recent X-ray studies (Pointecouteau, Arnaud \& Pratt 2005; Vikhlinin et al. 2006;
Voigt \& Fabian 2006; Zhang et al. 2006; Buote et al. 2007)
have shown good agreement between observational constraints at low
redshift and theoretical expectations.
By fitting 39 systems in the mass range between early-type galaxies up to
massive galaxy clusters, Buote et al. (2007) confirm with high significance
that the concentration decreases with increasing mass, as predicted
from CDM models, and require a $\sigma_8$, the dispersion of the mass
fluctuation within spheres of comoving radius of 8 $h^{-1}$ Mpc,
in the range $0.76-1.07$ (99\% confidence) definitely in contrast to 
the lower constraints obtained, for instance, from the analysis
of the {\it WMAP} 3 years data. Since it is based upon a selection
of the most relaxed systems, these results assumed a 10\%
upward early formation bias in the concentration parameter
for relaxed halos.
Using a sample of 34 massive, dynamically relaxed galaxy clusters
resolved with \chandra\ in the redshift range $0.06-0.7$,
Schmidt \& Allen (2007) highlight a possible tension between the observational
constraints and the numerical predictions, in the sense that either the relation
is steeper than previously expected or some redshift evolution has to  
be considered.
Comerford \& Natarajan (2007) compiled a large dataset of observed 
cluster concentration and masses, finding a normalization
higher by at least 20 per cent than the results from simulations.
In the sample, they use also strong lensing measurements of the concentration
concluding that these are systematically larger than the ones 
estimated in the X-ray band, and 55 per cent higher, on average, than 
the rest of the cluster population.
Recently, Wojtak \& {\L}okas (2010) analyze kinematic data of 41 nearby ($z<0.1$)
relaxed objects and find a normalization of the concentration -- mass relation
fully consistent with the amplitude of the power spectrum $\sigma_8$
estimated from WMAP1 data and within $1 \sigma$ from the constraint obtained
from WMAP5.

\begin{table*}
\caption{Sample of the galaxy clusters. 
}
\vspace*{-0.2cm}
\begin{center}
\begin{tabular}{l@{\hspace{.7em}} l@{\hspace{.8em}} c c c c c c}
\hline \\
  & & & & \\
 Cluster & Other name & z & Core Cl. & Entropy Cl. &  X-ray refs. \\ 
    &  \\
\hline \\
RXCJ0003.8+0203 & Abell2700 & 0.092 & ICC & MEC & Pr07, Cr08 \\
Abell3911 & -- & 0.097 & NCC & HEC & Sn08 \\
Abell3827 & -- & 0.098 & ICC & MEC & Sn08 \\
RXCJ0049.4-2931 & AbellS0084 & 0.108 & ICC & MEC & Si09 \\
Abell2034 & -- & 0.113 & NCC & HEC & Ke03, Ba07 \\
RXCJ1516.5-0056 & Abell2051 & 0.115 & NCC & HEC & Pr07, Cr08 \\
RXCJ2149.1-3041 & Abell3814 & 0.118 & CC & LEC & Cr08, Le08 \\
RXCJ1516.3+0005 & Abell2050 & 0.118 & NCC & HEC & Pr07, Cr08 \\
RXCJ1141.4-1216 & Abell1348 & 0.119 & CC & LEC & Pr07, Cr08 \\
RXCJ1044.5-0704 & Abell1084 & 0.132 & CC & LEC & Pr07, Cr08 \\
Abell1068 & RXCJ1040.7+3956 & 0.138 & CC & LEC & Wi04, Sn08 \\
RXCJ2218.6-3853 & Abell3856 & 0.138 & NCC & MEC & Pr07, Cr08 \\
RXCJ0605.8-3518 & Abell3378 & 0.141 & CC & LEC & Pr07, Cr08, Sn08 \\
RXCJ0020.7-2542 & Abell22 & 0.142 & NCC & HEC & Pr07, Cr08 \\
Abell1413 & RXCJ1155.3+2324 & 0.143 & ICC & MEC & Vi05, Ba07, Sn08, Ca09 \\
RXCJ2048.1-1750 & Abell2328 & 0.147 & NCC & HEC & Pr07, Cr08 \\
RXCJ0547.6-3152 & Abell3364 & 0.148 & NCC & HEC & Pr07, Cr08 \\
Abell2204 & RXC J1632.7+0534 & 0.152 & CC & LEC & Mo07, Sa09 \\
RXCJ0958.3-1103 & Abell907 & 0.153 & CC & LEC & Vi05, Cr08 \\
RXCJ2234.5-3744 & Abell3888 & 0.153 & NCC & HEC & Cr08 \\
RXCJ2014.8-2430 & RXCJ2014.8-24 & 0.161 & CC & LEC & Cr08 \\
RXCJ0645.4-5413 & Abell3404 & 0.167 & ICC & MEC & Cr08 \\
Abell2218 & -- & 0.176 & NCC & HEC & Go04, Ba07 \\
Abell1689 & -- & 0.183 & ICC & MEC & Pe98, An04, Ca09 \\
Abell383 & -- & 0.187 & CC & LEC & Vi05, Ca09, Zh10 \\
Abell209 & -- & 0.206 & NCC & MEC & Ca09 \\
Abell963 & -- & 0.206 & ICC & MEC & Sm05, Ba07, Ca09 \\
Abell773 & -- & 0.217 & NCC & HEC & Go04, Mo07, Ca09 \\
Abell1763 & -- & 0.223 & NCC & HEC & Du08, Ca09 \\
Abell2390 & -- & 0.228 & CC & LEC & Vi05, Mo07, Ca09, Zh10 \\
Abell2667 & -- & 0.230 & CC & LEC & Ca09 \\
RXCJ2129.6+0005 & -- & 0.235 & CC & LEC & Ca09, Zh10 \\
Abell1835 & -- & 0.253 & CC & LEC & Mo07, Zh10 \\
RXCJ0307.0-2840 & Abell3088 & 0.253 & CC & LEC & Fi05, Zh06 \\
Abell68 & -- & 0.255 & NCC & HEC & Zh10 \\
E1455+2232 & RXCJ1457.2+2220 & 0.258 & CC & LEC & Sn08 \\
RXCJ2337.6+0016 & -- & 0.273 & NCC & HEC & Fi05, Zh06, Zh10 \\
RXCJ0303.8-7752 & -- & 0.274 & NCC & HEC & Zh06 \\
RXCJ0532.9-3701 & -- & 0.275 & CC ? & MEC & Fi05, Zh06 \\
RXCJ0232.2-4420 & -- & 0.284 & Cool core remnant ? & MEC & Fi05, Zh06 \\
ZW3146 & RBS0864 & 0.291 & CC & LEC & Mo07 \\
RXCJ0043.4-2037 & Abell2813 & 0.292 & NCC & HEC & Zh06 \\
RXCJ0516.7-5430 & AbellS0520 & 0.295 & NCC & HEC & Zh06 \\
RXCJ1131.9-1955 & Abell1300 & 0.307 & NCC & HEC & Fi05, Zh06 \\
\hline \\
\end{tabular}

\end{center}
\label{tab:data}
\tablefoot{
We quote the name of the object, the redshift adopted
and the classification based on their X-ray properties. \\
({\it Core Cl.}): cool cores (CC), intermediate systems (ICC) and non-cool cores (NCC). \\
({\it Entropy Cl.}): as in Leccardi et al. (2010), low (LEC), medium (MEC) and high (HEC)
entropy cores characterizing clusters with stronger cooling cores and more relaxed structure (LEC),
more disturbed objects (HEC) and systems with intermediate properties (MEC). \\
({\it X-ray refs.}): Baldi et al. (2007, Ba07); Cavagnolo et al. (2009, Ca09);
Croston et al. (2008, Cr08); Finoguenov et al. (2005, Fi05); Govoni et al. (2004, Go04);
Kempner et al. (2003, Ke03); Morandi et al. (2007, Mo07); Pratt et al. (2007, Pr07);
Sanderson et al. (2009, Sa09); Sivanandam et al. (2009, Si09); Snowden et al. (2008, Sn08);
Vikhlinin et al. (2005, Vi05); Wise et al. (2004, Wi04); Zhang et al. (2006, Zh06); 
Zhang et al. (2010, Zh10).
}
\end{table*}

In this work, we use the results of the spectral analysis presented in 
Leccardi \& Molendi (2008) for a sample of 44 X-ray luminous galaxy clusters
located in the redshift range $0.1 - 0.3$ with the aim to (1) recover
their total and gas mass profiles, (2) constraining the cosmological parameters
$\sigma_8$ and $\Omega_{\rm m}$ through the analysis of the measured distribution
of $c_{200}$, $M_{200}$ and baryonic mass fraction in the mass 
range above $10^{14} M_{\odot}$.
We note that this is the statistically largest sample for which this study
has been carried on up to now between $z=0.1$ and $z=0.3$.

\begin{figure*}
\hbox{
 \epsfig{figure=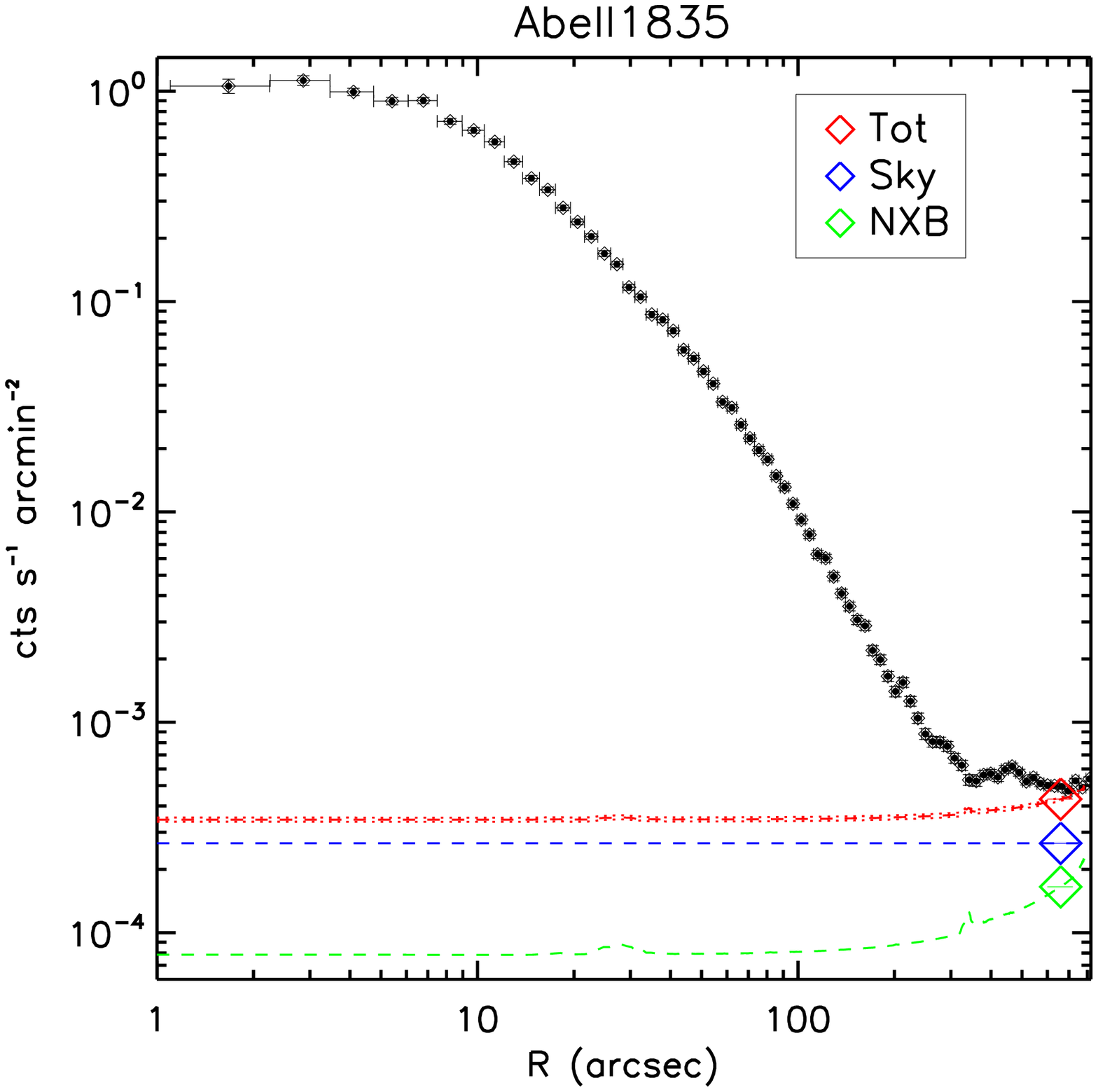,width=0.5\textwidth}
 \epsfig{figure=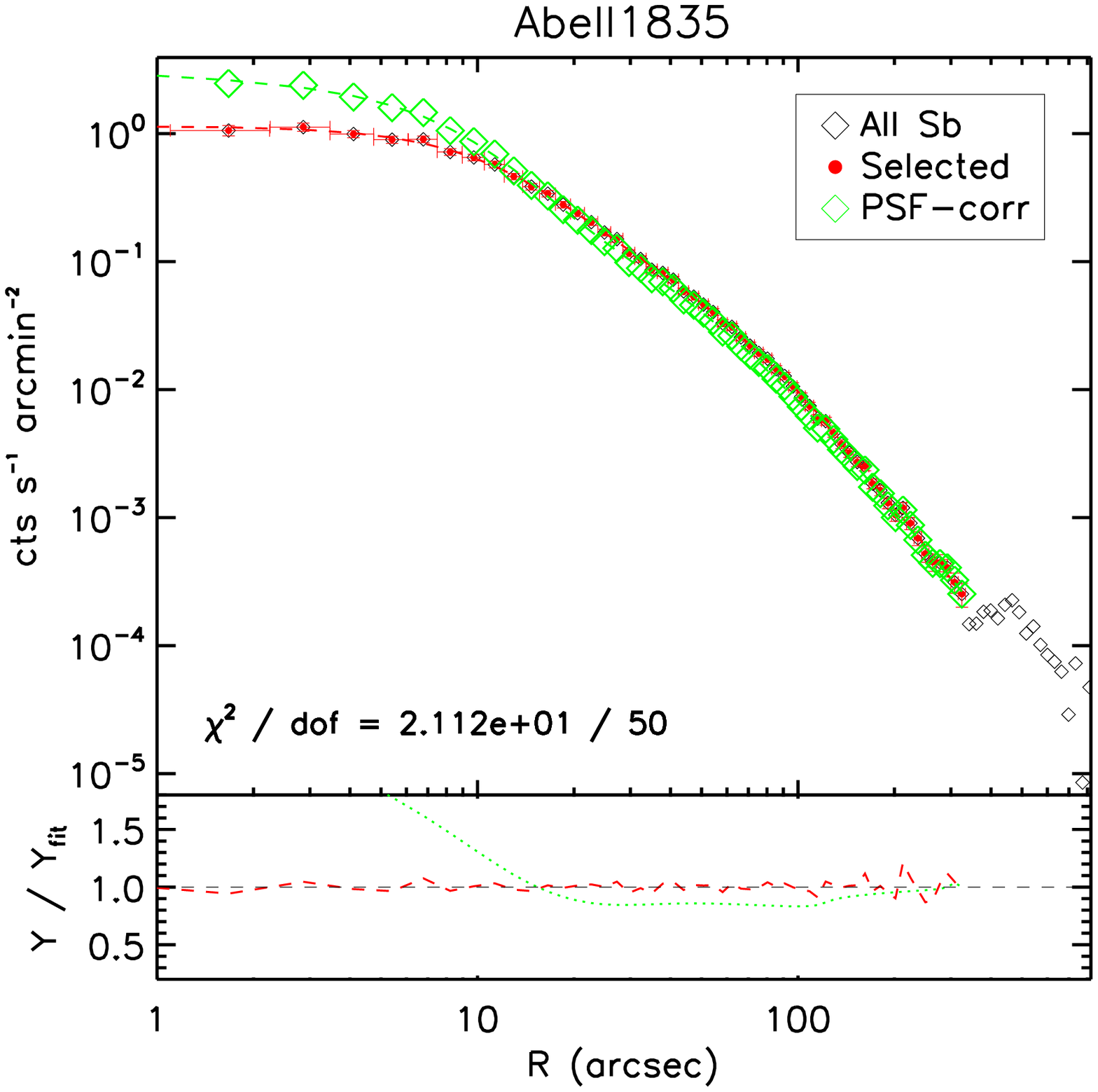,width=0.5\textwidth}
}
\caption{
{(\it Left)}
Surface brightness profile in the $0.7-1.2$ keV band (black filled circles) of Abell1835
compared with the profiles of the background components. The open diamonds show
the count rate predicted from the background spectral model in the annulus 10--12 arcmin
and rescaled for the mean vignetting correction of 0.472 at those radii:
the instrumental component (NXB; green), the photon component (CXB + galactic foregrounds; blue)
and the total background (sky + instrumental; red).
The dashed lines show the background profiles that we have used in our analysis:
the ``photon'' background (blue), which is constant and corresponds to the value
in the outer annulus rescaled to the center, and the instrumental background
profile (green), increasing with radius in order to consider the over-correction
of this component.
The red dashed line shows the total background that we have subtracted
from our source plus background profile, with its associated one $\sigma$ statistical error
(red dotted lines) obtained with a Monte Carlo simulation.
Note that the intensity of the background components and their relative contribution
vary significantly from cluster to cluster.
{(\it Right)}
Example of the PSF--corrected background--subtracted surface brightness profile
as obtained after the analysis outlined in Sect.~2.2.
This example refers to Abell1835, one of the objects
with the largest smearing effect due to the combination
of the telescope's PSF and the centrally peaked intrinsic profile.
} \label{fig:bkg_psf}
\end{figure*}

The outline of our work is the following.  
In Section~2, we describe the dataset of \xmm\ observations used in our analysis
to recover the gas and total mass profiles with the techniques 
presented in Section~3. 
In Section~4, we present a detailed discussion of the main 
systematic uncertainties that affect our measurements.
We investigate the $c_{200} - M_{200}$ relation in Section~5.
By using our measurements of $c_{200}$ and $M_{200}$, we constrain
the cosmological parameters $\sigma_8$ and $\Omega_{\rm m}$, breaking the
degeneracy between these parameters by adding the further cosmological
constraints from our estimates of the cluster baryon fraction, as
discussed in Section~6.
We summarize our results and draw the conclusion of the present study
in Section~7.
Throughout this work, if not otherwise stated, we plot and tabulate values
estimated by assuming a Hubble constant
$H_0= 70 h_{70}^{-1}$ km s$^{-1}$ Mpc$^{-1}$
and $\Omega_{\rm m}=1-\Omega_{\Lambda}=0.3$, and quote
errors at the 68.3 per cent ($1 \sigma$) level of confidence.

We list here in alphabetic order, with the adopted acronyms, 
the work to which we will refer more often in the present study:
Bullock et al. (2001 -- B01); 
Dolag et al. (2004 -- D04);
Eke, Navarro \& Steinmetz (2001 -- E01); 
Leccardi \& Molendi (2008 -- LM08);
Macci\`o et al. (2008 -- M08);
Navarro, Frenk \& White (1997 -- NFW);
Neto et al. (2007 -- N07).

\begin{figure*}
\hbox{
 \epsfig{figure=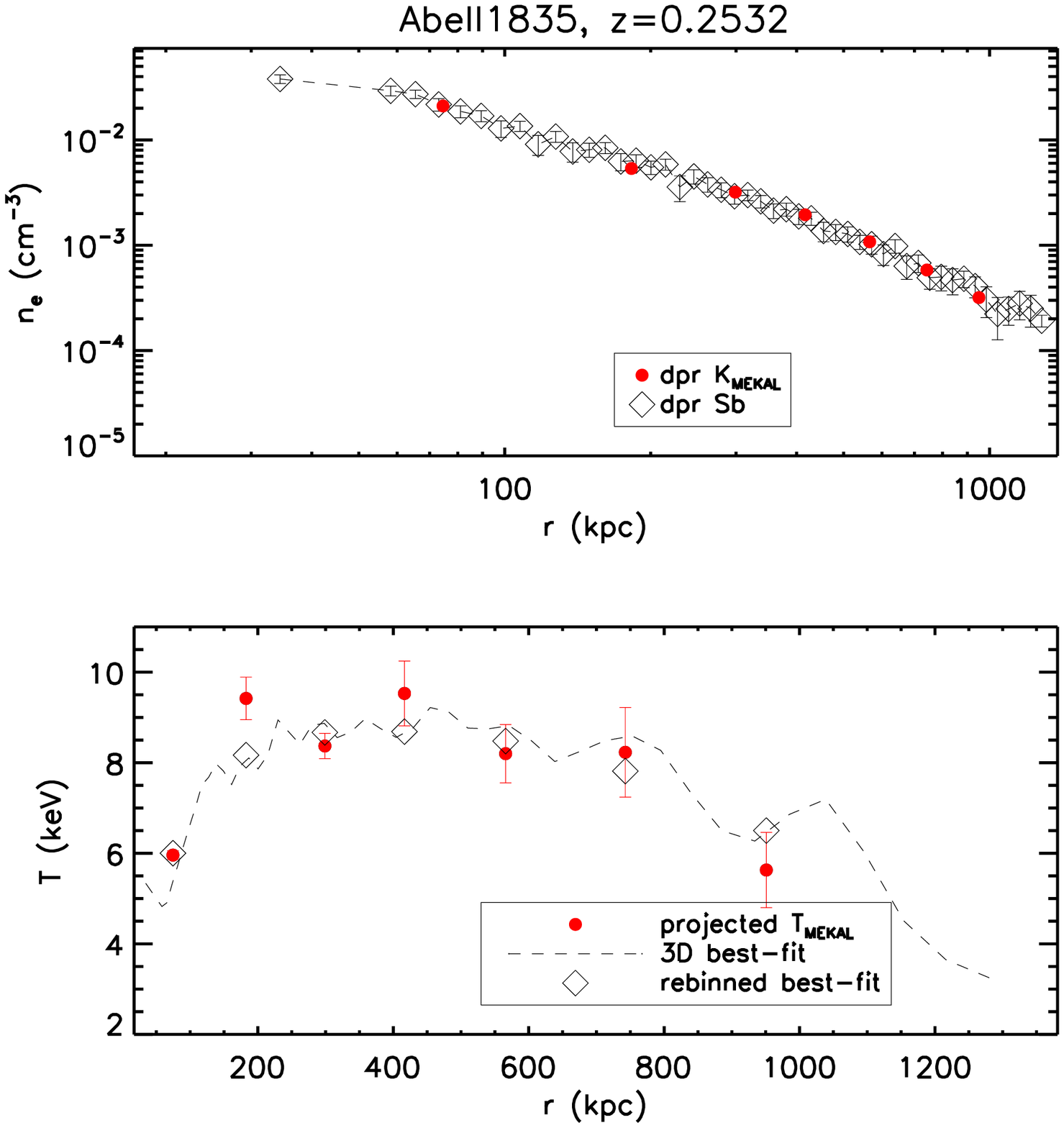,width=0.5\textwidth}
 \epsfig{figure=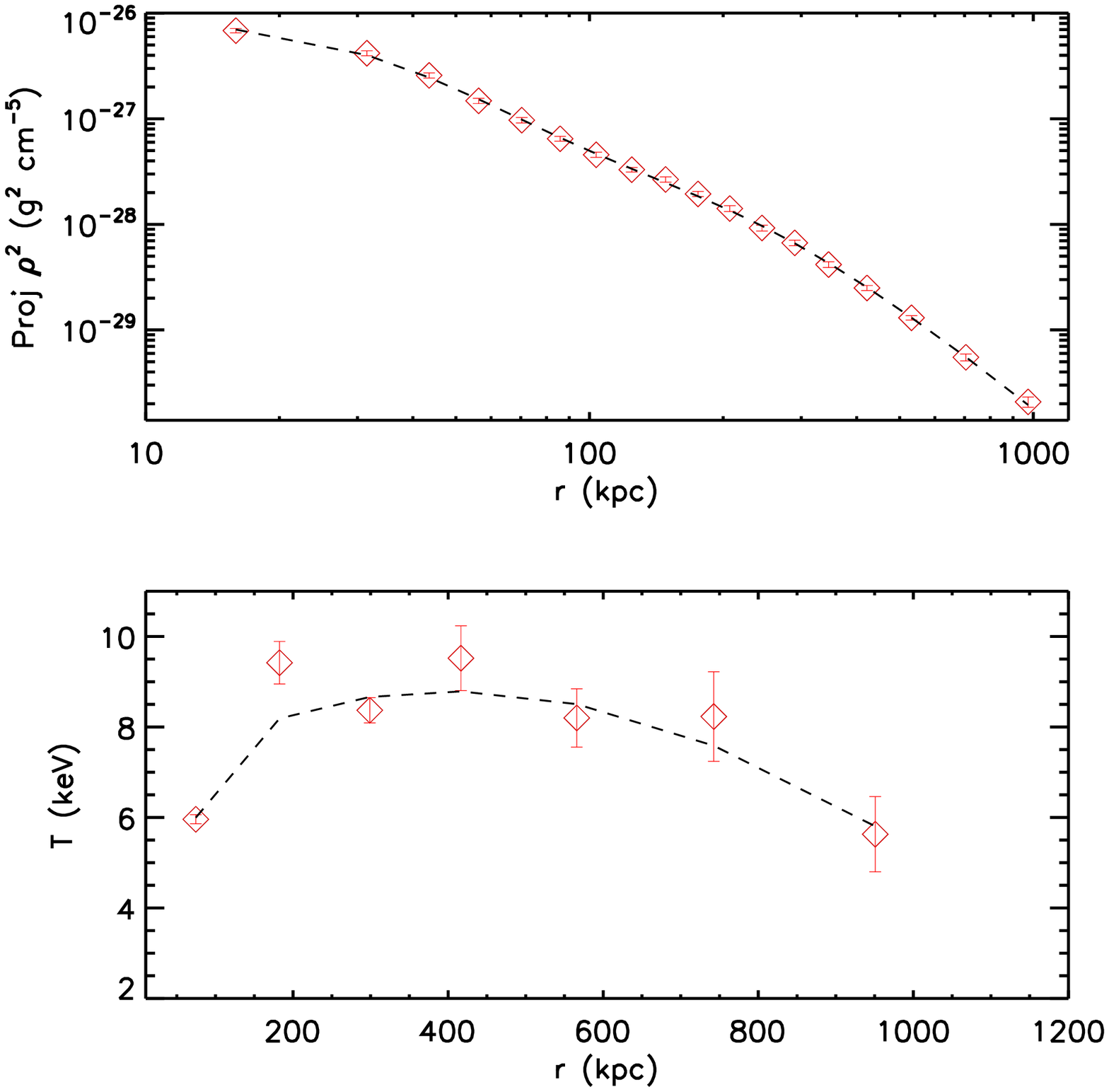,width=0.5\textwidth}
} \hbox{
 \epsfig{figure=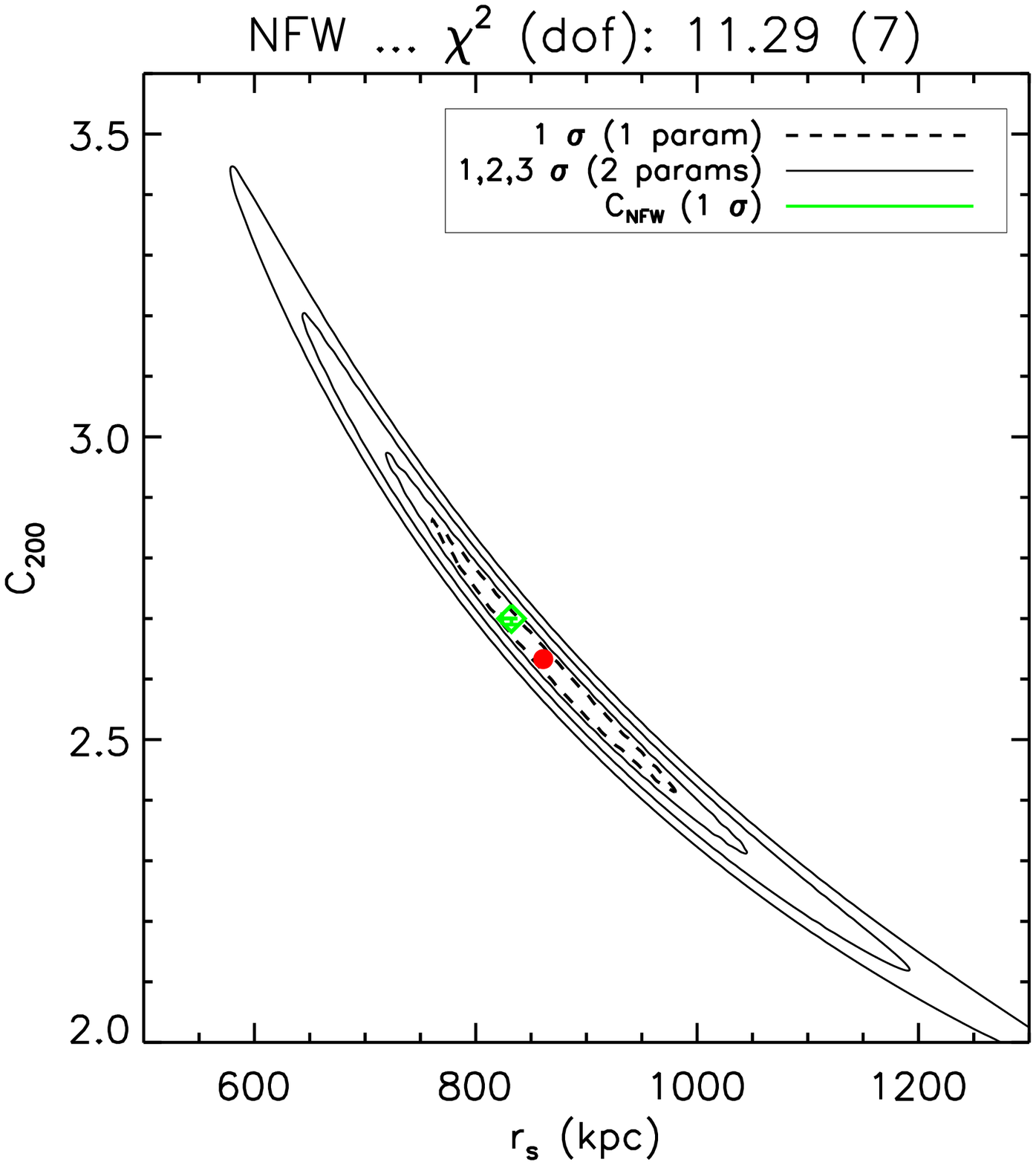,width=0.5\textwidth}
 \epsfig{figure=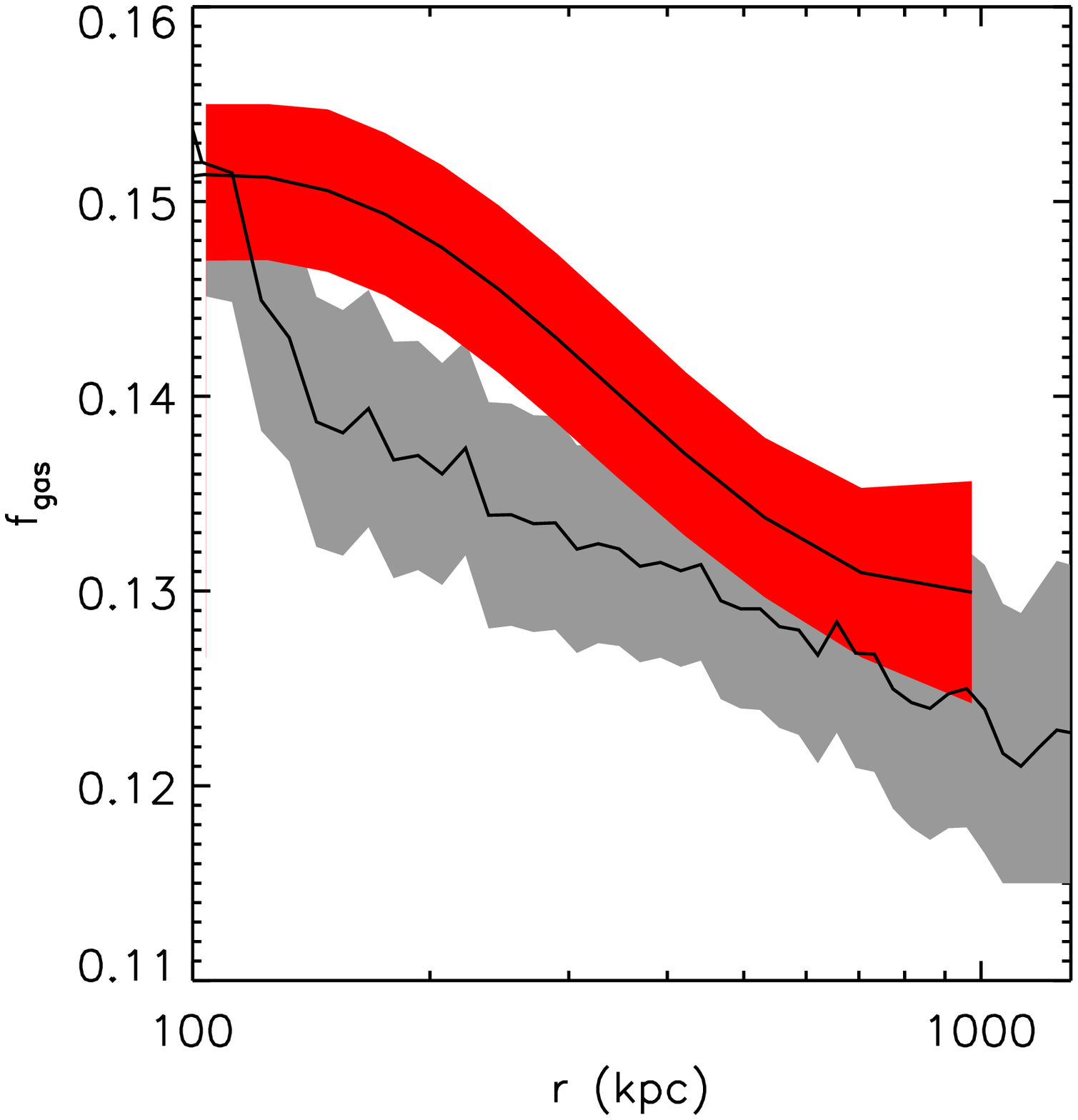,width=0.5\textwidth}
}
\caption{
Example of the results of the two analyses adopted for the mass reconstruction.
{\it (Top and middle panels, left)} Gas density profile as obtained from the deprojection of the
surface brightness profile compared to the one recovered from the deprojection
of the normalizations of the thermal model in the spectral analysis; 
observed temperature profile with overplotted the best-fit model (from \meti).
{\it (Top and middle panels, right)} Data (diamonds) and models (dashed lines) of the
projected gas density squared and temperature (from \metii).
{\it (Bottom, left)} Constraints in the $r_{\rm s}-c$ plane with the prediction 
(in green) obtained by imposing the relation $c_{200} = 4.305/(1+z) \times 
\left(M_{200}/10^{14} h_{100}^{-1} M_{\odot} \right)^{-0.098}$ from M08.
{\it (Bottom, right)} Gas mass fraction profile obtained from \meti\ (gray) and
\metii\ (red).
} \label{fig:meth1}
\end{figure*}

\section{The dataset}

Leccardi \& Molendi (2008) have retrieved from the \xmm\ archive all observations 
of clusters available at the end of May 2007 (and performed before March 2005,
when the CCD6 of EPIC-MOS1 was switched off) and satisfying the selection criteria 
to be hot ($kT > 3.3$ keV), at intermediate redshift ($0.1 < z <0.3$), 
and at high galactic latitude ($|b| > 20^o$).
Upper and lower limits to the redshift range are determined, respectively, by
the cosmological dimming effect and the size of the EPIC field
of view ($15'$ radius).
Out of 86 observations, 23 were excluded because they are highly affected by
soft proton flares (see Table~1 in LM08) 
and have cleaned exposure time less than 16 ks when summing MOS1 and MOS2.
Furthermore, 15 observations were excluded because they show evidence of recent 
and strong interactions (see Table~2 in LM08).
The spectral analysis of the remaining 48 exposures, for a total 
of 44 clusters, is presented in LM08 and summarized
in the next subsection.
In Table~\ref{tab:data}, we present the list of the clusters analyzed 
in the present work.

\subsection{Spatially resolved spectral analysis}

We use gas temperature profiles measured by LM08.
A detailed description of how the profiles were obtained and
tested against systematic uncertainties can be found in
their paper. Here we briefly review some of the most important points.
Unlike most temperature estimates the one reported in LM08
have been secured by performing background modelling rather than background
subtraction. Great care and considerable effort has gone into building
an accurate model of the EPIC background, both in terms of its instrumental
and cosmic components. Unfortunately the impossibility of performing an adequate
monitoring of the {\it pn} instrumental background during source observation
resulted in the exclusion of this detector from the analysis.
Therefore, we adopt the measurements obtained from the two MOS instruments
(M1 and M2, hereafter) independently in the following analysis.

The impact of small errors in the background estimates on temperature
and normalization estimates was tested both by performing Monte-Carlo
simulations (a-priori tests) and by checking how results varied
for different choices of key parameters (a-posteriori tests).
The detailed analysis allowed to track systematic errors and provide
an error budget including both statistical and systematic uncertainties.

The two profiles have been analyzed both independently and after they were
combined as described below.
M1 and M2 are cross-calibrated to about 5\% (Mateos et al 2009).
The largest discrepancy appears to be in the high energy range (above 4.5
keV), leading to a general tendency where M2 returns slightly softer
spectra than M1.  Since a similar comparison between M2 and {\sl pn}
shows that the latter returns even softer spectra, the M2 experiment
may be viewed as returning spectra which are intermediate between M1 and {\sl pn}
in the $0.7-10$ keV band.
As consequence of that, a systematic shift between the M1 and M2 temperature profiles
is present, meaning that an higher measurements is obtained with M1.
This shift is not very sensitive to the value of the temperature, but
instead manifests itself as a difference between M1 and M2 in the
shape of the radial temperature profile.
Using as reference the value of gas temperature measured with M2,
we estimate the median deviation in the different radial bins to be
4.8 \% in the inner bin, 8.9 \% in the following 4 bins, 10 \% from the 5th bin
upwards.
The two profiles are then combined by a weighted mean and a further
systematic error is added, as described in Leccardi \& Molendi (2008; see Section~5.3): 
i.e. 2, 3 and 5 per cent increases are considered at $0.3 < r/R_{180} < 0.36$,
$0.36 < r/R_{180} < 0.45$ and $r/R_{180} > 0.45$, respectively.
For this purpose, $R_{180}$ as defined in Tab.~3 in Leccardi \& Molendi (2008)
is considered. An error of the same amount is propagated in quadrature
with the statistical error.

On the other hand, no significant effect is noticed when the values of the 
normalization of the thermal model $K$ obtained from the two different instruments
are compared. The combined profile is then the direct result of the
weighted mean of the two estimates.  

Unlike in LM08, where the focus was on the measure of the $T_{\rm gas}$ profile in outer regions, 
here we need to recover a detailed description of both the $T_{\rm gas}$ 
and surface brightness $S_{\rm b}$ profiles at large and small radii. 
A significant improvement compared to the treatment by LM08
has been the correction of the spectral mixing between different annuli
caused by the finite PSF of the MOS instruments. We adopted the 
cross-talk modification of the ancillary region file (ARF) generation software
(using the {\it crossregionarf} parameter of the {\tt argen} task of SAS), treating 
the cross-talk contribution to the spectrum of a given annulus from a nearby
annulus as an additional model component (see Snowden et al. 2008).
This is a thermal model with parameters linked to the thermal spectrum
fitted to the nearby annulus and associated to the appropriate ARF
file of that region (i.e., the usual ARF familiar to X-ray astronomers).
We found the correction to be important in particular to the first two annuli 
used in the analysis.
The annuli have been therefore fitted jointly in XSPEC version 12 (Arnaud 1996), 
which allows to associate different models to different RMF and ARF files.
A comparison of the values obtained with this modelling and the values quoted
in Snowden et al. (2008) for the 16 clusters in common with our sample give
results in agreement within the errors.

\begin{table*}
\caption{Results on the mass reconstruction.}
\vspace*{-0.2cm}
\begin{center}
\begin{tabular}{l@{\hspace{.8em}} c@{\hspace{.7em}} c@{\hspace{.7em}} c 
 c@{\hspace{.5em}} c@{\hspace{.5em}} c@{\hspace{.5em}} c@{\hspace{.5em}}
 c@{\hspace{.5em}} c@{\hspace{.5em}} c@{\hspace{.5em}} c@{\hspace{.5em}} c@{\hspace{.5em}} }
\hline \\
  & & & \multicolumn{4}{c}{{\it Method 1}} & \multicolumn{5}{c}{{\it Method 2}} \\
 Cluster & $R_{\rm sp}$ & $R_{\rm xsp}$ & & $r_{\rm s}$ & $c_{200}$ & $M_{200}$ & $\chi_T^2$ (N) 
       & $r_{\rm s}$ & $c_{200}$ & $M_{200}$ & $\chi_T^2$ (N) & $\chi_n^2$ (N) \\
    & kpc & kpc & & kpc & & $10^{14} M_{\odot}$ &  & kpc & & $10^{14} M_{\odot}$ &  \\
\hline \\
RXCJ0003.8+0203 & $605$ & $414$ & & $143^{+36}_{-28}$ & $8.06^{+1.52}_{-1.30}$ & $1.90\pm0.23$ & $2.3 (7)$ & $227\pm77$ & $5.67\pm1.49$ & $2.66\pm0.76$ & $2.7 (6)$ & $7.1 (11)$ \\
Abell3911 & $836$ & $754$ & & $261^{+108}_{-59}$ & $5.59^{+1.33}_{-1.39}$ & $3.88\pm0.50$ & $10.2 (9)$ & $517\pm211$ & $3.18\pm0.79$ & $5.56\pm1.73$ & $10.8 (8)$ & $9.1 (13)$ \\
Abell3827 & $792$ & $767$ & & $390^{+89}_{-64}$ & $4.47^{+0.67}_{-0.64}$ & $6.61\pm0.73$ & $3.4 (9)$ & $345\pm66$ & $4.82\pm0.94$ & $5.74\pm0.86$ & $2.5 (8)$ & $19.8 (25)$ \\
RXCJ0049.4-2931 & $386$ & $371$ & & $71^{+30}_{-19}$ & $12.77^{+3.80}_{-3.18}$ & $0.94\pm0.16$ & $5.7 (6)$ & $104\pm17$ & $9.54\pm0.84$ & $1.26\pm0.16$ & $6.2 (5)$ & $7.6 (6)$ \\
Abell2034 & $690$ & $866$ & & $979^{+7}_{-317}$ & $2.46^{+0.81}_{-0.06}$ & $17.64\pm2.17$ & $7.7 (9)$ & $436\pm110$ & $4.16\pm0.68$ & $7.62\pm1.47$ & $9.4 (8)$ & $62.1 (22)$ \\
RXCJ1516.5-0056 & $411$ & $502$ & & $563^{+0}_{-114}$ & $2.75^{+0.49}_{-0.06}$ & $4.73\pm0.42$ & $8.1 (7)$ & $245\pm139$ & $4.91\pm2.00$ & $2.22\pm0.95$ & $4.4 (6)$ & $5.6 (4)$ \\
RXCJ2149.1-3041 & $663$ & $513$ & & $251^{+41}_{-28}$ & $4.79^{+0.43}_{-0.49}$ & $2.21\pm0.21$ & $24.6 (7)$ & $335\pm22$ & $4.08\pm0.26$ & $3.28\pm0.24$ & $8.0 (6)$ & $17.2 (11)$ \\
RXCJ1516.3+0005 & $624$ & $514$ & & $185^{+67}_{-42}$ & $7.06^{+1.64}_{-1.54}$ & $2.84\pm0.41$ & $2.8 (7)$ & $240\pm92$ & $5.81\pm1.05$ & $3.49\pm0.98$ & $2.7 (6)$ & $15.5 (7)$ \\
RXCJ1141.4-1216 & $676$ & $519$ & & $496^{+60}_{-36}$ & $3.15^{+0.19}_{-0.24}$ & $4.88\pm0.37$ & $13.0 (7)$ & $422\pm19$ & $3.50\pm0.13$ & $4.12\pm0.22$ & $3.3 (6)$ & $6.6 (6)$ \\
RXCJ1044.5-0704 & $847$ & $566$ & & $286^{+23}_{-27}$ & $4.56^{+0.34}_{-0.25}$ & $2.86\pm0.18$ & $13.3 (7)$ & $388\pm63$ & $3.71\pm0.42$ & $3.88\pm0.58$ & $8.0 (6)$ & $14.8 (9)$ \\
Abell1068 & $998$ & $1026$ & & $564^{+66}_{-49}$ & $3.02^{+0.20}_{-0.22}$ & $6.40\pm0.48$ & $8.0 (9)$ & $432\pm8$ & $3.59\pm0.05$ & $4.86\pm0.13$ & $5.0 (8)$ & $17.4 (18)$ \\
RXCJ2218.6-3853 & $764$ & $587$ & & $597^{+184}_{-166}$ & $3.16^{+0.85}_{-0.55}$ & $8.76\pm1.62$ & $9.9 (7)$ & $524\pm175$ & $3.41\pm0.78$ & $7.46\pm2.29$ & $7.6 (6)$ & $1.1 (7)$ \\
RXCJ0605.8-3518 & $893$ & $598$ & & $369^{+47}_{-39}$ & $4.10^{+0.34}_{-0.34}$ & $4.51\pm0.36$ & $17.8 (7)$ & $380\pm43$ & $4.04\pm0.32$ & $4.73\pm0.46$ & $9.4 (6)$ & $16.6 (8)$ \\
RXCJ0020.7-2542 & $695$ & $603$ & & $473^{+245}_{-154}$ & $4.17^{+1.41}_{-1.07}$ & $10.03\pm2.67$ & $17.7 (7)$ & $599\pm205$ & $3.54\pm0.82$ & $12.50\pm3.79$ & $5.2 (6)$ & $18.2 (10)$ \\
Abell1413 & $1360$ & $793$ & & $287^{+23}_{-32}$ & $5.83^{+0.57}_{-0.35}$ & $6.12\pm0.32$ & $6.3 (8)$ & $280\pm40$ & $5.84\pm0.60$ & $5.73\pm0.65$ & $5.3 (7)$ & $6.7 (14)$ \\
RXCJ2048.1-1750 & $806$ & $619$ & & $742^{+80}_{-370}$ & $2.23^{+1.63}_{-0.21}$ & $5.96\pm1.12$ & $2.7 (7)$ & $1365\pm578$ & $1.37\pm1.40$ & $8.59\pm3.01$ & $3.0 (6)$ & $22.6 (12)$ \\
RXCJ0547.6-3152 & $847$ & $624$ & & $443^{+253}_{-71}$ & $4.10^{+0.59}_{-1.17}$ & $7.89\pm1.51$ & $4.7 (7)$ & $445\pm158$ & $3.95\pm0.95$ & $7.18\pm2.20$ & $2.9 (6)$ & $10.9 (16)$ \\
Abell2204 & $858$ & $837$ & & $816^{+137}_{-0}$ & $2.81^{+0.02}_{-0.28}$ & $15.93\pm1.19$ & $58.3 (8)$ & $696\pm30$ & $3.09\pm0.09$ & $13.19\pm0.64$ & $33.7 (7)$ & $4.4 (19)$ \\
RXCJ0958.3-1103 & $1088$ & $639$ & & $872^{+260}_{-183}$ & $2.39^{+0.42}_{-0.39}$ & $11.94\pm2.02$ & $3.8 (7)$ & $942\pm223$ & $2.21\pm0.37$ & $12.03\pm3.28$ & $1.9 (6)$ & $6.4 (7)$ \\
RXCJ2234.5-3744 & $745$ & $640$ & & $506^{+261}_{-220}$ & $4.28^{+2.31}_{-1.16}$ & $13.42\pm4.15$ & $1.4 (7)$ & $648\pm271$ & $3.52\pm1.23$ & $15.72\pm6.11$ & $1.7 (6)$ & $6.7 (13)$ \\
RXCJ2014.8-2430 & $999$ & $878$ & & $462^{+59}_{-25}$ & $3.86^{+0.15}_{-0.30}$ & $7.56\pm0.53$ & $28.6 (8)$ & $589\pm70$ & $3.28\pm0.44$ & $9.57\pm1.02$ & $13.1 (7)$ & $12.0 (10)$ \\
RXCJ0645.4-5413 & $1287$ & $904$ & & $380^{+135}_{-89}$ & $4.58^{+1.06}_{-0.96}$ & $7.08\pm1.12$ & $7.7 (8)$ & $323\pm174$ & $5.01\pm1.76$ & $5.69\pm2.14$ & $12.4 (7)$ & $4.3 (13)$ \\
Abell2218 & $1024$ & $716$ & & $243^{+95}_{-79}$ & $6.26^{+2.46}_{-1.48}$ & $4.76\pm0.74$ & $11.3 (7)$ & $404\pm129$ & $4.14\pm0.62$ & $6.35\pm1.57$ & $10.0 (6)$ & $6.3 (10)$ \\
Abell1689 & $999$ & $974$ & & $211^{+22}_{-19}$ & $8.31^{+0.64}_{-0.63}$ & $7.36\pm0.44$ & $16.3 (8)$ & $238\pm28$ & $7.51\pm0.66$ & $7.88\pm0.75$ & $9.2 (7)$ & $25.3 (21)$ \\
Abell383 & $740$ & $589$ & & $435^{+95}_{-0}$ & $3.40^{+0.03}_{-0.42}$ & $4.43\pm0.37$ & $27.0 (6)$ & $505\pm81$ & $3.18\pm0.68$ & $5.71\pm0.91$ & $11.5 (5)$ & $4.1 (4)$ \\
Abell209 & $1317$ & $1069$ & & $604^{+272}_{-133}$ & $3.03^{+0.67}_{-0.77}$ & $8.60\pm1.23$ & $8.8 (8)$ & $504\pm311$ & $3.35\pm0.92$ & $6.77\pm2.87$ & $8.5 (7)$ & $12.3 (16)$ \\
Abell963 & $995$ & $813$ & & $377^{+107}_{-83}$ & $4.35^{+0.94}_{-0.76}$ & $6.17\pm0.83$ & $5.7 (7)$ & $233\pm58$ & $6.19\pm1.06$ & $4.20\pm0.72$ & $4.6 (6)$ & $7.7 (11)$ \\
Abell773 & $977$ & $846$ & & $605^{+408}_{-233}$ & $3.27^{+1.49}_{-1.05}$ & $10.94\pm3.12$ & $6.6 (7)$ & $489\pm188$ & $3.70\pm0.61$ & $8.42\pm2.88$ & $5.1 (6)$ & $12.0 (9)$ \\
Abell1763 & $1553$ & $1136$ & & $192^{+194}_{-49}$ & $7.50^{+2.30}_{-3.41}$ & $4.25\pm0.74$ & $8.1 (8)$ & $165\pm91$ & $8.23\pm1.76$ & $3.62\pm0.77$ & $3.2 (7)$ & $9.2 (12)$ \\
Abell2390 & $1322$ & $1156$ & & $1258^{+0}_{-95}$ & $2.06^{+0.12}_{-0.04}$ & $24.71\pm1.16$ & $11.6 (8)$ & $2973\pm13$ & $1.13\pm0.03$ & $53.68\pm3.48$ & $5.3 (7)$ & $15.6 (12)$ \\
Abell2667 & $966$ & $885$ & & $993^{+0}_{-48}$ & $2.24^{+0.08}_{-0.02}$ & $15.88\pm0.45$ & $4.8 (7)$ & $867\pm116$ & $2.43\pm0.27$ & $13.51\pm1.92$ & $2.1 (6)$ & $1.7 (6)$ \\
RXCJ2129.6+0005 & $883$ & $702$ & & $418^{+68}_{-37}$ & $3.71^{+0.27}_{-0.38}$ & $5.40\pm0.44$ & $3.4 (6)$ & $350\pm55$ & $4.14\pm0.46$ & $4.39\pm0.61$ & $2.0 (5)$ & $5.1 (7)$ \\
Abell1835 & $1281$ & $950$ & & $866^{+46}_{-143}$ & $2.64^{+0.34}_{-0.09}$ & $17.53\pm1.41$ & $10.9 (7)$ & $1122\pm157$ & $2.14\pm0.16$ & $20.29\pm2.89$ & $9.8 (6)$ & $6.2 (12)$ \\
RXCJ0307.0-2840 & $691$ & $951$ & & $611^{+297}_{-175}$ & $3.15^{+0.88}_{-0.78}$ & $10.44\pm2.39$ & $5.0 (7)$ & $301\pm69$ & $5.18\pm0.66$ & $5.57\pm0.94$ & $5.2 (6)$ & $2.5 (5)$ \\
Abell68 & $634$ & $746$ & & $834^{+0}_{-257}$ & $2.65^{+0.82}_{-0.06}$ & $15.96\pm1.97$ & $4.4 (6)$ & $1262\pm234$ & $1.94\pm0.25$ & $21.78\pm4.20$ & $5.0 (5)$ & $12.6 (7)$ \\
E1455+2232 & $946$ & $752$ & & $214^{+26}_{-22}$ & $6.32^{+0.53}_{-0.51}$ & $3.66\pm0.29$ & $2.7 (6)$ & $210\pm35$ & $6.25\pm0.54$ & $3.35\pm0.54$ & $1.3 (5)$ & $3.9 (7)$ \\
RXCJ2337.6+0016 & $803$ & $1004$ & & $332^{+342}_{-154}$ & $4.99^{+3.52}_{-2.18}$ & $6.81\pm1.91$ & $1.1 (7)$ & $499\pm490$ & $3.52\pm1.11$ & $8.13\pm5.31$ & $2.3 (6)$ & $1.5 (4)$ \\
RXCJ0303.8-7752 & $906$ & $1007$ & & $1115^{+14}_{-497}$ & $1.85^{+1.04}_{-0.09}$ & $13.21\pm2.33$ & $5.7 (7)$ & $563\pm525$ & $3.07\pm1.58$ & $7.82\pm5.26$ & $4.9 (6)$ & $2.5 (5)$ \\
RXCJ0532.9-3701 & $781$ & $787$ & & $278^{+170}_{-98}$ & $5.97^{+2.43}_{-1.82}$ & $6.88\pm1.83$ & $3.3 (6)$ & $325\pm195$ & $5.22\pm1.36$ & $7.39\pm4.17$ & $3.1 (5)$ & $10.2 (6)$ \\
RXCJ0232.2-4420 & $1099$ & $1032$ & & $1172^{+0}_{-409}$ & $1.80^{+0.66}_{-0.04}$ & $14.28\pm1.90$ & $12.4 (7)$ & $515\pm257$ & $3.15\pm0.64$ & $6.53\pm2.71$ & $13.0 (6)$ & $3.0 (6)$ \\
ZW3146 & $1210$ & $820$ & & $510^{+61}_{-31}$ & $3.37^{+0.15}_{-0.26}$ & $7.79\pm0.49$ & $27.5 (6)$ & $719\pm86$ & $2.64\pm0.26$ & $10.48\pm1.30$ & $18.1 (5)$ & $4.3 (15)$ \\
RXCJ0043.4-2037 & $940$ & $823$ & & $186^{+196}_{-81}$ & $7.80^{+5.05}_{-3.51}$ & $4.70\pm1.24$ & $10.7 (6)$ & $142\pm85$ & $9.16\pm2.67$ & $3.44\pm0.88$ & $7.7 (5)$ & $2.4 (7)$ \\
RXCJ0516.7-5430 & $821$ & $1061$ & & $785^{+405}_{-472}$ & $2.41^{+2.82}_{-0.75}$ & $10.44\pm2.88$ & $1.8 (7)$ & $462\pm261$ & $3.49\pm0.84$ & $6.46\pm1.45$ & $2.1 (6)$ & $2.3 (7)$ \\
RXCJ1131.9-1955 & $1285$ & $1091$ & & $797^{+494}_{-309}$ & $2.43^{+1.16}_{-0.76}$ & $11.31\pm2.50$ & $6.9 (7)$ & $1839\pm629$ & $1.27\pm0.84$ & $19.82\pm5.26$ & $3.0 (6)$ & $4.1 (5)$ \\
\hline \\
\end{tabular}

\end{center}
\label{tab:all}
\tablefoot{
We quote the name of the object, the upper limit of the radial range investigated 
in the spatial ($R_{\rm sp}$) and spectral analysis ($R_{\rm xsp}$),
the best-fit values of the scale radius,
the concentration parameters, $M_{200}$ and
minimum $\chi^2$ with the corresponding degrees of freedom.
In the case of \metii, we quote two minimum $\chi^2$, corresponding
to the minima obtained from the simultaneous fits of the temperature ($\chi_T^2$)
and gas density ($\chi_n^2$) profiles.
}
\end{table*}

\subsection{Analysis of the surface brightness profile}

We extend the spectral analysis presented in LM08 with
a spatial analysis of the combined exposure--corrected M1-M2
images.
 
We extract surface brightness profiles from MOS images in the energy band 
$0.7-1.2$ keV, in order to keep the background as low as possible with respect 
to the source. For this reason, we avoid the intense fluorescent instrumental 
lines of Al ($\sim 1.5$ keV) and Si ($\sim 1.8$ keV) (LM08). 
To correct for the vignetting, we divide the images by the corresponding exposure maps.
From the surface brightness profiles, we subtract the background that is estimated
starting from the spectral modelling of the background components in the external ring 
10--12 arcmin (see LM08 for details on the adopted models). 
We recall here that in the procedure of LM08 the normalizations of the background 
components are the only free parameters of the fit and that the galactic foreground 
emission, the cosmic X-ray background and the cosmic ray induced continuum give 
a significant contribution in the $0.7-1.2$ keV energy range.
The intensities of the background components in the annulus 10--12 arcmin 
are given by the count rates predicted by the best fit spectral model in this region. 
In order to associate errors to these count rates, we perform a simulation within XSPEC: 
we allow the normalizations of the background components to vary randomly within 
their errors, we obtain the count rates associated to this fake model and we iterate 
this procedure. 
The error on the level of the background components is the width of 
the distribution of the simulated count rates.
Using these values in the outer annulus, we reconstruct the background profile at all radii. 
The ``photon'' components (CXB and galactic foreground) are affected by vignetting 
in the same way as the source photons and, therefore, dividing by the exposure map 
effectively corrects also these background components for the vignetting.
In order to  reconstruct the ``photon'' background profile, 
it is thus sufficient to rescale the count rate for the mean vignetting in the 
outer annulus (constant blue profile in Fig.~\ref{fig:bkg_psf}). 
On the contrary, the instrumental background does not suffer from vignetting 
and, therefore, dividing the image by the exposure map ``mis-corrects'' this component. 
In order to consider this effect, we divide the corresponding count rate by 
the vignetting profile (that we derive from the exposure map in the $0.7-1.2$ keV), 
obtaining the growing green curve in Fig.~\ref{fig:bkg_psf}. 
The total background profile (red line in Fig.~\ref{fig:bkg_psf}) is the sum 
of the photon (blue) and instrumental (green) profiles. 

The surface brightness profiles $S_b(r)$ have been first extracted from the 
combined images and binned by requiring a fixed number of 200 counts in each radial bin
to preserve all the spatial information available.
After the background subtraction, they have been corrected for the PSF smearing.
For this purpose, a sum of a cusped $\beta-$model and of a $\beta-$model 
(Cavaliere \& Fusco-Femiano 1978)
with seven free parameters, $f_m(r) = a_0 \times \left[
x_1^{-a_2} \times \left( 1+x_1^2 \right)^{0.5-3 a_3+a_2/2} +a_4 \left( 1+x_2^2 \right)^{0.5-3 a_6}
\right]$ (with $x_1 = r/a_1$ and $x_2 = r/a_2$),
is convolved with the predicted PSF (Ghizzardi 2001)
and fitted to the observed profile background--subtracted $S_b(r)$ to obtain the best-fit
convolved model $f_c({\bar{a_i}}; r)$.
Finally, to correct $S_b(r)$ for the PSF-convolution, we apply 
a correction at each radius $\hat{r}$ where $S_b(r)$ is measured 
equal to the ratio $f_m({\bar{a_i}}; \hat{r}) / f_c({\bar{a_i}}; \hat{r})$.
An example of the results of the procedure is shown in Fig.~\ref{fig:bkg_psf}.
These corrected profiles are, finally, used in the following analysis
up to the radial limit, $R_{\rm sp}$, beyond which the ratio between the profile and the 
error on it (including the estimated uncertainty on the measurement of the background) is below 2.

\begin{figure*}
\hbox{
 \epsfig{figure=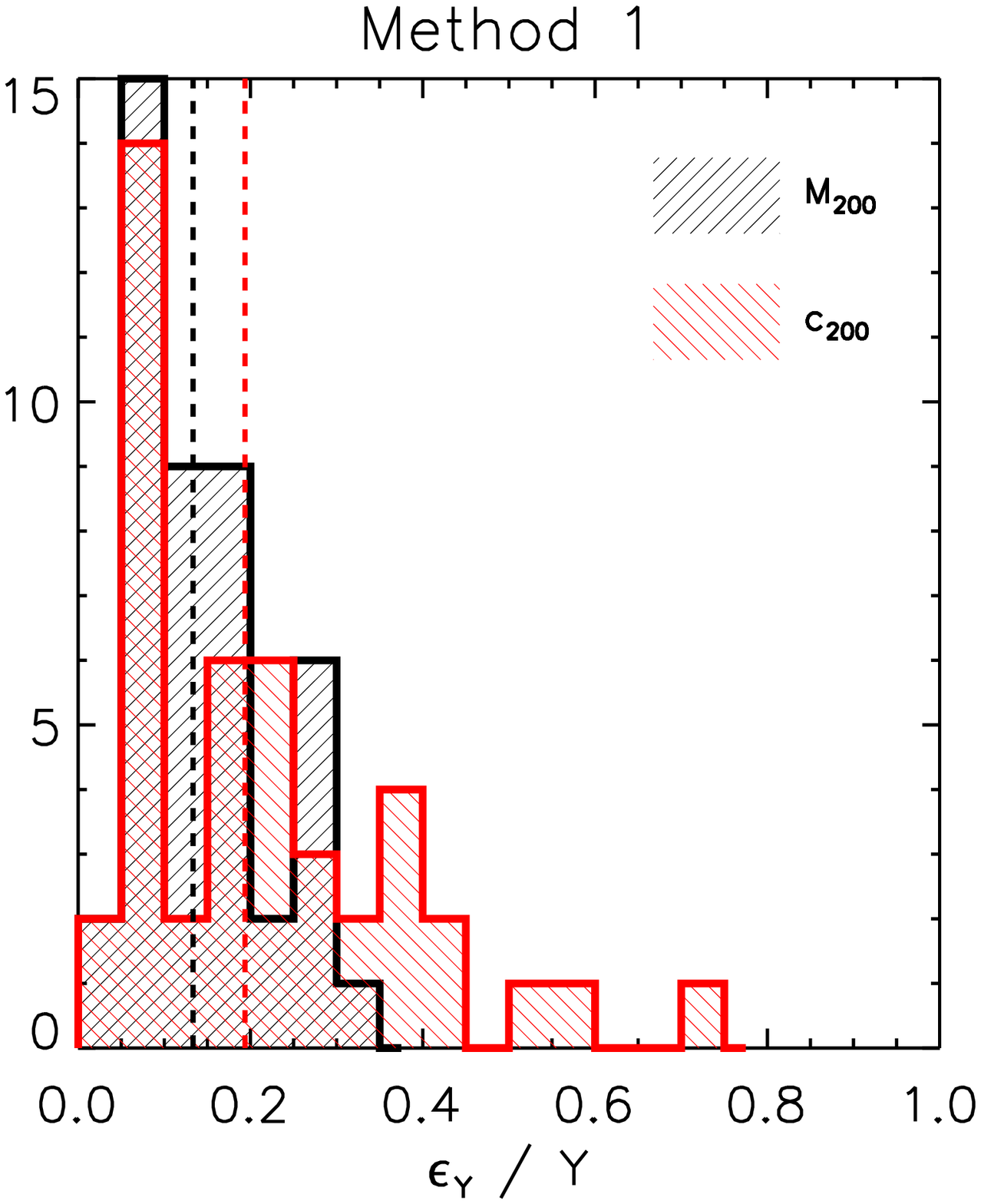,width=0.25\textwidth}
 \epsfig{figure=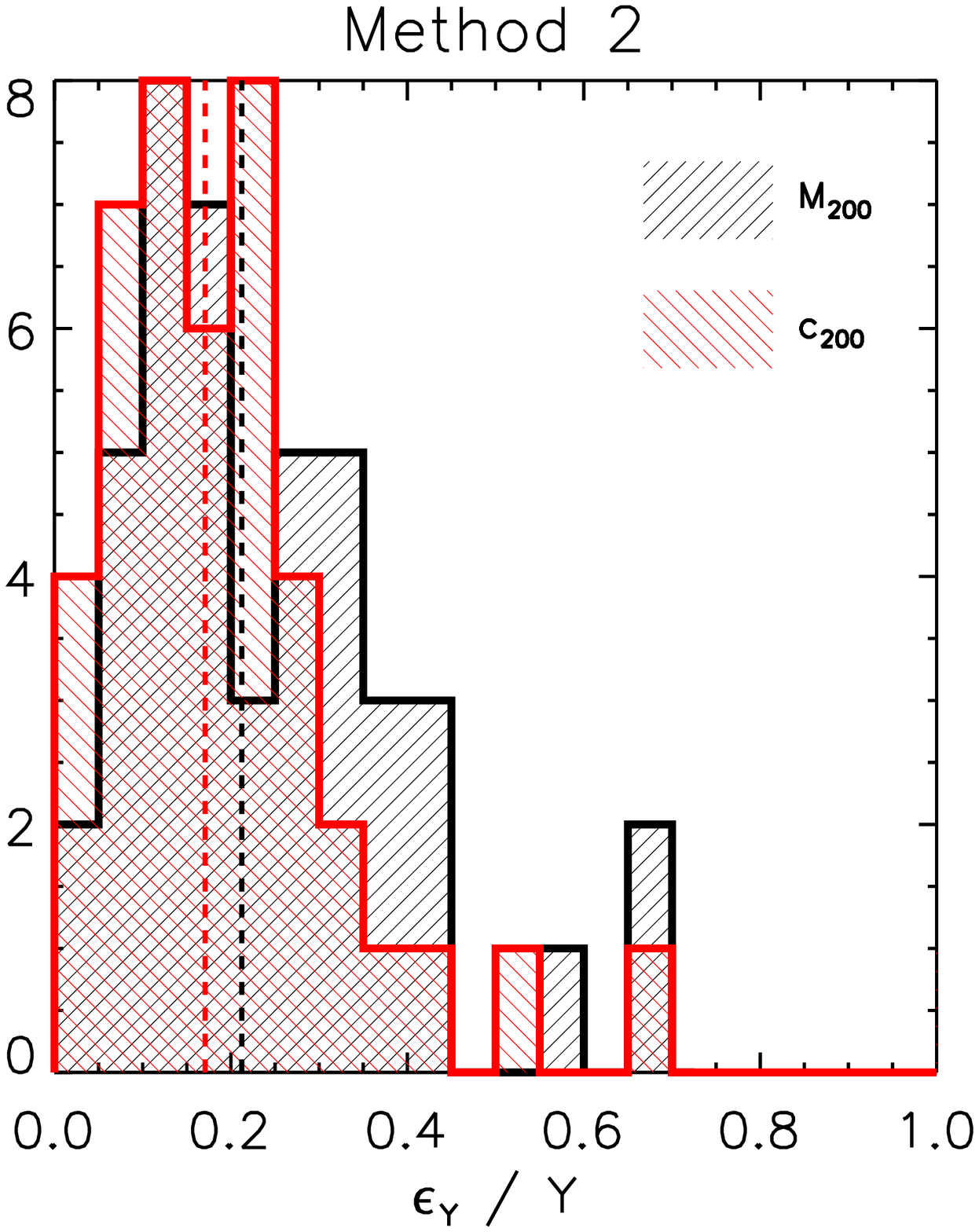,width=0.25\textwidth}
%} \hbox{
 \epsfig{figure=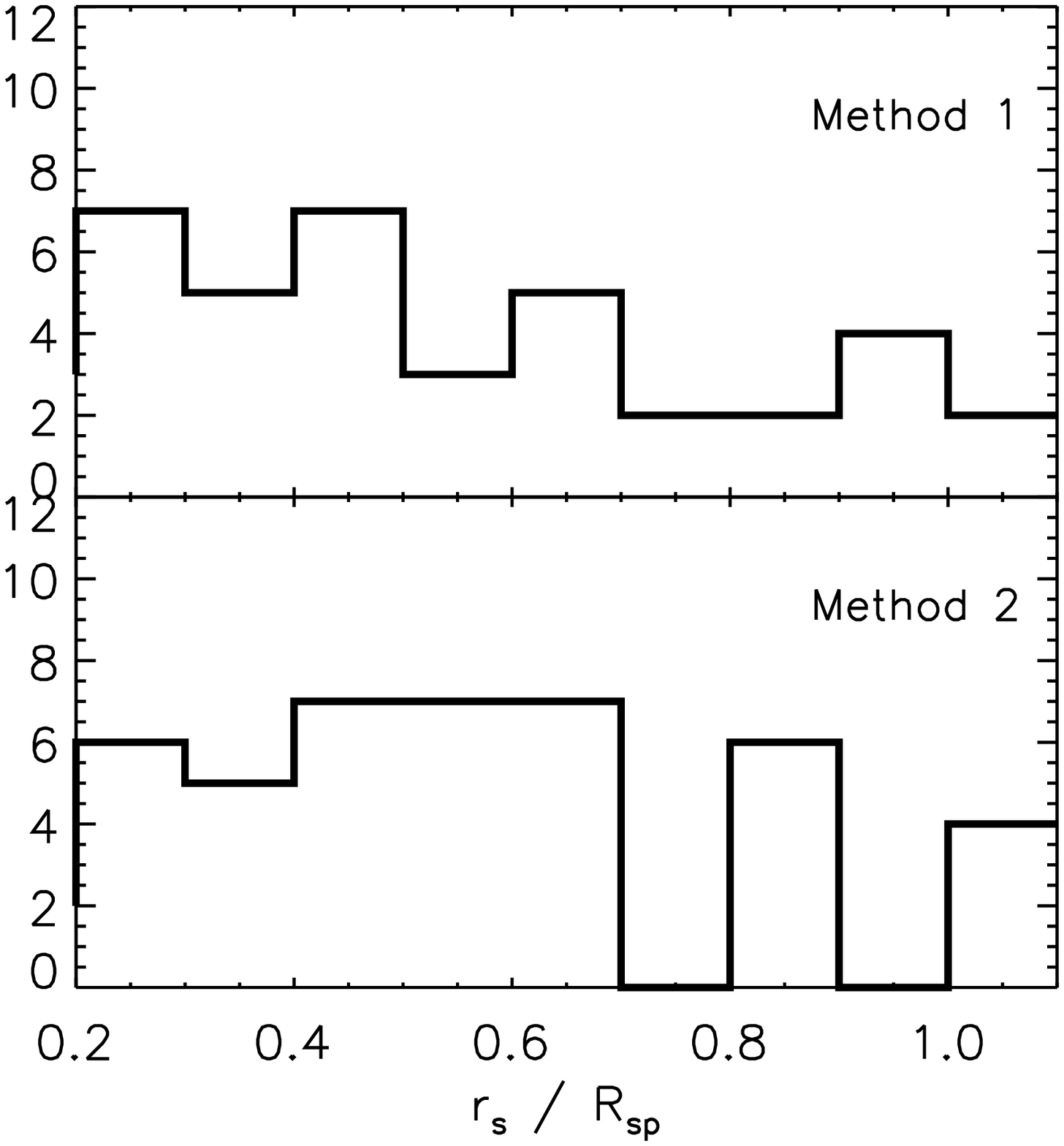,width=0.25\textwidth}
 \epsfig{figure=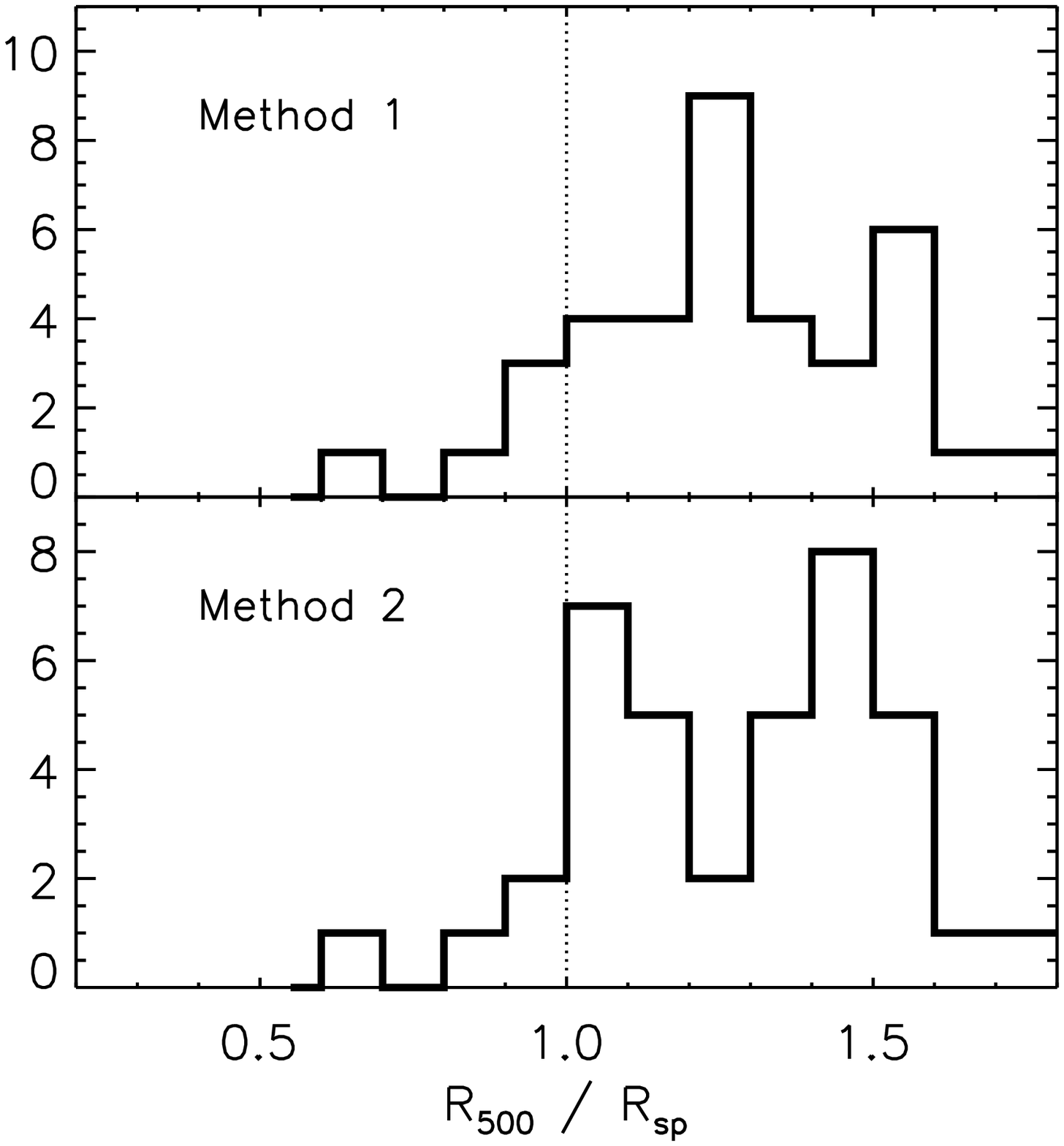,width=0.25\textwidth}
} \caption{
%({\it Upper panels})
({\it First 2 panels on the left}) Relative errors on $M_{200}$ (black) and $c_{200}$ (red)
estimated by the two methods. The median values are indicated by a dashed line.
({\it 3rd and 4th panel from left}) Ratios between the best-fit result on the scale radius $r_{\rm s}$
({\it 3rd panel}) and on $R_{500}$ ({\it 4th panel}) and outermost radius reached with the spatial analysis.
} \label{fig:erry}
\end{figure*}

\begin{table*}
\caption{Estimates of $R_{200}$, $R_{500}$ and the gas mass fraction.}
\vspace*{-0.2cm}
\begin{center}
\begin{tabular}{l@{\hspace{.8em}} c@{\hspace{.7em}} c@{\hspace{.7em}} c@{\hspace{.7em}} 
  c@{\hspace{.7em}} c@{\hspace{.7em}} c@{\hspace{.7em}} }
\hline \\
  & \multicolumn{3}{c}{{\it Method 1}} & \multicolumn{3}{c}{{\it Method 2}} \\
 Cluster & $R_{200}$ & $R_{500}$ & $f_{\rm gas}$ & $R_{200}$ & $R_{500}$ & $f_{\rm gas}$ \\
    & kpc & kpc & $<R_{500}$  & kpc & kpc & $<R_{500}$ \\
\hline \\
RXCJ0003.8+0203 & $1231\pm65$ & $824\pm38$ & $0.117\pm0.049$ & $1360\pm122$ & $899\pm50$ & $0.097\pm0.010$ \\
Abell3911 & $1589\pm88$ & $1044\pm41$ & $0.146\pm0.017$ & $1773\pm155$ & $1130\pm75$ & $0.126\pm0.015$ \\
Abell3827 & $1894\pm84$ & $1228\pm45$ & $0.140\pm0.012$ & $1823\pm87$ & $1184\pm40$ & $0.147\pm0.008$ \\
RXCJ0049.4-2931 & $980\pm59$ & $666\pm37$ & $0.143\pm0.020$ & $1071\pm39$ & $721\pm29$ & $0.123\pm0.008$ \\
Abell2034 & $2491\pm140$ & $1569\pm75$ & $0.073\pm0.007$ & $1957\pm108$ & $1267\pm114$ & $0.123\pm0.016$ \\
RXCJ1516.5-0056 & $1668\pm65$ & $1039\pm38$ & $0.105\pm0.009$ & $1309\pm159$ & $845\pm98$ & $0.120\pm0.015$ \\
RXCJ2149.1-3041 & $1298\pm52$ & $846\pm30$ & $0.131\pm0.027$ & $1452\pm36$ & $942\pm26$ & $0.101\pm0.006$ \\
RXCJ1516.3+0005 & $1416\pm95$ & $940\pm54$ & $0.142\pm0.074$ & $1502\pm107$ & $991\pm81$ & $0.122\pm0.015$ \\
RXCJ1141.4-1216 & $1635\pm55$ & $1047\pm31$ & $0.086\pm0.012$ & $1551\pm27$ & $1003\pm23$ & $0.095\pm0.005$ \\
RXCJ1044.5-0704 & $1399\pm35$ & $923\pm19$ & $0.146\pm0.009$ & $1531\pm72$ & $996\pm21$ & $0.119\pm0.007$ \\
Abell1068 & $1772\pm57$ & $1140\pm31$ & $0.091\pm0.007$ & $1645\pm13$ & $1061\pm7$ & $0.105\pm0.002$ \\
RXCJ2218.6-3853 & $1991\pm159$ & $1275\pm84$ & $0.099\pm0.018$ & $1900\pm167$ & $1222\pm132$ & $0.107\pm0.024$ \\
RXCJ0605.8-3518 & $1613\pm64$ & $1057\pm29$ & $0.133\pm0.012$ & $1643\pm49$ & $1071\pm25$ & $0.125\pm0.006$ \\
RXCJ0020.7-2542 & $2023\pm228$ & $1329\pm124$ & $0.062\pm0.016$ & $2182\pm200$ & $1415\pm82$ & $0.060\pm0.009$ \\
Abell1413 & $1837\pm64$ & $1207\pm21$ & $0.161\pm0.010$ & $1809\pm58$ & $1188\pm28$ & $0.167\pm0.007$ \\
RXCJ2048.1-1750 & $1792\pm155$ & $1110\pm80$ & $0.132\pm0.044$ & $2008\pm269$ & $1187\pm109$ & $0.114\pm0.020$ \\
RXCJ0547.6-3152 & $1921\pm161$ & $1251\pm85$ & $0.105\pm0.057$ & $1882\pm168$ & $1219\pm81$ & $0.116\pm0.013$ \\
Abell2204 & $2450\pm79$ & $1549\pm44$ & $0.115\pm0.008$ & $2319\pm33$ & $1477\pm47$ & $0.126\pm0.007$ \\
RXCJ0958.3-1103 & $2183\pm174$ & $1366\pm87$ & $0.086\pm0.013$ & $2191\pm174$ & $1363\pm106$ & $0.087\pm0.014$ \\
RXCJ2234.5-3744 & $2237\pm293$ & $1474\pm164$ & $0.079\pm0.067$ & $2377\pm294$ & $1542\pm159$ & $0.085\pm0.025$ \\
RXCJ2014.8-2430 & $1935\pm56$ & $1245\pm32$ & $0.136\pm0.014$ & $2067\pm70$ & $1323\pm16$ & $0.120\pm0.004$ \\
RXCJ0645.4-5413 & $1919\pm133$ & $1243\pm65$ & $0.161\pm0.020$ & $1811\pm183$ & $1174\pm83$ & $0.177\pm0.022$ \\
Abell2218 & $1671\pm120$ & $1100\pm53$ & $0.159\pm0.019$ & $1820\pm120$ & $1122\pm66$ & $0.154\pm0.016$ \\
Abell1689 & $1892\pm40$ & $1279\pm24$ & $0.156\pm0.008$ & $1946\pm54$ & $1304\pm21$ & $0.151\pm0.005$ \\
Abell383 & $1577\pm79$ & $1015\pm39$ & $0.121\pm0.042$ & $1697\pm100$ & $1090\pm17$ & $0.101\pm0.005$ \\
Abell209 & $2006\pm125$ & $1267\pm57$ & $0.146\pm0.015$ & $1873\pm197$ & $1196\pm54$ & $0.160\pm0.013$ \\
Abell963 & $1750\pm95$ & $1153\pm50$ & $0.137\pm0.015$ & $1586\pm74$ & $1049\pm36$ & $0.164\pm0.011$ \\
Abell773 & $2100\pm257$ & $1350\pm130$ & $0.116\pm0.041$ & $1959\pm170$ & $1140\pm92$ & $0.156\pm0.019$ \\
Abell1763 & $1644\pm105$ & $1079\pm52$ & $0.212\pm0.025$ & $1575\pm88$ & $1028\pm39$ & $0.213\pm0.012$ \\
Abell2390 & $2735\pm63$ & $1695\pm36$ & $0.108\pm0.013$ & $3484\pm67$ & $2026\pm57$ & $0.079\pm0.005$ \\
Abell2667 & $2374\pm36$ & $1478\pm22$ & $0.114\pm0.018$ & $2259\pm103$ & $1417\pm72$ & $0.118\pm0.013$ \\
RXCJ2129.6+0005 & $1711\pm60$ & $1099\pm30$ & $0.165\pm0.012$ & $1619\pm63$ & $1042\pm16$ & $0.177\pm0.006$ \\
Abell1835 & $2433\pm86$ & $1540\pm46$ & $0.120\pm0.012$ & $2539\pm100$ & $1583\pm34$ & $0.109\pm0.006$ \\
RXCJ0307.0-2840 & $2030\pm199$ & $1302\pm103$ & $0.105\pm0.017$ & $1695\pm78$ & $1114\pm59$ & $0.147\pm0.017$ \\
Abell68 & $2293\pm127$ & $1457\pm71$ & $0.079\pm0.008$ & $2549\pm165$ & $1489\pm155$ & $0.082\pm0.020$ \\
E1455+2232 & $1484\pm46$ & $980\pm26$ & $0.160\pm0.013$ & $1445\pm59$ & $954\pm14$ & $0.163\pm0.006$ \\
RXCJ2337.6+0016 & $1779\pm192$ & $1178\pm96$ & $0.148\pm0.027$ & $1894\pm278$ & $1225\pm173$ & $0.141\pm0.033$ \\
RXCJ0303.8-7752 & $2191\pm179$ & $1347\pm93$ & $0.116\pm0.016$ & $1888\pm301$ & $1203\pm122$ & $0.148\pm0.024$ \\
RXCJ0532.9-3701 & $1784\pm179$ & $1186\pm102$ & $0.141\pm0.027$ & $1835\pm233$ & $1207\pm105$ & $0.136\pm0.019$ \\
RXCJ0232.2-4420 & $2230\pm141$ & $1380\pm71$ & $0.123\pm0.013$ & $1798\pm167$ & $1152\pm137$ & $0.178\pm0.032$ \\
ZW3146 & $1875\pm49$ & $1206\pm26$ & $0.159\pm0.010$ & $2040\pm77$ & $1293\pm22$ & $0.135\pm0.005$ \\
RXCJ0043.4-2037 & $1604\pm157$ & $1068\pm82$ & $0.176\pm0.032$ & $1472\pm95$ & $982\pm133$ & $0.199\pm0.033$ \\
RXCJ0516.7-5430 & $2029\pm246$ & $1273\pm114$ & $0.127\pm0.022$ & $1767\pm112$ & $1135\pm57$ & $0.157\pm0.013$ \\
RXCJ1131.9-1955 & $2121\pm206$ & $1325\pm93$ & $0.155\pm0.023$ & $2513\pm271$ & $1475\pm97$ & $0.120\pm0.015$ \\
\hline \\
\end{tabular}

\end{center}
\label{tab:fgas}
\tablefoot{These estimates refer to the mass models obtained with two different methods 
(see Table~\ref{tab:all}) and are evaluated at the ovedensities determined 
from the {\it total} (i.e. dark$+$gas) mass profiles.
All the quoted errors are at $1 \sigma$ level.
}
\end{table*}

\begin{figure*}
 \epsfig{figure=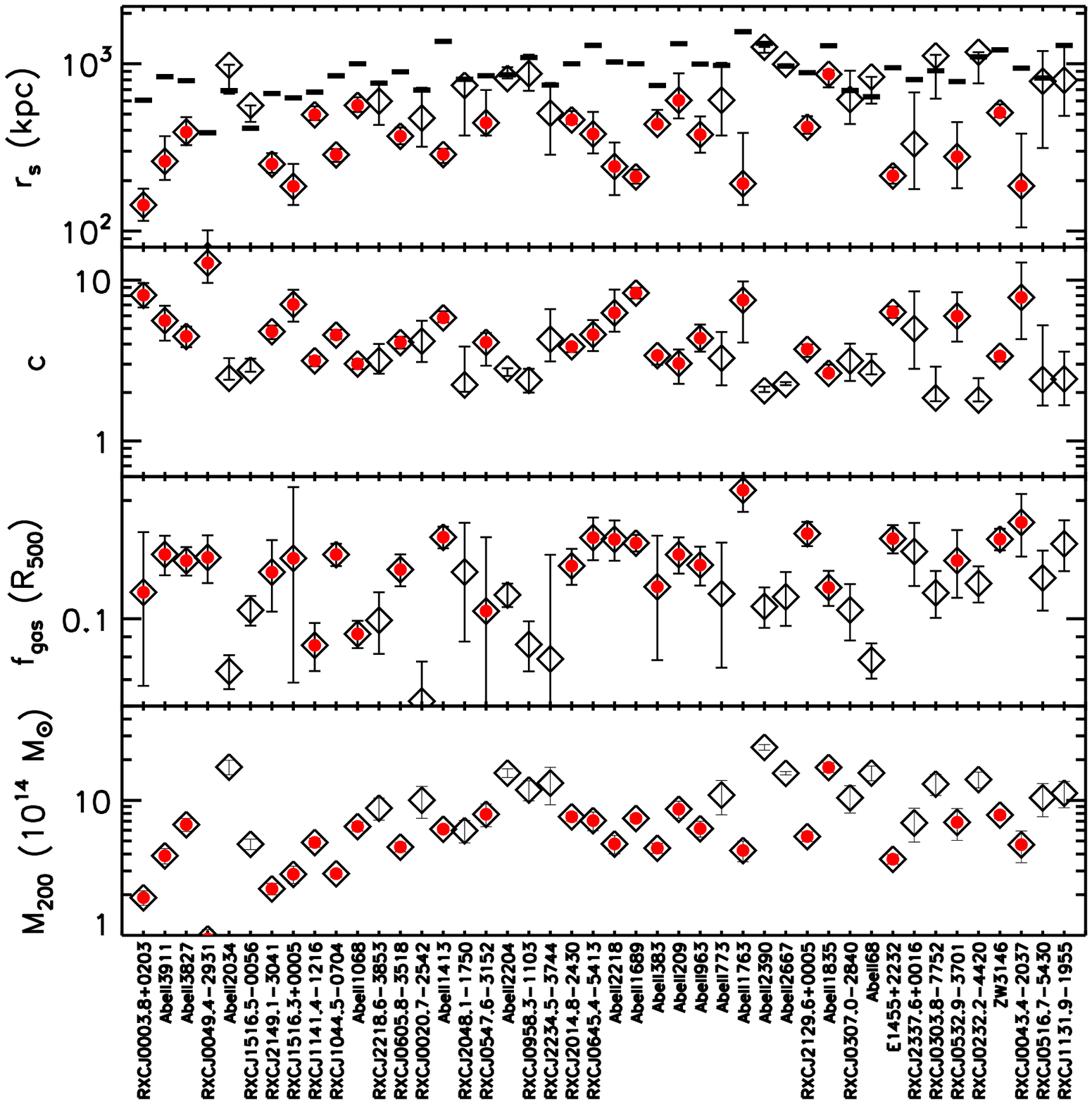,width=\textwidth}
\caption{Best-fit values of $r_{\rm s}$, $c$, gas mass fraction $f_{\rm gas}$ at $R_{500}$
and dark mass $M_{\rm DM}$ within $R_{200}$ as obtained from \meti\
(diamonds; the ones including red points indicate the objects
where the condition $(r_{\rm s}+\epsilon_{r_{\rm s}}) < R_{\rm sp}$
is satisfied by both methods.
$R_{\rm sp}$ here plotted as horizontal line in the upper panel).
}
\label{fig:cxo}
\end{figure*}

\section{Estimates of the mass profiles}

We use the profiles of the spectroscopically determined ICM temperature and 
of the PSF--corrected surface brightness estimated, as
described in the previous section, to recover the X-ray gas,
the dark and the total mass profiles, 
under the assumptions of the spherical geometry distribution
of the Intracluster Medium (ICM) and that the hydrostatic equilibrium 
holds between ICM and the underlying gravitational potential.
We apply the two following different methods: 

\begin{itemize}

\item (\meti)
This technique is described in Ettori et al. (2002) and has been widely used to recover
the mass profiles in recent X-ray studies of both observational (e.g. Morandi et al. 2007; 
Donnarumma et al. 2009, 2010) and simulated datasets 
(e.g. Rasia et al. 2006; Meneghetti et al. 2010)
against which it has been thoroughly tested.

We summarize here the algorithm adopted and how it uses the observed measurements.
Starting from the X-ray surface brightness profile and the radially resolved
spectroscopic temperature measurements, this method
puts constraints on the parameters of the functional form 
describing the dark matter $M_{\rm DM}$, defined as the total mass minus
the gas mass (we neglect the marginal contribution from the mass in stars that 
amounts to about 10-15 \% of the gas mass in massive systems 
--see, e.g., discussion in Ettori et al. 2009 and Andreon 2010--, 
and is here formally included in the $M_{\rm DM}$ term).
In the present work, we adopt a NFW profile: 
\begin{eqnarray}
\lefteqn{ M_{\rm DM}(<r) = M_{\rm tot}(<r) - M_{\rm gas}(<r) = 
4 \pi \ r_{\rm s}^3 \ \rho_{\rm s} \ f(x), }
\nonumber \\
\lefteqn{ \rho_{\rm s} =  \rho_{c, z} \frac{200}{3} \frac{c^3}
{\ln (1+c) -c/(1+c)} },  \nonumber \\
\lefteqn{ f(x) =  \ln (1+x) - \frac{x}{1+x} } , 
\label{eq:mass_nfw}
\end{eqnarray}
where $x = r/r_{\rm s}$, $\rho_{c, z} = 3 H_z^2 / 8\pi G$ is the critical density 
at the cluster's redshift $z$, $H_z = H_0 \times \left[ \Omega_{\Lambda} +\Omega_{\rm m} 
(1+z)^3 \right]^{1/2}$ is the Hubble constant at redshift $z$ 
for an assumed flat Universe ($\Omega_{\rm m}+\Omega_{\Lambda}=1$),
and the relation $R_{200} = c_{200} \times r_{\rm s}$ holds.

The two parameters $(r_{\rm s}, c_{200})$ are constrained by minimizing
a $\chi^2$ statistic defined as 
\begin{equation}
\chi^2_T = \sum_i\frac{ \left(T_{\rm data, i} - T_{\rm model, i} \right)^2}{\epsilon_{T, i}^2}
\end{equation}
where the sum is done over the annuli of the spectral analysis; 
$T_{\rm data}$ are the either deprojected or observed temperature
measurements obtained in the spectral analysis; $T_{\rm model}$ are
either the three-dimensional or projected values of the estimates of $T_{\rm gas}$ 
recovered from the inversion of the hydrostatic equilibrium equation 
(see below) for a given gas density and total mass profiles;
$\epsilon_T$ is the error on the spectral measurements.
The gas density profile, $n_{\rm gas}$, is estimated from the geometrical deprojection
(Fabian et al. 1981, Kriss et al. 1983, McLaughlin 1999, Buote 2000, Ettori et al. 2002)
of either the measured X-ray surface brightness or the estimated normalization
of the thermal model fitted in the spectral analysis (see Fig.~\ref{fig:meth1}).
In the present study, we consider the observed spectral values of the temperature and
evaluate $T_{\rm model}$ by projecting the estimates of $T_{\rm gas}$ over 
the annuli adopted in the spectral analysis accordingly to the recipe 
in Mazzotta et al. (2004) and using the gas density profile
obtained from the deprojection of the PSF--deconvolved surface brightness profile
(see Sect.~2.2). 
We exclude the deprojected data of the gas density within a cutoff radius
of 50 kpc because the influence of the central galaxy is expected to be 
not negligible, in particular for strong low-entropy core systems.
The values of $T_{\rm gas}$ are then obtained from 
\begin{eqnarray}
-G \mu m_{\rm a} \frac{n_{\rm gas} M_{\rm tot}(<r)}{r^2} =
\frac{d\left(n_{\rm gas} \times T_{\rm gas} \right)}{dr},  
\label{eq:mtot}
\end{eqnarray}
where $G$ is the universal gravitational constant, $m_{\rm a}$ is the atomic mass unit and
$\mu$=0.61 is the mean molecular weight in atomic mass unit.
To solve this differential equation, we need to define a boundary condition that is here
fixed to the value of the pressure measured in the outermost point of the gas density 
profile, $P_{\rm out} = P_{\rm gas}(R_{\rm sp}) = n_{\rm gas}(R_{\rm sp}) \times
T_{\rm gas}(R_{\rm sp})$, where $T_{\rm gas}(R_{\rm sp})$ is estimated by linear 
extrapolation in the logarithmic space, if required. The systematic uncertainties
introduced by this assumption on $P_{\rm out}$ are discussed in the next section.
Note that by applying \meti\ the errors on the gas density 
do not propagate into the estimates of the parameters of the mass profile 
and are used both to define the range of the accepted values of $P_{\rm out}$ 
and to evaluate the uncertainties on the gas mass profiles.
The allowed range at $1 \sigma$ of the two interesting parameters, 
$r_{\rm s}$ and $c_{200}$, is defined from the minimum and the maximum 
of the values that permit $\chi^2_T$ to be lower or equal to $min(\chi^2_T)+1$.
The average error on the mass is then the mean of the upper and lower limit 
obtained at each radius from the allowed ranges at $1 \sigma$ of $r_{\rm s}$ and $c_{200}$.
Only for the purpose of estimating the profile of $M_{\rm gas}(<r)$, and eventually to provide
the extrapolated values, the deprojected gas density profile 
is fitted with the generic functional form described in Ettori et al. (2009) 
and adapted from the one described in Vikhlinin et al. (2006), 
$n_{\rm gas} = n_{\rm gas, 0} \, (r/r_{\rm c,0})^{-\alpha_0} 
\times \left(1 + (r/r_{\rm c,0})^2 \right)^{-1.5 \alpha_1 +\alpha_0/2} 
 \times \left(1 + (r/r_{\rm c,1})^{\alpha_2} \right)^{-\alpha_3/\alpha_2}$.

\item (\metii)
The second method follows the approach described in Humphrey et al. (2006)
and Gastaldello et al. (2007) where further details of this technique (
in particular in Appendix B of Gastaldello et al. 2007) are provided. We assume
parametrizations for the gas density and mass profiles to calculate the gas
temperature assuming hydrostatic equilibrium,

\begin{equation}
T(r)=
T_0\frac{n_{\rm gas, 0}}{n_{\rm gas}(r)} - \frac{\mu m_{\rm a} G}{k_B n_{\rm gas}(r)}
\int_{r_0}^{r}\frac{n_{\rm gas} M_{\rm tot}\, dr}{r^2},
\label{eqn_hydrostatic_t}
\end{equation}

where $n_{\rm gas}$ is the gas density, $n_{\rm gas, 0}$ and $T_0$ are density and temperature
at some ``reference'' radius $r_0$ and $k_B$ is Boltzmann's constant.
The $n_{\rm gas}$ and $T(r)$ profiles are fitted simultaneously to the
data to constrain the parameters of the gas density and mass
models. The parameters of the mass model are obtained from fitting the gas 
density and temperature data and goodness-of-fit for any mass model can be
assessed directly from the residuals of the fit. The quality of the data, 
in particular of the temperature profile, motivated the use of this approach rather
than the default approach of parametrizing the temperature and mass profiles
to calculate the gas density used in Gastaldello et al. (2007).
We projected the parametrized models of the three-dimensional quantities, 
$n_{\rm gas}^2$ and $T$, and fitted these projected emission-weighted models 
to the results obtained from our analysis of the data projected on the sky. 
With respect to the paper cited above, the XSPEC normalization have been
derived converting the XMM surface brightness in the $0.7-1.2$ keV band using 
the effective area and observed projected temperature and metallicity obtained
in the wider radial bins used for spectral extraction. 
The models have been integrated over each radial bin (rather than only evaluating at a
single point within the bin) to provide a consistent comparison.
We considered an NFW profile of eq.~\ref{eq:mass_nfw} for fitting 
the total mass and two models for fitting the gas density profile: 
the $\beta$ model (Cavaliere \& Fusco-Femiano 1978)
%\begin{equation}
%  n_{\rm gas} = n_{\rm gas, 0} \Big[1+\Big(\frac{r}{r_c}\Big)^2\Big]^{-\frac{3}{2}\beta},
%\end{equation}
%where $n_{\rm gas, 0}$ is the central gas density, $r_c$ the core radius,
%and the asymptotic slope is $-3\beta$; and 
a double $\beta$ model in which a common value of beta is assumed, and
a cusped $\beta$ model (Pratt \& Arnaud 2002; Lewis et al. 2003).
%the latter of which is a modified $\beta$
%model allowing for steepening of the profile in the inner regions ($r<r_c$),
%\begin{equation}
%   n_{\rm gas} = n_{\rm gas, c} 2^{3\beta/2-\alpha_c/2}\left(\frac{r}{r_c}\right)^{-\alpha_c}
%   \left[1+\left(\frac{r}{r_c}\right)^2\right]^{-\frac{3}{2}\beta+\frac{\alpha_c}{2}},
%\label{cusp_beta}
%\end{equation}
%where the exponent $\alpha_c$ is the slope of the power-law cusp at
%small radii and $n_{\rm gas, c}=n_{\rm gas}(r_c)$.
The last two models have been introduced to account for the sharply peaked
surface brightness in the centers of relaxed X-ray systems and they provide the
necessary flexibility to parametrize adequately the shape of the gas density 
profiles of the objects in our sample when the traditional $\beta$ model fails
in fitting the data.

\end{itemize}

\begin{figure*}
\hbox{
 \epsfig{figure=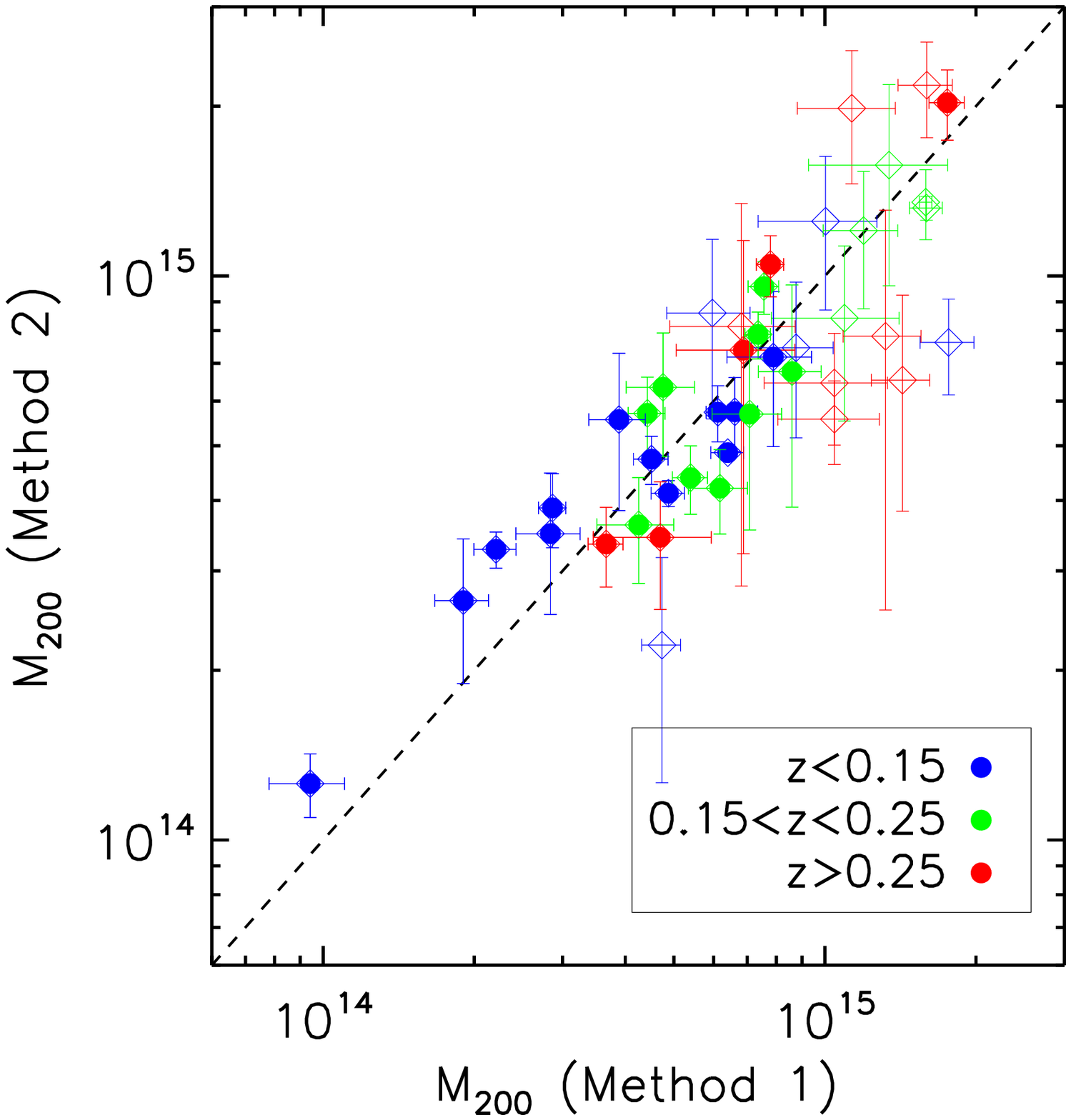,width=0.33\textwidth}
 \epsfig{figure=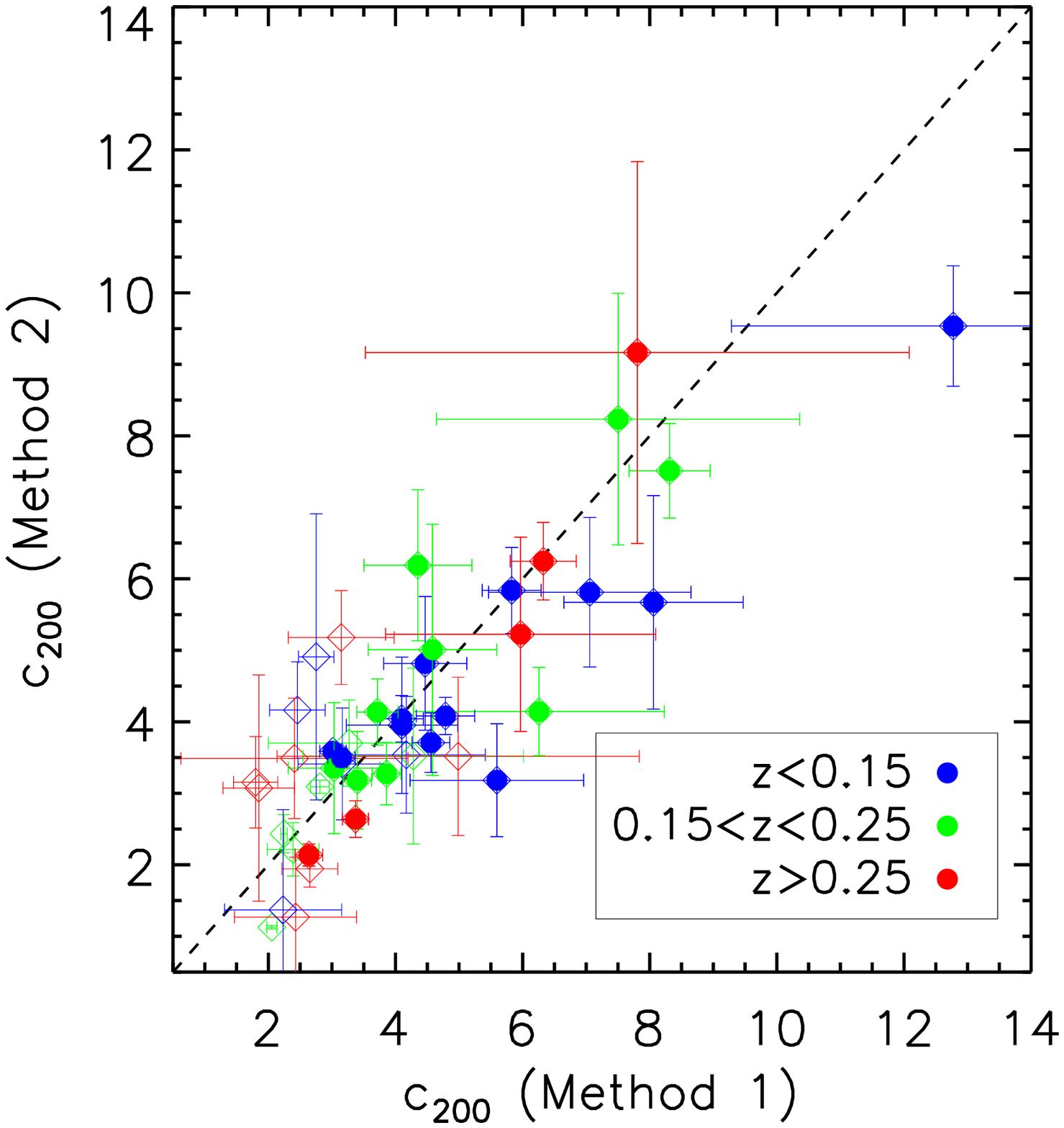,width=0.33\textwidth}
 \epsfig{figure=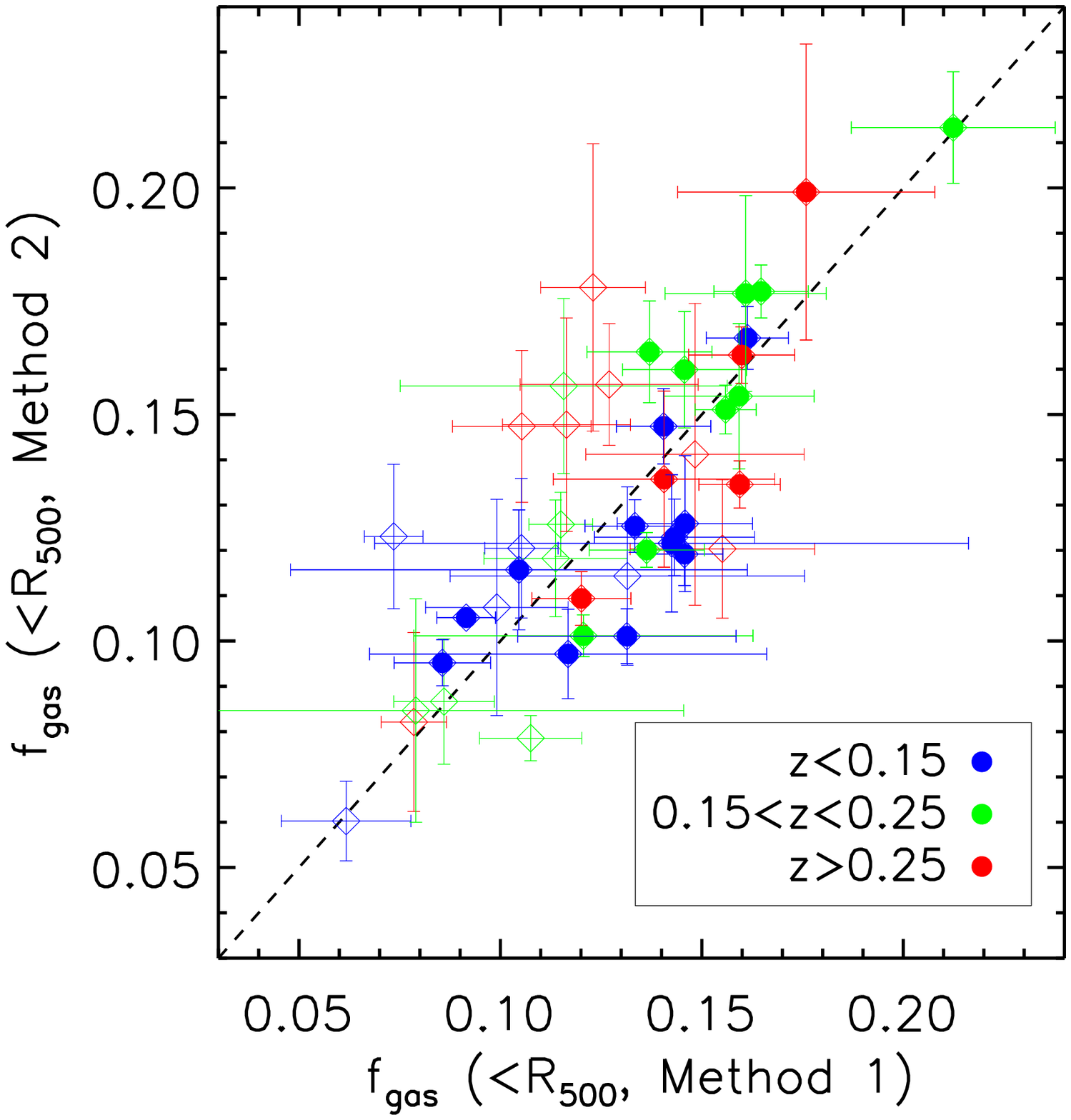,width=0.33\textwidth}
}
\hbox{
 \epsfig{figure=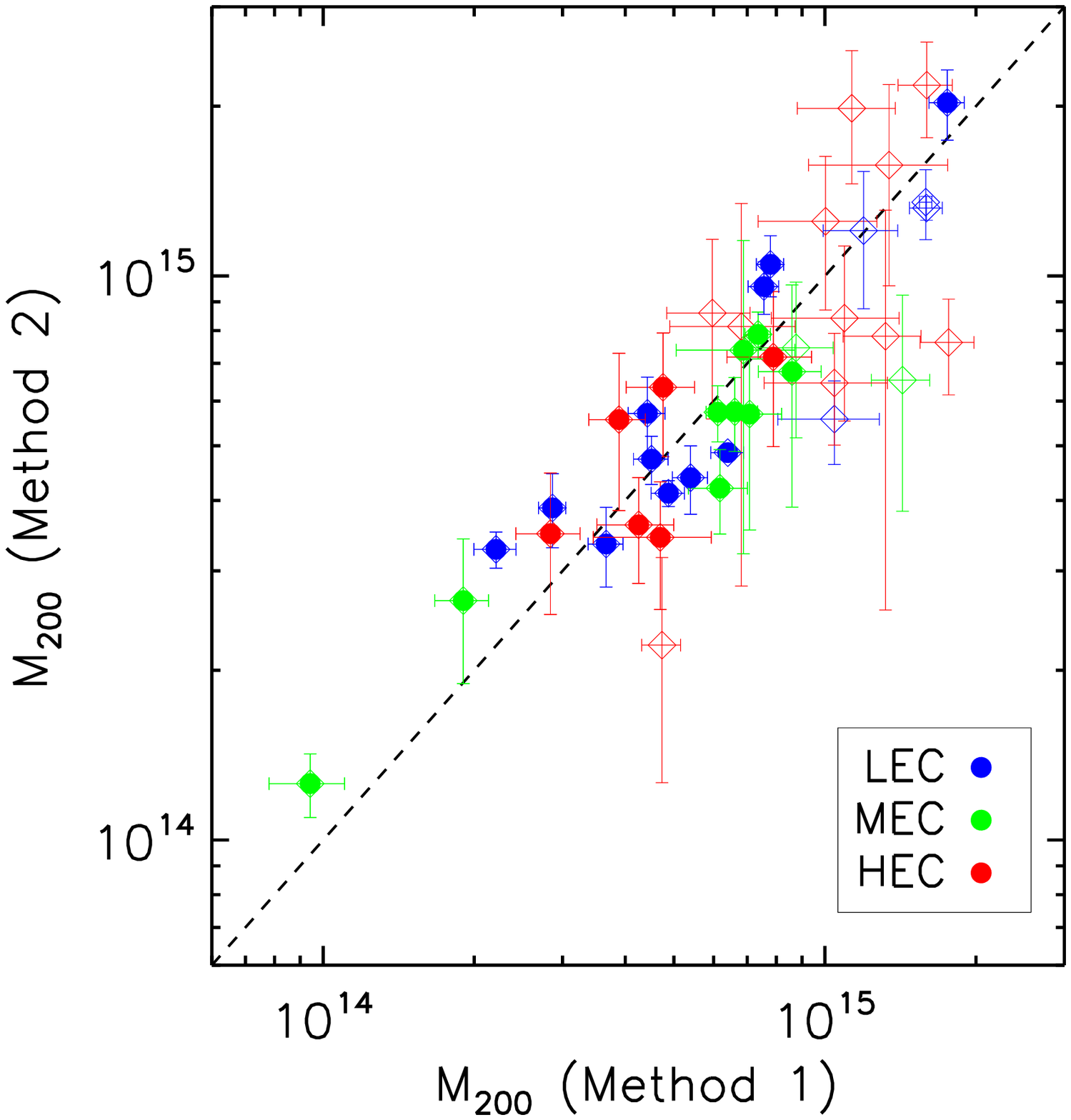,width=0.33\textwidth}
 \epsfig{figure=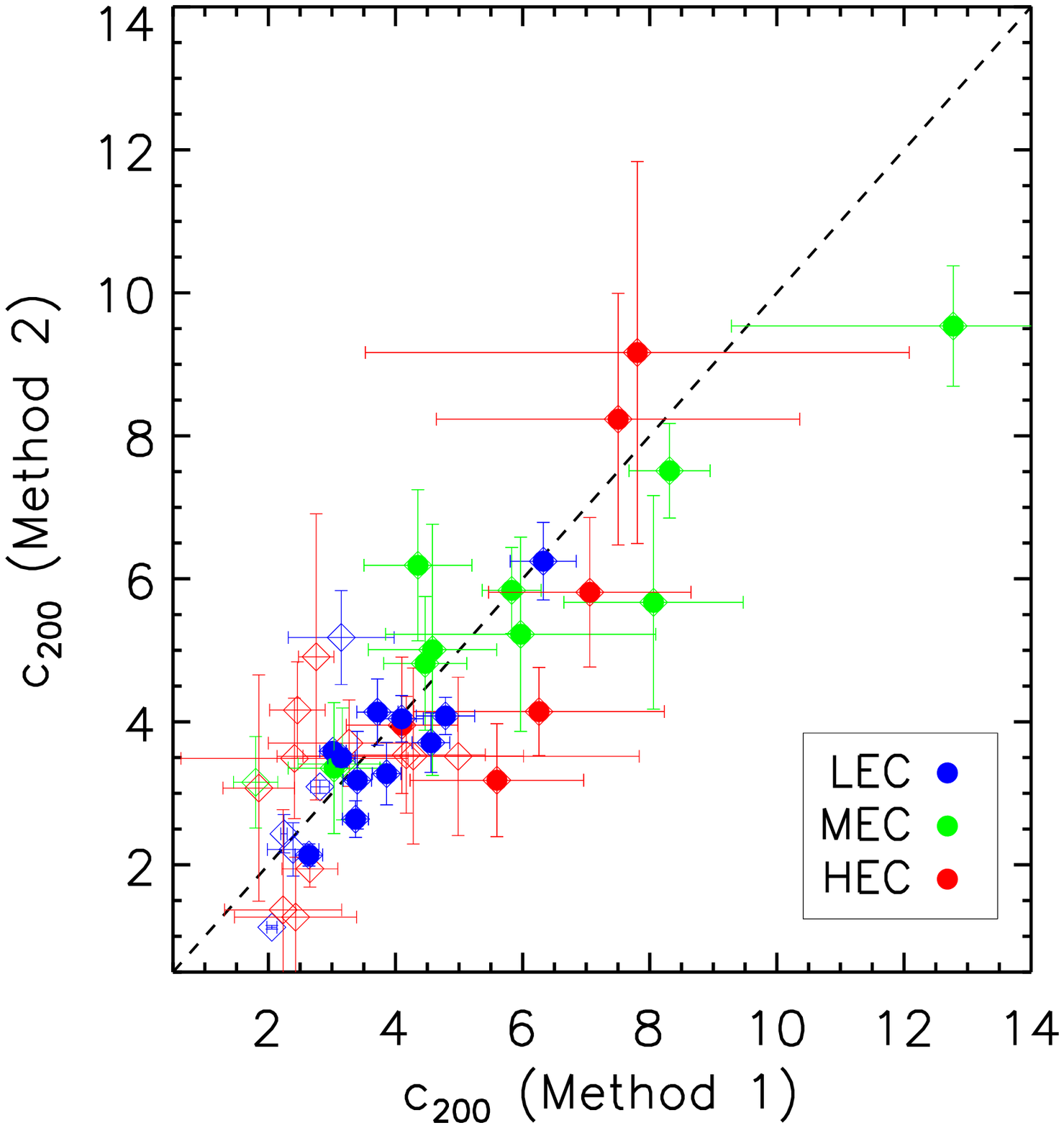,width=0.33\textwidth}
 \epsfig{figure=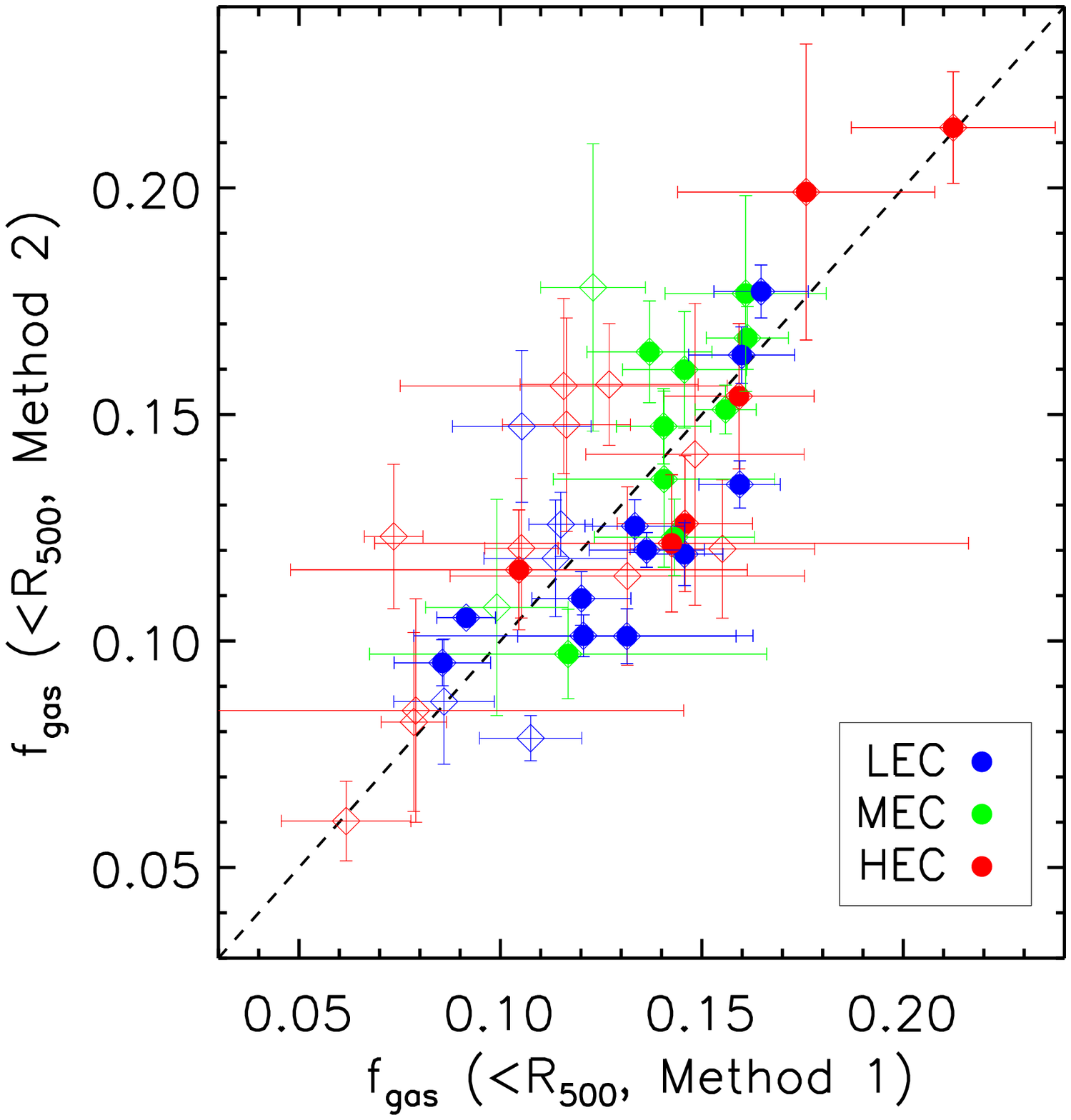,width=0.33\textwidth}
}
\caption{Estimates of $M_{200}$ ({\it left}),
$c_{200}$ ({\it center}) and $f_{\rm gas}(<R_{500})$ ({\it right})
with the two methods.
({\it Upper panels}) 
The color code indicates the objects at $z<0.15$ (blue), in the range
$0.15<z<0.25$ (green) and at $z>0.25$ (red).
({\it Lower panels}) 
Distribution of Low (LEC), Medium (MEC), High (HEC) Entropy Core
systems.
} \label{fig:mcf}
\end{figure*}

Hereafter, we define $M_{\Delta} = M_{\rm DM}(<R_{\Delta})$ (i.e. $M_{200}$
is the dark matter enclosed within a sphere where the mean cluster overdensity
in dark matter only is 200 times the critical density at the cluster's redshift)
and $f_{\rm gas}(<R_{\Delta})$ is the ratio between the gas mass, $M_{\rm gas}$,
and the total mass, $M_{\rm tot} = M_{\rm DM} + M_{\rm gas}$, estimated
within $R_{\Delta}$, where the overdensity is here estimated by using
the {\it total} (i.e. dark$+$gas) mass profile.

The best-fit values obtained for an assumed NFW dark matter mass 
profiles are quoted in Table~\ref{tab:all}.
In Table~\ref{tab:fgas}, we present our estimates of $R_{200}$, $R_{500}$ and the gas mass
fraction $f_{\rm gas} = M_{\rm gas} / M_{\rm tot}$, that is hereafter considered within 
$R_{500}$ to avoid a problematic extrapolation of the data up to $R_{200}$.
In Fig.~\ref{fig:erry}, we show the relative errors provided from the two methods 
on the estimates of $c_{200}$ and $M_{200}$. The distribution of the statistical 
uncertainties is comparable, with median values of 15--20\% on both $c_{200}$ and $M_{200}$
with \meti\ and \metii.
Also the distributions of the measurements of $c_{200}$ and $M_{200}$ are very similar,
with 1st--3rd quartile range of $2.70-5.29$ and $4.7-11.1 \times 10^{14} M_{\odot}$
with \meti\ and $3.17-5.09$ and $4.3-9.1 \times 10^{14} M_{\odot}$ with \metii.

Moreover, the two methods show a good agreement between the two estimates
of the gas mass fraction $f_{\rm gas} (<R_{500})$, as shown in Fig.~\ref{fig:mcf}.
We measure a median (1st, 3rd quartile) of $0.131 (0.106, 0.147)$, and a median relative 
error of 12\%,  with \meti\ and $0.124 (0.108, 0.155)$, 
and a relative error of 10\%, with \metii.

As shown in the last two panels of Fig.~\ref{fig:erry}, we note that the large majority 
of our data is able to define a scale radius $r_{\rm s}$ well within the radial
range investigated in the spectral and spatial analysis, 
allowing a quite robust constraints of the fitted parameters.

To rely on the best estimates of the concentration and mass, we define in the 
following analysis a further subsample by collecting the clusters that 
satisfy the criterion that the upper value at $1 \sigma$
of the scale radius, as estimated from the 2 methods,
is lower than the upper limit of the spatial extension of the detected 
X-ray emission,
i.e. $(r_{\rm s} +\epsilon_{r_{\rm s}}) < R_{\rm sp}$.
Imposing this condition, we select the 26 clusters
where a more robust (i.e. with well defined and constrained
free parameters) mass reconstruction is achievable.

\begin{table*}
\caption{Median deviations measured in the distribution of $c_{200}$, $M_{\rm DM}$
and $f_{\rm gas}$.
}
\vspace*{-0.2cm}
\begin{center}
\begin{tabular}{l@{\hspace{.8em}} c c c }
\hline \\
 Dataset & $(\hat{c}_{200} - c_{200})/c_{200}$ & $(\hat{M}_{\rm DM} - M_{\rm DM})/M_{\rm DM}$ & $(\hat{f}_{\rm gas} - f_{\rm gas})/f_{\rm gas}$ \\
 & \\
\hline \\
{\it Method 2} & $-0.013$ & $+0.008$ & $+0.036$ \\
M2 & $+0.010$ & $-0.017$ & $+0.009$ \\
$T_{\rm 3D}$ & $-0.048$ & $-0.036$ & $+0.024$ \\
fit $n_{\rm gas}$ & $+0.001$ & $+0.011$ & $+0.000$ \\
$P_{\rm out}$ & $-0.011$ & $+0.030$ & $-0.014$ \\
 & \\
 {\bf at $R_{200}$} & {\bf $(-0.048,+0.010)$} & {\bf $(-0.036,+0.030)$} & {\bf $(-0.014,+0.036)$} \\ 
 & \\
 & \\
{\it Method 2} & $-$ & $-0.015$ & $+0.035$ \\
M2 & $-$ & $-0.018$ & $+0.010$ \\
$T_{\rm 3D}$ & $-$ & $-0.046$ & $+0.025$ \\
fit $n_{\rm gas}$ & $-$ & $+0.012$ & $-0.008$ \\
$P_{\rm out}$ & $-$ & $+0.028$ & $-0.013$ \\
 & \\
 {\bf at $R_{500}$} & $-$ & {\bf $(-0.046,+0.028)$} & {\bf $(-0.013,+0.035)$} \\ 
 & \\
 & \\
{\it Method 2} & $-$ & $-0.073$ & $+0.032$ \\
M2 & $-$ & $-0.013$ & $+0.008$ \\
$T_{\rm 3D}$ & $-$ & $-0.059$ & $+0.028$ \\
fit $n_{\rm gas}$ & $-$ & $+0.004$ & $+0.000$ \\
$P_{\rm out}$ & $-$ & $+0.020$ & $-0.009$ \\
 & \\
 {\bf at $R_{2500}$} & $-$ & {\bf $(-0.073,+0.020)$} & {\bf $(-0.009,+0.032)$} \\ 
\hline \\
\end{tabular}

\end{center}
\label{tab:syst}
\tablefoot{The deviations are measured with respect to the estimates
obtained from the combined {\it M1}+{\it M2} profile with the \meti\ for the
whole sample of 44 clusters.
Dataset: (\metii) \metii\ is used for mass reconstruction;
(M2) only the $T(r)$ profile from M2 is used;
($T_{3D}$) the deprojected spectral measurements of $T(r)$ are used in \meti\ instead
of the projected estimates of $T_{\rm model}$ (see Sect.~3);
(fit $n_{\rm gas}$) a model fitted to the gas density profile is used in \meti;
($P_{\rm out}$) the outer value of the pressure is not fixed.
}
\end{table*}

%\begin{figure*}
%\hbox{
% \epsfig{figure=compare_data2.ps,width=0.5\textwidth}
% \epsfig{figure=compare_data3.ps,width=0.5\textwidth}
%}
%\caption{Box and whisker plot of the values of $\hat{c}_{200}/c_{200}-1$
%(left) and $\hat{M}_{200}/M_{200}-1$ (right) 
%with respect to the results obtained with \meti.
%The box encloses the interquartile range defined between the first ($Q_1$, 25\%) and
%third ($Q_3$, 75\%) quartile of the distribution of the sorted values.
%The whiskers extend out to the maximum or
%minimum value of the data, or to the 1.5 times either the $Q_3$ or $Q_1$,
%if there is data beyond this range. Outliers are identified with small circles.
%} \label{fig:cm_syst}
%\end{figure*}

\section{Systematics in the measurements of $c_{200}$, $M_{200}$ and $f_{\rm gas}$}

The derived quantities $c_{200}$, $M_{200}$ and $f_{\rm gas}(<R_{500})$ 
are measured with a relative statistical error of about 20, 15 and 10 \%, 
respectively (see Section~3 and Figures~\ref{fig:erry} and ~\ref{fig:mcf}).
Here, we investigate the main uncertainties affecting our techniques
that will be treated as systematic effects in the following analysis.

We consider two main sources of systematic errors: 
(i) the analysis of our dataset, both for what concerns the 
estimates of the gas temperature and the reconstructed gas density profile;
(ii) the limitations and assumptions in the techniques adopted 
for the mass reconstruction.
 
In Table~\ref{tab:syst}, we summarize our findings tabulated as relative 
median difference with respect to the estimates obtained with \meti.
% Box-plots of the distribution of these deviations are shown in Fig.~\ref{fig:cm_syst}.
Overall, we register systematic uncertainties of
$(-5, +1)$\% on $c_{200}$, $(-4, +3)$\% on $M_{200}$ and
$(-1,+4)$\% on $f_{\rm gas}(<R_{500})$, where these ranges
represent the minimum and maximum estimated in the dataset investigated
and quoted in Table~\ref{tab:syst}. 

\subsection{Systematics from the spectral analysis}

The ICM properties of the present dataset have been studied 
through spatially resolved spectroscopic measurements of the 
gas temperature profile and deprojected, PSF--corrected 
surface brightness profile as accessible to \xmm\
(see Section~2).

To assess the systematics propagated through the
temperature measurements, we present the results obtained with
{\it M2} only, i.e. before any correction introduced
from the harder spectra observed with {\it M1} (see Sect.~2.1).
Overall, the systematics are in the order of a few per cent, with the largest offset
of about $4$ per cent on the concentration and mass measurements at
$R_{500}$ and beyond.

When the deprojected spectral values of the gas temperature, instead of the
projected ones, are compared with the predictions from the model, 
we measure differences below $5$\% (see dataset labelled ``$T_{\rm 3D}$'').

On the gas density profile, we investigate the role played from the use
of a functional form instead of the values obtained directly from deprojection.
To this purpose,
we use a revised form of the one introduced from Vikhlinin et al. (2006)
to fit the gas density profile and, then, we adopt
it as representative of the gas density profile to be put
in hydrostatic equilibrium with the gravitational potential
in equation~\ref{eq:mtot}. The measurements obtained are
labelled ``fit $n_{\rm gas}$'' and show discrepancies in the order
of 1 per cent or less.

\subsection{Systematics from the mass reconstruction methods}

With the intention to assess the the bias affecting the reconstructed
mass values, we make use of the gas temperature and density profiles
through two independent techniques (labelled \meti\ and \metii), as
described in Section~3.
With respect to \meti, \metii\ provides differences on $M_{\rm DM}$ 
that are lower than 10 per cent, increasing from about 1 per cent
at $R_{200}$ up to 7 per cent at $R_{2500}$ (see Table~\ref{tab:syst}).
The bias on $f_{\rm gas}$ remains stable around 3--4 per cent, 
suggesting that some systematics affect also the estimate of $M_{\rm gas}$.
This is due to the application of two different functional forms
in \meti\ and \metii\
over a radial range that extends beyond the observational limit 
(see, e.g., Fig.~\ref{fig:erry}).

The mass reconstruction of \meti\ depends upon the boundary condition 
on the gas pressure profile. In particular, to solve the differential 
equation~\ref{eq:mtot}, an outer value on the pressure is fixed to
the product of the observed estimate of the gas density profile at 
the outermost radius and an extrapolated measurement of the gas temperature.
Using a grid of values for the pressure obtained from the best-fit results 
of the gas density and temperature profiles, we evaluate a systematic
bias on the mass of about 3 per cent, on the gas mass fraction 
of 1 per cent, and on $c_{200}$ of about 1 per cent
(see dataset labelled ``$P_{\rm out}$'').

\begin{figure*}
\hbox{
 \epsfig{figure=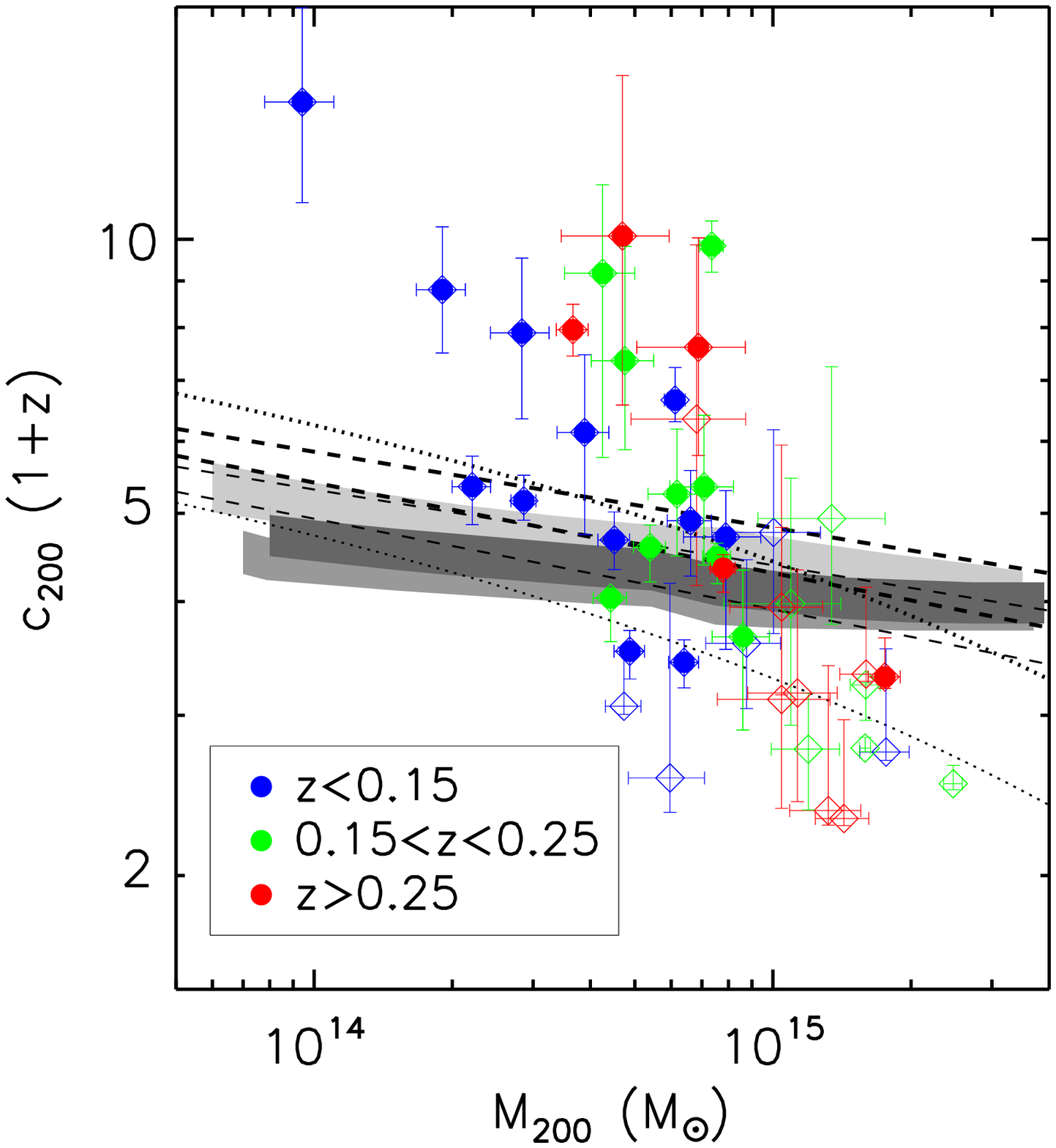,width=0.5\textwidth}
 \epsfig{figure=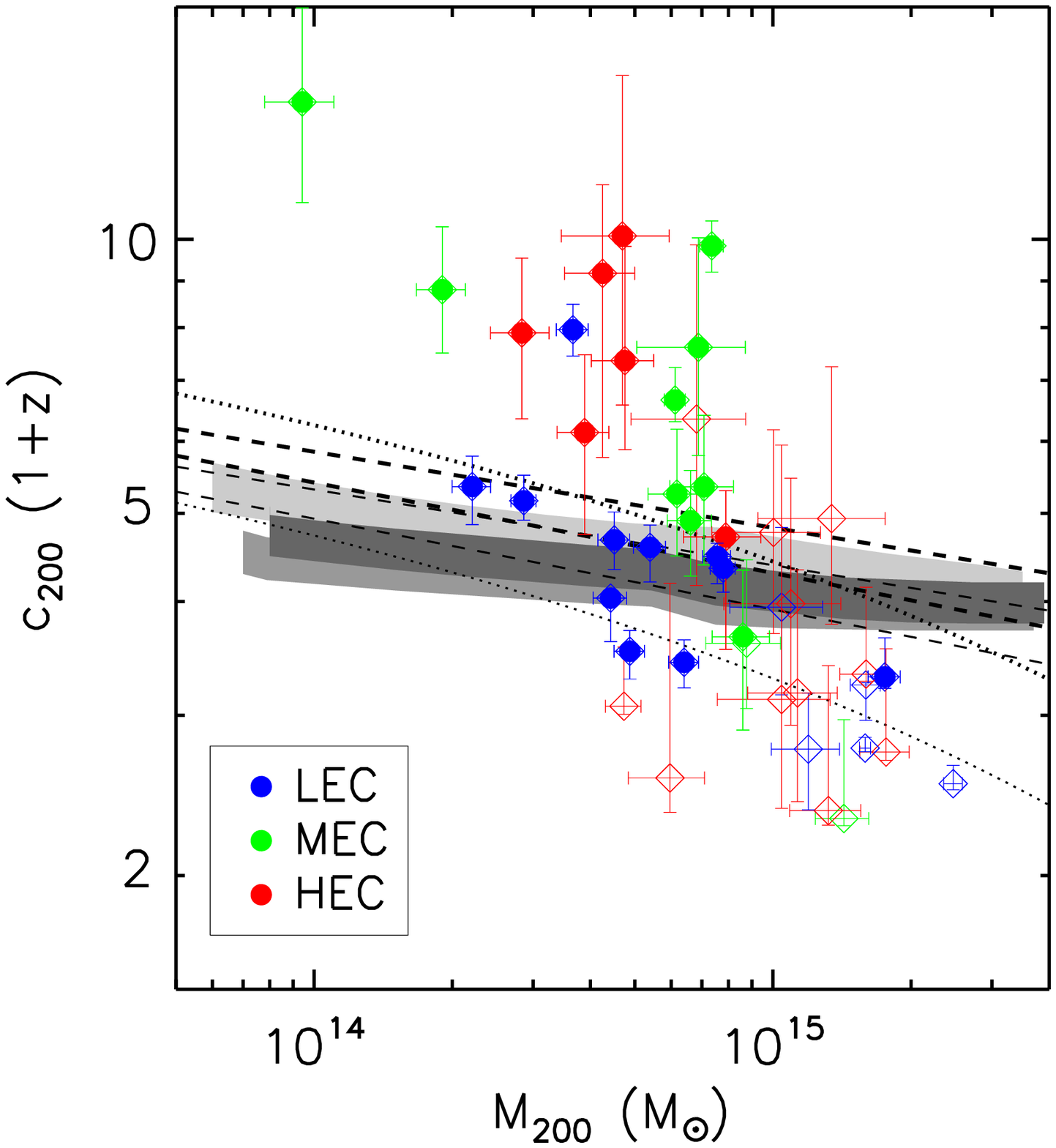,width=0.5\textwidth}
}
\caption{Data in the plane $(c_{200}, M_{200})$ used to constrain the cosmological 
parameters $(\Omega_{\rm m}, \sigma_8)$.
The dotted lines show the predicted relations from Eke et al. (2001)
for a given $\Lambda$CDM cosmological model at $z=0$ (from top to bottom:
$\sigma_8=0.9$ and $\sigma_8=0.7$).
% $\sigma=0.87, z=0.1$; $\sigma=0.87, z=0.3$; $\sigma=0.8, z=0.3$).
The shaded regions show the predictions in the redshift range $0.1-0.3$
for an assumed cosmological model in agreement with 
WMAP-1, 5 and 3 years (from the top to the bottom, respectively) from 
Bullock et al. (2001; after Macci\`o et al. 2008).
The dashed lines indicate the best-fit range at $1\sigma$ obtained for relaxed halos in a 
WMAP-5 years cosmology from Duffy et al. (2008; thin lines: $z=0.1$, thick lines: $z=0.3$).
Color codes and symbols as in Fig.~\ref{fig:mcf}.
} 
\label{fig:cm}
\end{figure*}

\section{The $c_{200} - M_{200}$ relation}

In this section, we investigate the $c_{200} - M_{200}$ relation.
We note that our sample has not been selected to be representative
of the cluster population in the given redshift range and, in the mean time,
does not include only relaxed systems.
Therefore,
the results here presented on the $c_{200} - M_{200}$ relation
have to be just considered for a qualitative comparison with
the predictions from numerical simulations and to assess
differences or similitude with previous work on this topic.

As we show in Fig.~\ref{fig:cm} using the measurements obtained with \meti,
the median relation between concentration and total masses for CDM halos
as function of redshift is represented well from the analytic algorithms, as in 
N97, E01 and B01. 
These models relate the halo properties to the physical mechanism of halo formation.
Considering the weak dependence of the halo concentrations on the mass and redshift,
Dolag et al. (2004) introduced a two-parameter functional form, 
$c = c_0 M^B / (1+z)$.
We consider this relation in its logarithmic form and fit linearly to our data
the expression:
\begin{equation}
\log_{10} \left( c_{200} \times (1+z) \right) = A +B \times 
\log_{10} \left( \frac{M_{200}}{10^{15} M_{\odot}} \right).
\label{eq:cm}
\end{equation}
%where the normalization for the mass, $10^{15} h_{70}^{-1} M_{\odot}$,
%is chosen being approximately the middle point in the logarithmic $X$ axis. 
A minimum in the $\chi^2$ distribution is looked for by 
taking into account the errors on both the coordinates
(we use the routine {\sl FITEXY} in IDL). The errors are
assumed to be Gaussian in the logarithmic space, although 
they are properly measured as Gaussian in the linear space.

We also express our results in term of the concentration $c_{15}$ expected
for a dark matter halo of $10^{15} h_{70}^{-1} M_{\odot}$ and
equal to $10^A$ once the parameters in equation~\ref{eq:cm} are used.
We convert to $c_{15}$ even the results from literature obtained, for instance, 
at different overdensity, as described in the Appendix.

\begin{table*}
\caption{Best-fit values of the $c_{200}-M_{200}$ relation.
}
\vspace*{-0.2cm}
\begin{center}
\begin{tabular}{l c c c c}
\hline \\
 Dataset & $c_{15}$ & $A$ & $B$ & $\sigma_{\log_{10} c}$ \\
 & \\
\hline \\
 & \\
 \multicolumn{5}{c}{All objects (44 clusters)} \\
 & \\
 {\it Method 1} - {\sl Weighted Mean} & $3.60^{+0.05}_{-0.05}$ & $0.556\pm0.006$ & $-0.1$ & $0.193/0.116$ \\
 {\it Method 1} - {\sl FITEXY} & $3.62\pm0.07$ & $0.558\pm0.008$ & $-0.451\pm0.023$ & $0.135/0.116$ \\
 {\it Method 1} - {\sl BCES} & $3.78\pm0.18$ & $0.577\pm0.021$ & $-0.544\pm0.071$ & $0.132/0.116$ \\
 {\it Method 1} - {\sl LINMIX} & $3.79\pm0.21$ & $0.579\pm0.025$ & $-0.444\pm0.077$ & $0.132/0.116$ \\
 & \\
 {\it Method 2} - {\sl Weighted Mean} & $3.42^{+0.03}_{-0.03}$ & $0.534\pm0.004$ & $-0.1$ & $0.203/0.119$ \\
 {\it Method 2} - {\sl FITEXY} & $3.21\pm0.05$ & $0.507\pm0.006$ & $-0.466\pm0.015$ & $0.146/0.119$ \\
 {\it Method 2} - {\sl BCES} & $3.51\pm0.16$ & $0.545\pm0.020$ & $-0.612\pm0.084$ & $0.133/0.119$ \\
 {\it Method 2} - {\sl LINMIX} & $3.72\pm0.20$ & $0.571\pm0.024$ & $-0.493\pm0.067$ & $0.131/0.119$ \\
 & \\
 \multicolumn{5}{c}{Selected objects (26 clusters)} \\
 & \\
 {\it Method 1} - {\sl Weighted Mean} & $4.61^{+0.09}_{-0.09}$ & $0.664\pm0.008$ & $-0.1$ & $0.165/0.092$ \\
 {\it Method 1} - {\sl FITEXY} & $4.06\pm0.17$ & $0.608\pm0.018$ & $-0.321\pm0.050$ & $0.143/0.092$ \\
 {\it Method 1} - {\sl BCES} & $3.84\pm0.38$ & $0.584\pm0.043$ & $-0.586\pm0.116$ & $0.138/0.092$ \\
 {\it Method 1} - {\sl LINMIX} & $4.24\pm0.47$ & $0.628\pm0.048$ & $-0.370\pm0.125$ & $0.132/0.092$ \\
 & \\
 {\it Method 2} - {\sl Weighted Mean} & $4.00^{+0.05}_{-0.05}$ & $0.602\pm0.005$ & $-0.1$ & $0.181/0.079$ \\
 {\it Method 2} - {\sl FITEXY} & $2.87\pm0.13$ & $0.458\pm0.020$ & $-0.576\pm0.063$ & $0.169/0.079$ \\
 {\it Method 2} - {\sl BCES} & $3.18\pm0.57$ & $0.502\pm0.078$ & $-0.782\pm0.259$ & $0.150/0.079$ \\
 {\it Method 2} - {\sl LINMIX} & $3.93\pm0.45$ & $0.594\pm0.050$ & $-0.438\pm0.131$ & $0.130/0.079$ \\
 & \\
 \multicolumn{5}{c}{only LEC objects (11 clusters)} \\
 & \\
 {\it Method 1} - {\sl Weighted Mean} & $4.14^{+0.09}_{-0.09}$ & $0.617\pm0.009$ & $-0.1$ & $0.091/0.033$ \\
 {\it Method 1} - {\sl FITEXY} & $3.68\pm0.15$ & $0.565\pm0.017$ & $-0.297\pm0.051$ & $0.081/0.033$ \\
 {\it Method 1} - {\sl BCES} & $3.27\pm0.52$ & $0.514\pm0.070$ & $-0.472\pm0.229$ & $0.093/0.033$ \\
 {\it Method 1} - {\sl LINMIX} & $3.75\pm0.45$ & $0.574\pm0.052$ & $-0.279\pm0.150$ & $0.081/0.033$ \\
 & \\
 {\it Method 2} - {\sl Weighted Mean} & $3.81^{+0.05}_{-0.05}$ & $0.581\pm0.005$ & $-0.1$ & $0.096/0.046$ \\
 {\it Method 2} - {\sl FITEXY} & $3.13\pm0.11$ & $0.496\pm0.015$ & $-0.376\pm0.054$ & $0.080/0.046$ \\
 {\it Method 2} - {\sl BCES} & $3.21\pm0.75$ & $0.506\pm0.102$ & $-0.450\pm0.303$ & $0.074/0.046$ \\
 {\it Method 2} - {\sl LINMIX} & $3.39\pm0.36$ & $0.530\pm0.046$ & $-0.377\pm0.133$ & $0.071/0.046$ \\
 & \\
 \multicolumn{4}{c}{Simulations} \\
 & \\
 B01  & $4.29$ & $0.632$ & $-0.102$ \\
 D04  & $4.01$ & $0.603$ & $-0.130$ \\
 S06 -- all, relaxed & $4.64, 4.86$ & $0.667, 0.687$ & $-0.120, -0.160$ \\
 N07 -- all, relaxed & $3.77, 4.33$ & $0.576, 0.636$ & $-0.110, -0.100$ \\
 M08 / WMAP-1 -- all, relaxed & $3.47, 4.18$ & $0.540, 0.621$ & $-0.119, -0.104$ \\
 M08 / WMAP-3 -- all, relaxed & $2.94, 3.41$ & $0.469, 0.533$ & $-0.088, -0.083$ \\
 M08 / WMAP-5 -- all, relaxed & $2.98, 3.56$ & $0.474, 0.551$ & $-0.110, -0.098$ \\
\hline \\
\end{tabular}

\end{center}
\label{tab:cm}
\tablefoot{The best-fit values refer to equation~\ref{eq:cm}
and are obtained by using
(i) the linear least--squares fitting with errors in both variables
({\sl FITEXY}), (ii) the linear regression method {\sl BCES},
(iii) a Bayesian linear regression method ({\sl LINMIX}).
In the last column, the total ($\sigma_{\rm tot} = \sum^N_i (y_i - A -B x_i)^2 / N$)
and statistical ($\sigma_{\rm stat} = \sum^N_i \epsilon_{y_i}^2 / N$) scatters are quoted,
where $y_i = \log_{10} \left( c_{200} (1+z) \right)$, $x_i = \log_{10} M_{200}$,
$\epsilon_{y_i}$ is the statistical error on $y_i$ and $N$ is the number of objects.
}
\end{table*}

We measure $A \approx 0.6$ and $B$ systematically lower than $-0.1$,
with the best-fit results obtained through \meti\ that prefer,
with respect to \metii, a relation with slightly higher normalization 
(by $\sim10$ per cent) and flatter (by $10-30$ per cent) distribution in mass.
In both cases, a total scatter of $\sigma_{\log_{10} c} 
\approx 0.13$ is measured both in the whole sample of 44 objects, where
the statistical scatter related to the observed uncertainties is still dominant,
and in the subsample of 26 selected clusters. 

When a slope $B=-0.1$ is assumed, as measured in numerical simulations
over one order of magnitude in mass almost independently from the underlying
cosmological model (see e.g. Dolag et al. 2004, Macci\`o et al. 2008), 
the measured normalizations of the $c_{200}-M_{200}$ relation 
fall into the range of the estimated values for samples of simulated clusters
(see Table~\ref{tab:cm}).

All the values of normalization and slope are confirmed, within the estimated
errors, with both the {\sl BCES} bisector method (as described in Akritas \& Bershady 1996
and implemented in the routines made available from M.A.~Bershady)
and a Bayesian method that accounts for measurement errors in linear
regression, as implemented in the IDL routine {\sl LINMIX\_ERR}
by B.C.~Kelly (see Kelly 2007).
As we quote in Table~\ref{tab:cm}, with these linear regression methods 
(and after $10^6$ bootstrap resampling of the data in {\sl BCES}), 
we measure a typical error that is larger by a factor $2-3$ in normalization
and up to $6$ in the slope than the corresponding values obtained 
through the covariance matrix of the {\sl FITEXY} method.

These values compare well with the measurements obtained from numerical
simulations of DM-only galaxy clusters, although these simulations sample, on average, 
mass ranges lower than the ones investigated here.
Recent work from Shaw et al. (2006) and Macci\`o et al. (2008) summarize 
the findings.
The slope of the relation, as previously obtained from B01 and D04,
lies in the range $(-0.160,-0.083)$, with a preferred value of
about $-0.1$. The normalizations for low-density Universe with a relatively
higher $\sigma_8$, as from WMAP-1, are more in agreement with the
observed constraints on, e.g., $c_{15}$. For instance, 
M08 find $c_{15} = 4.18, 3.41, 3.56$ for relaxed objects in a background cosmology
that matches WMAP-1, 3 and 5 year data, respectively\footnote{
We refer to Appendix A for a detailed discussion of the conversions adopted}.
Shaw et al. (2006) measure $c_{15} = 4.64$ using a flat Universe 
with $\Omega_{\rm m}=0.3$ and $\sigma_8=0.95$.
D04 for a $\Lambda$CDM with $\sigma_8=0.9$ require $c_{15} = 4.29$.
All these values show the sensitivity of the normalization 
to the assumed cosmology, that is further discussed in the section
where constraints on the cosmological parameters $(\Omega_{\rm m}-\sigma_8)$
will be obtained through the measured $c-M$ relation.
Neto et al. (2007) study the statistics of the halo concentrations at $z=0$ 
in the {\it Millennium Simulation} (with an underlying cosmology of
$\Omega_{\rm m}=1-\Omega_{\Lambda}=0.25, \sigma_8=0.9$) and find that 
a power-law with $B=-0.10$ and $c_{15} = 4.33$ fits fairly well the relation 
for relaxed objects, with an intrinsic logarithmic scatter for the most massive 
objects of 0.092 (see their Fig.~7).

We note, however, that, while the normalizations we measure for a fixed slope $B=-0.1$ 
are well in agreement with the results from numerical simulations, a systematic lower value of the 
slope is measured, when it is left to vary. 
To test the robustness of this evidence, we have implemented
Monte-Carlo runs using the best-fit central values estimated in N-body simulations 
(see Appendix B for details). With almost no dependence upon the input values from
numerical simulations and using the {\sl FITEXY} technique that provides the results
with the most significant deviations from $B \approx -0.1$, 
we measure in the 3 samples here considered
(i.e. all 44 objects, the selected 26 objects, and the only 11 LEC objects)
a probability of about 0.5 (1), 20 (42) and 26 (46) per cent, respectively,
to obtain a slope lower than the measured $1 (3) \sigma$ upper limit.
These result confirm that the systematic uncertainties present in the measurements
of the concentration and dark mass within $R_{200}$ are still affecting the sample
of 44 objects, whereas they are significantly reduced in the selected subsamples.

Our best-fit results are in good agreement also with previous constraints 
obtained from X-ray measurements in the same cosmology.
Pointecouteau et al. (2005) measure $c_{15} \approx 4.5$ and $B =-0.04 \pm 0.03$
in a sample of ten nearby ($z<0.15$) and relaxed objects observed with \xmm\
in the temperature range $2-9$ keV.
Zhang et al. (2006) measure a steeper slope of $-1.5\pm0.2$, probably
affected from few outliers, in the REFLEX-DXL sample of 13 X-ray luminous
and distant ($z\sim0.3$) clusters observed with \xmm, that, they claim, 
are however not well reproduced from a NFW profile.
Voigt \& Fabian (2006) show a good agreement with B01 results and
$B\approx -0.2$ for their estimates of 12 mass profiles of X-ray luminous
objects observed with \chandra\ in the redshift range $0.02-0.45$.
A good match with the results in D04, and within the scatter
found in simulations, is obtained with 13 low-redshift relaxed systems with
$T_{\rm gas}$ in the range $0.7-9$ keV as measured with \chandra\ in 
Vikhlinin et al. (2006).
Schmidt \& Allen (2007), using \chandra\ observations of 34 massive relaxed
galaxy clusters, measure $B = -0.45 \pm 0.12$ (95\% c.l.), significantly steeper
than the value predicted from CDM simulations. Leaving free the redshift dependence
that they estimate to be consistent with the $(1+z)^{-1}$ expected evolution,
they measure a normalization $c_{15} \approx 5.4 \pm 0.6$ (95\% c.l.), 
definitely higher than our best-fit parameter.
Buote et al. (2007) fit the $c-M$ relation from 39 systems in the mass range
$0.06-20 \times 10^{14} M_{\odot}$ selected from \chandra\ and \xmm\ archives
to be relaxed. Analysing the tabulated values of 
the 20 galaxy clusters with $M_{200}>10^{14} M_{\odot}$,
that include the most massive systems from the \xmm\ study of Pointecouteau 
et al. (2005) and the \chandra\ analysis in Vikhlinin et al. (2006),
we measure $B=-0.08 \pm 0.05$ and $c_{15} \approx 5.16 \pm 0.36$. 
%(see, f.e., Bayesian method in Table~\ref{tab:cm}).

Overall, we conclude however that the slope of the $c-M$ relation 
cannot be reliably determined from the fitting over a narrow mass range
as the one considered in the present work and that, once the slope
is fixed to the expected value of $B=-0.1$, the normalization,
with estimates of $c_{15}$ in the range $3.8-4.6$, agrees with 
results of previous observations and simulations for a calculations
in a low density Universe.

\subsection{The subsample of Low-Entropy-Core objects}

Following Leccardi et al. (2010), we have employed the pseudo-entropy ratio
($\sigma \equiv (T_\mathrm{IN}/T_\mathrm{OUT}) 
\times (E\!M_\mathrm{IN}/E\!M_\mathrm{OUT})^{-1/3}$,
where $\mathrm{IN}$ and $\mathrm{OUT}$ define regions within 
$\approx$0.05~$R_{180}$ and encircled in the annulus with bounding 
radii 0.05-0.20~$R_{180}$, respectively, and $T$ and $E\!M$ are
the cluster temperature and emission measure) to classify our sample 
of 44 galaxy clusters accordingly to their core properties.
We identify 17 High-Entropy-Core (HEC), 11 Medium-Entropy-Core (MEC) 
and 16 Low-Entropy-Core (LEC; see Table~\ref{tab:data}) systems.
While the MEC and HEC objects are progressively more disturbed 
(about 85 per cent of the merging clusters are HEC)
and with a core that presents less evidence in the literature of a temperature decrement 
and a peaked surface brightness profile (intermediate, ICC, and no cool core, NCC, systems),
the LEC objects represent the prototype of a relaxed cluster with a well defined cool core 
(CC in Table~\ref{tab:data}) at low entropy (see also Cavagnolo et al. 2009). 
These systems are predicted from numerical simulations to have higher concentrations 
for given mass, by about 10 per cent, and lower scatter, by about 15-20 per cent, 
in the $c-M$ relation (e.g. M08, Duffy et al. 2008). 

Out of 16, eleven LEC objects are selected under the condition that their scale radius
is within the radial coverage of our data.
We measure their $c-M$ relation to have slightly lower normalization 
($A \approx 0.5-0.6$, $c_{15} \approx 3.2-3.7$) and flatter distribution ($B = -0.4 \pm 0.2$)
than the one observed in the selected subsample of 26 objects, with a dispersion
around the logarithmic value of the concentration of 0.08, that is about 40 per cent lower
than the similar value observed in the latter sample.
This is consistent in a scenario where disturbed systems have an estimated concentration
through the hydrostatic equilibrium equation that is biased higher (and with larger scatter)
than in relaxed objects up to a factor of 2 due to the action of the ICM motions 
(mainly the rotational term in the inner regions and the random gas term above $R_{500}$), 
as discussed in Lau et al. (2009; see also Fang et al. 2009, Meneghetti et al. 2010) 
for galaxy clusters extracted from high-resolution Eulerian cosmological simulations.

\section{Cosmological constraints from the measurements of
$c_{200}, M_{200},$ and $f_{\rm gas}$}
\label{sect:cosmo}

$N-$body simulations have provided theoretical fitting functions
that are able to reproduce the distribution of the concentration
parameter of the NFW density profile as function of halo mass
and redshift (e.g. NFW, E01, B01, N07). 
Basically, all these semi-empirical prescriptions 
provide the expected values of the concentration parameter
for a given set of cosmological parameters (essentially, the cosmic
matter density, $\Omega_{\rm m}$, and the normalization of the power spectrum
on clusters scale, $\sigma_8$) for a given mass (the estimated cluster dark
mass, $M_{200}$, in our case) at the measured redshift of the analyzed object.
They assume that the concentration reflects the background density 
of the Universe at the formation time of a given halo.
The cosmological model influences the concentration and virial mass because of
the cosmic background density and the evolution of structure formation.
For instance, the NFW model uses two free parameters, $(f, C)$,
to define the collapse redshift at which half of the final mass
$M$ is contained in progenitors of mass $\ge f M$, with $C$ representing
the ratio between the characteristic overdensity and the mean density
of the Universe at the collapse redshift.
We use $(f, C) = (0.1, 3000)$.

B01 assume, instead, an alternative model to improve the agreement 
between the predicted redshift dependence of the concentrations and
the results of the numerical simulations by using two free parameters,
$F$ and $K$, where $F$ is still a fixed fraction ($0.01$ in our study)
of a halo mass at given redshift and $K$ indicates the concentration of the
halo at the collapse redshift. $K$ has to be calibrated with numerical
simulations and is fixed here to be equal to $4$ (see also Buote et al. 2007
for a detailed discussion on the role played from the parameters $F$ and $K$
on the prediction of the concentrations as function of the background
cosmology and halo masses).
M08 have revised this model by assuming that the characteristic density of the halo,
that in B01 scales as $(1+z)^3$, is independent of redshift.
This correction propagates into the growth factor of the concentration
parameter that becomes shallower with respect to the mass dependence 
at masses higher than $10^{13} h^{-1} M_{\odot}$, permitting
larger concentrations at the high-mass end than the original B01 formulation.

The prescription in E01 defines with the only parameter $C_{\sigma}$ (equal to 28,
in our analysis, as suggested in their original work) the collapse 
redshift $z_{\rm c}$ through the relation 
$D(z_{\rm c}) \sigma_{\rm eff}(M_{\rm s}) = C_{\sigma}^{-1}$,
where $D(z)$ is the linear growth factor, $\sigma_{\rm eff}$ is the 
effective amplitude of the linear power spectrum at $z=0$ and 
$M_{\rm s}$ is the total mass within the radius at which the 
circular velocity of an NFW halo reaches its maximum and
that is equal to 2.17 times the scale radius, $r_{\rm s}$.

As tested in high-resolution numerical simulations (see, e.g., N07, 
M08, Duffy et al. 2008), these 3 formulations provide different predictions 
over different mass range and redshift: for massive systems a $z<1$, as
the ones under investigation in the present analysis, the original B01
tends to underestimate the concentration at fixed halo mass; its revised version
after M08 partially compensate for this difference but still shows some tension
with numerically simulated objects (see, e.g., figure~5 in M08); 
NFW overestimates the concentration, whereas E01 provide good estimates
(see, e.g. figure~2 in Duffy et al. 2008) also considering its simpler
and more robust formulation, being dependent upon a single parameter
that does not need an independent calibration from simulations
evolved with a given background cosmology (note, indeed, that as pointed
out in M08, both NFW and B01 models have normalizations that, ideally, have to be 
determined empirically for each assumed cosmology with a dedicated
numerical simulation).

Hereafter, we consider E01 as the model of reference and use the other
prescriptions as estimate of the systematics affecting our constraints. 

In particular, to constrain the cosmological parameters 
of interest, $\sigma_8$ and $\Omega_{\rm m}$, 
we calculate first the concentration $c_{200, ijk} = 
c_{200}(M_i, \Omega_{\rm m, j}, \sigma_{8, k})$
predicted from the model investigated at each cluster redshift
for a given grid of values in mass, $M_i$,  
cosmic density parameter, $\Omega_{\rm m, j}$,
and power spectrum normalization, $\sigma_{8, k}$.

\begin{figure*} 
\hbox{
 \epsfig{figure=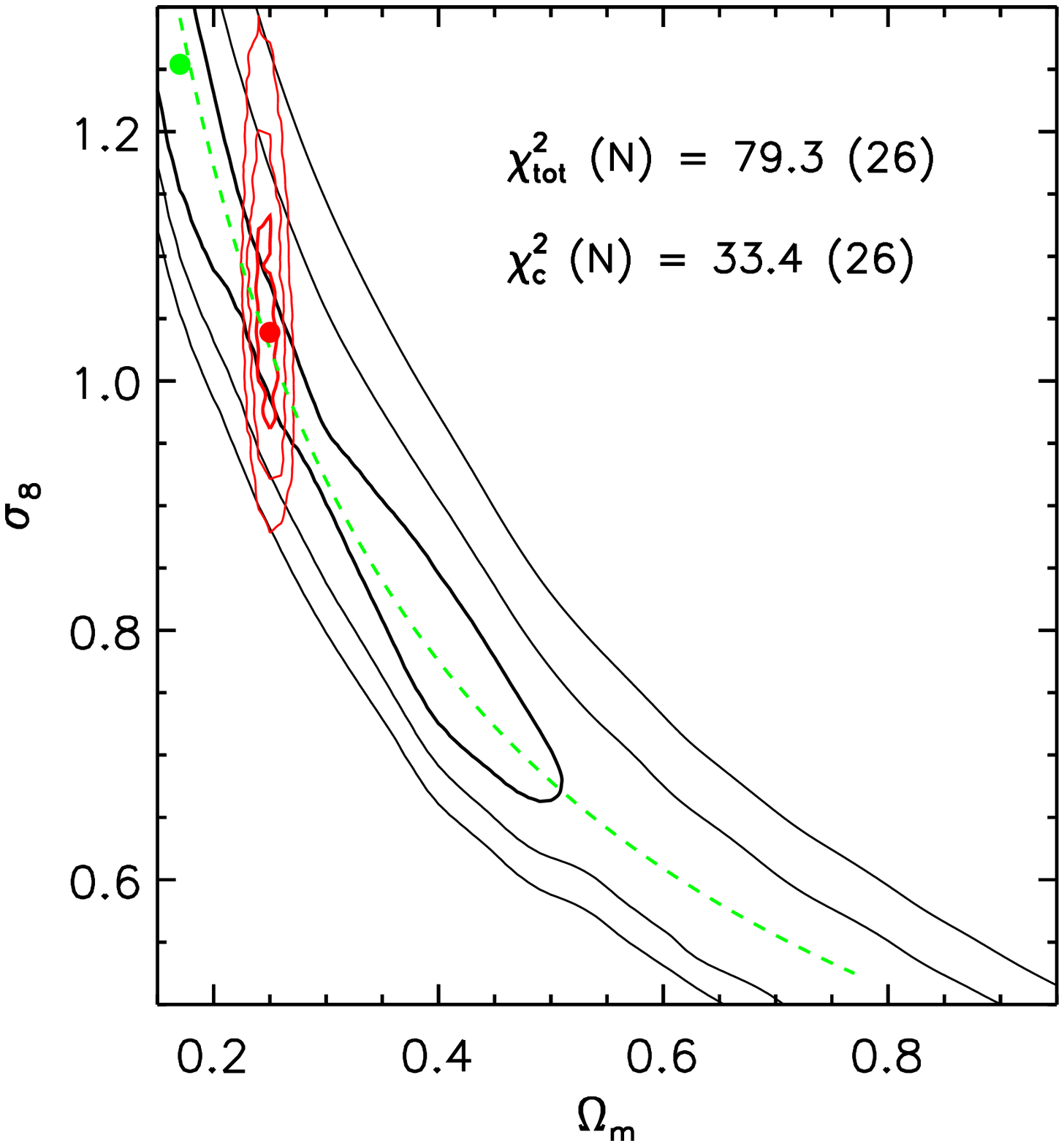,width=0.33\textwidth}
 \epsfig{figure=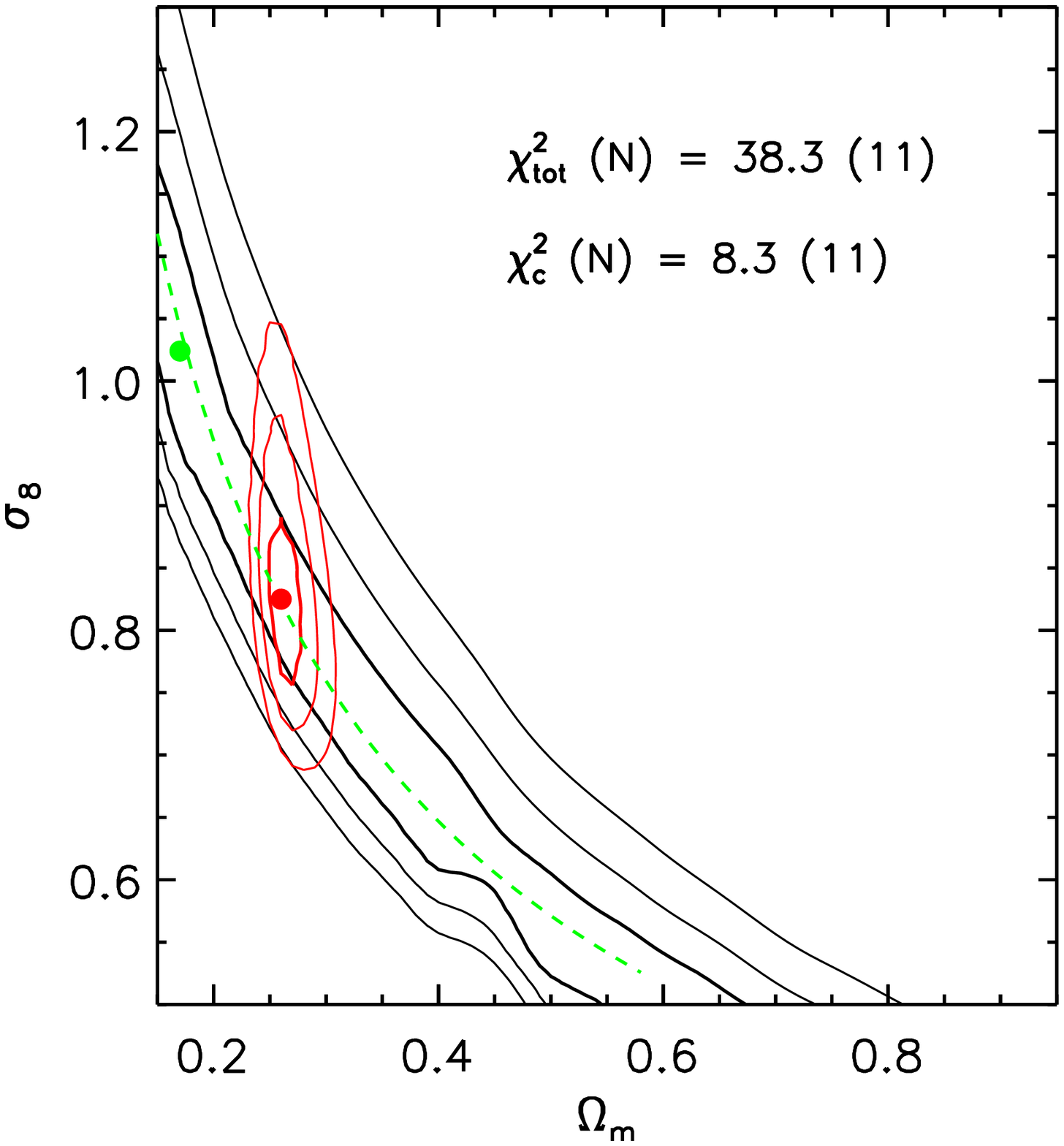,width=0.33\textwidth}
 \epsfig{figure=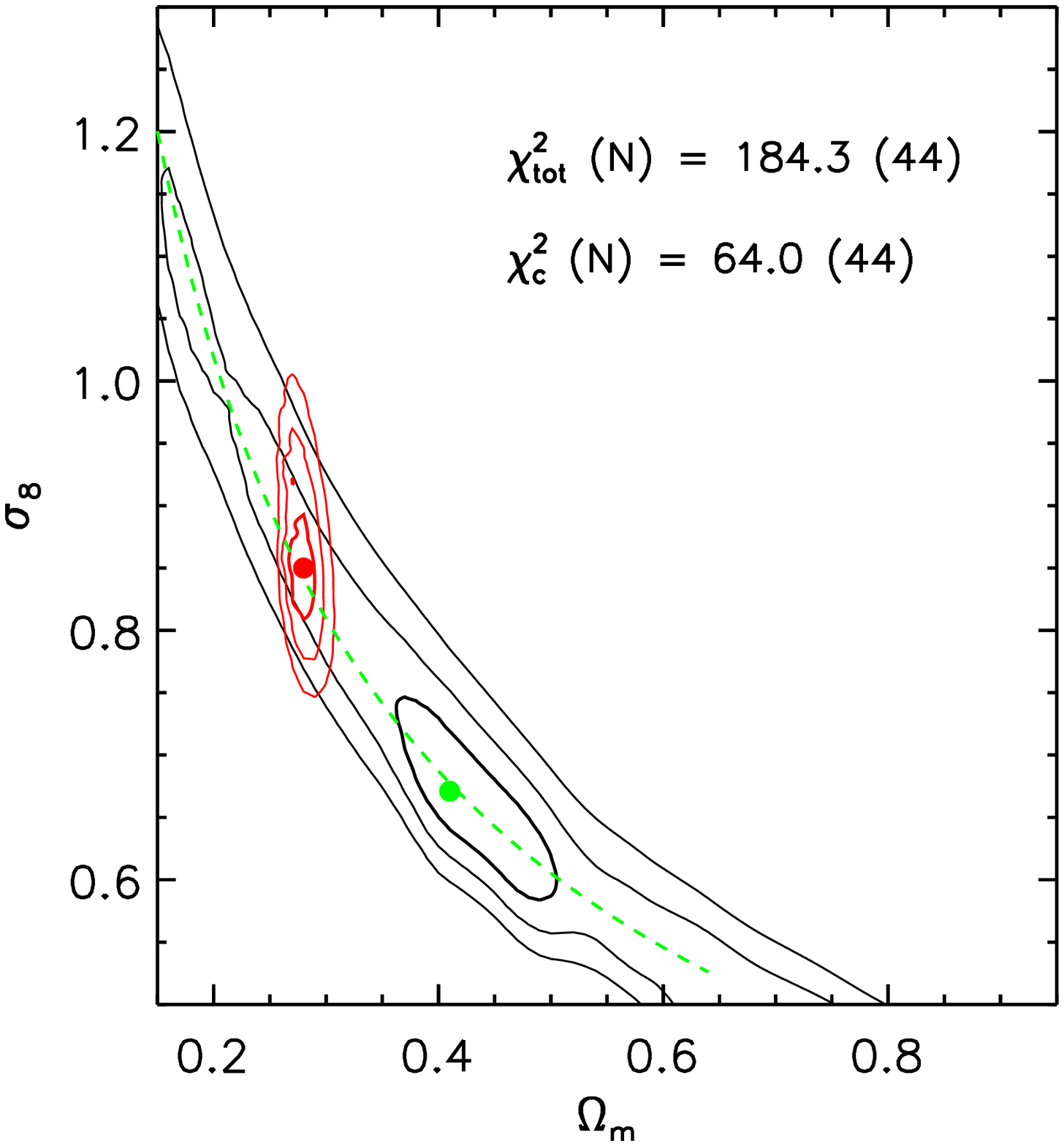,width=0.33\textwidth}
}
\caption{Cosmological constraints in the $(\Omega_{\rm m}, \sigma_8)$ plane
obtained from equations~\ref{eq:chi2c} and \ref{eq:chi2f}
by using predictions from the model by Eke et al. (2001).
The confidence contours at $1, 2, 3 \sigma$ on 2 parameters (solid contours) are displayed.
The combined likelihood with the probability distribution provided from the
cluster gas mass fraction method is shown in red.
The dashed green line indicates the power-law fit 
$\sigma_8 \, \Omega_{\rm m}^{\gamma} = \Gamma$.
The best-fit results are quoted in Table~\ref{tab:cosmo}.
A relative logarithmic scatter of 0.139 (see Sect.~\ref{sect:cosmo})
is considered in the models.
Systematic uncertainties on $c_{200}$ and $f_{\rm gas}(<R_{500})$ as
quoted in Table~\ref{tab:syst} are also propagated.
({\it Left}) From the subsample of 26 clusters satisfying
the condition $(r_{\rm s}+\epsilon_{r_{\rm s}}) < R_{\rm sp}$;
({\it center}) from the subsample of the LEC objects;
({\it right}) from all the 44 clusters.
} \label{fig:s8om} 
\end{figure*}

Then, we proceed with the following analysis:

\begin{enumerate}

\item a new mass $M_{200, j}$ and concentration $c_{200, j}$ are estimated 
from the X-ray data for given $\Omega_{\rm m, j}$;

\item we perform a linear interpolation on the theoretical prediction of $c_{200, ijk}$
to associate a concentration $\hat{c}_{200, jk}$ to the new mass $M_{200, j}$
for given $\Omega_{\rm m, j}$ and $\sigma_{8, k}$;

\item we evaluate the merit function $\chi_c^2$
\begin{equation}
\chi_c^2 = \chi_c^2(\Omega_{\rm m, j}, \sigma_{8, k}) =
\sum_{{\rm data}, i} \frac{\left(c_{200, i} - \hat{c}_{200, jk}\right)^2}
{\epsilon_{200, i}^2 +\sigma_c^2},
\label{eq:chi2c}
\end{equation}  
where $\epsilon_{200, i}$ is the $1 \sigma$ uncertainty related to the measured $c_{200, i}$
and $\sigma_c$ is the scatter intrinsic to the mean predicted value $\hat{c}_{200, jk}$
as evaluated in Neto et al. (2007; see their fig.~7 and relative discussion).
They estimate in the mass bin $10^{14.25}-10^{14.75} h^{-1} M_{\odot}$ a logarithmic mean
value of the concentration parameter of $0.663$, with a dispersion of $0.092$, corresponding
to a relative uncertainty of $0.139$.
We take into account these estimates by associating to the expectation of $\hat{c}_{200, jk}$
a scatter equals to $10^{\log \hat{c}_{200, jk} \pm \epsilon_c}$, where 
$\epsilon_c = 0.139 \times \log \hat{c}_{200, jk}$;

\item a minimum in the $\chi_c^2$ distribution, $\chi_{c, {\rm min}}^2$, 
is evaluated and the regions encompassing $\chi_{c, {\rm min}}^2 + (2.3,6.17,11.8)$ 
are estimated to constrain the best-fit values and the $1, 2, 3 \sigma$ 
intervals in the $(\Omega_{\rm m}, \sigma_8)$ plane shown in Fig.~\ref{fig:s8om}. 
To represent the observed degeneracy in the $\sigma_8 - \Omega_{\rm m}$ plane,
we quote in Table~\ref{tab:cosmo} (and show with a dashed line in Fig.~\ref{fig:s8om})
the best-fit values of the power-law fit
$\sigma_8 \, \Omega_{\rm m}^{\gamma} = \Gamma$, obtained by fitting this function
on a grid of values estimated, at each assigned $\Omega_{\rm m}$,
the best-fit result, and associated $1 \sigma$ error, of $\sigma_8$. 

\item A further constraint on the $\Omega_{\rm m}$ parameter that allows
us to break the degeneracy in the $\sigma_8 - \Omega_{\rm m}$ plane
(as highlighted from the banana-shape of the likelihood contours plotted in 
Fig.~\ref{fig:s8om}) is provided from the gas mass fraction distribution. 
We use our estimates of $f_{\rm gas} (<R_{500}) = f_{500}$ from \meti\
quoted in Table~\ref{tab:fgas}. We follow the procedure described in 
Ettori et al. (2009) and assume:
(i) $\Omega_{\rm b} h_{70}^2 = 0.0462 \pm 0.0012$ and 
$H_0 = 70.1 \pm 1.3$ from the best-fit results of the joint analysis in Komatsu et al. (2008),
(ii) a depletion parameter at $R_{500}$ $b_{500} = 0.874 \pm 0.023$, 
(iii) a contribution of cold baryons to the total budget 
$f_{\rm cold} = 0.18 (\pm 0.05) f_{\rm gas}$.
All the quoted errors are at $1 \sigma$ level.
Then, we look for a minimum in the function $\chi_f^2 = \chi_f^2(\Omega_{\rm m, j})$
\begin{equation}
\chi_{f}^2 = \sum_{{\rm data}, i} \frac{\left[ f_{500, i}
(1+ f_{\rm cold})/b_{500} - \hat{f}_{{\rm bar}, j} \right]^2}
{\epsilon_{f, i}^2},
\label{eq:chi2f}
\end{equation}
where $\hat{f}_{{\rm bar}, j} = \Omega_{\rm b}/\Omega_{{\rm m}, j}$ and
$\epsilon_{f, i}$ is given from the sum in quadrature of all the statistical errors,
namely, on $f_{500}, f_{\rm cold}, H_0, b_{500}$ and $\Omega_{\rm b}$. 

\item We combine the two $\chi^2$ distribution, $\chi_{\rm tot}^2 = \chi_c^2 + \chi_{f}^2$,
and plot in Fig.~\ref{fig:s8om} the constraints obtained from both $\chi_c^2$ only 
and $\chi_{\rm tot}^2$, quoting the best-fit results in Table~\ref{tab:cosmo}.

\item The effect of the systematic uncertainties, assumed to be normally
distributed, is also considered by propagating
them in quadrature to the measurements of $c_{200}$ and $f_{500}$, as obtained 
from the analysis summarized in Table~\ref{tab:syst}.
The constraints obtained after this further correction are indicated
with label ``$+$syst'' in Table~\ref{tab:cosmo}.

\end{enumerate}

\begin{table*}
\caption{Cosmological constraints on $\sigma_8$ and $\Omega_{\rm m}$.}
\vspace*{-0.2cm}
\begin{center}
\begin{tabular}{l c  c c c  c c c}
\hline \\
 Model & N data & $\gamma$ & $\Gamma$ & $\chi^2_c$  &  $\sigma_8$ & $\Omega_{\rm m}$ & $\chi^2_{\rm tot}$ \\ 
  & &  & & & \multicolumn{3}{c}{adding $f_{\rm gas}$} \\
\hline \\
E01 & 26 & $0.596 \pm 0.030$ & $0.449 \pm 0.012$ & 33.4 & $1.039_{-0.106}^{+0.124}$ & $0.25_{-0.01}^{+0.01}$ & 79.3 \\
(LEC) E01 & 11 & $0.558 \pm 0.042$ & $0.388 \pm 0.018$ & 8.3 & $0.825_{-0.083}^{+0.114}$ & $0.26_{-0.01}^{+0.02}$ & 38.3 \\
(all) E01 & 44 & $0.569 \pm 0.026$ & $0.408 \pm 0.012$ & 64.0 & $0.850_{-0.056}^{+0.087}$ & $0.28_{-0.01}^{+0.01}$ & 184.0 \\
 & \\
B01+M08 & 26 & $0.668 \pm 0.040$ & $0.547 \pm 0.014$ & 33.4 & $1.260_{-0.076}^{+0.040}$ & $0.25_{-0.01}^{+0.01}$ & 79.1 \\
NFW & 26 & $0.718 \pm 0.086$ & $0.344 \pm 0.036$ & 35.4 & $0.940_{-0.150}^{+0.252}$ & $0.25_{-0.01}^{+0.01}$ & 81.8 \\
E01 ($b_c$) & 26 & $0.574 \pm 0.032$ & $0.418 \pm 0.014$ & 36.8 & $0.939_{-0.082}^{+0.108}$ & $0.25_{-0.01}^{+0.01}$ & 82.7 \\
E01 ($b_M$) & 26 & $0.591 \pm 0.030$ & $0.458 \pm 0.012$ & 33.3 & $1.003_{-0.089}^{+0.145}$ & $0.26_{-0.01}^{+0.02}$ & 79.5 \\
E01 ($b_c, b_M$) & 26 & $0.576 \pm 0.032$ & $0.423 \pm 0.014$ & 36.8 & $0.936_{-0.109}^{+0.102}$ & $0.26_{-0.01}^{+0.02}$ & 82.9 \\
E01 ($M_{\rm tot}$) & 26 & $0.588 \pm 0.030$ & $0.441 \pm 0.012$ & 28.7 & $1.006_{-0.081}^{+0.116}$ & $0.25_{-0.01}^{+0.01}$ & 74.9 \\
\hline \\
\end{tabular}

\end{center}
\label{tab:cosmo}
\tablefoot{These cosmological contraints are obtained
from equations~\ref{eq:chi2c} and \ref{eq:chi2f}
corresponding to the confidence contours shown in Fig.~\ref{fig:s8om}.
To represent the observed degeneracy, we quote the best-fit values of the power-law
$\sigma_8 \, \Omega_{\rm m}^{\gamma} = \Gamma$.
Errors at $2 \sigma$ (95.4\%) level of confidence are indicated.
}
\end{table*}

The cosmological constraints we obtain with 3 different 
analytic models (E01, B01+M08, NFW) are summarized in Table~\ref{tab:cosmo}
and likelihood contours for the model of reference E01 are
plotted in Fig.~\ref{fig:s8om}.
To represent the observed degeneracy, we constrain the parameters
of the power-law fit $\sigma_8 \, \Omega_{\rm m}^{\gamma} = \Gamma$.
As expected from the properties of the prescriptions, 
E01 provides constraints on $\sigma_8$, for given $\Omega_{\rm m}$, 
that lie between the other two, with 
$\gamma = 0.60 \pm 0.04$ and $\Gamma=0.44\pm0.02$ (at $2\sigma$ level;
statistical only).
We break the degeneracy of the best-fit values in the ($\sigma_8, \Omega_{\rm m}$) plane
by assuming that the cluster baryon fraction represents the cosmic value well.
We obtain that $\sigma_8 = 1.0 \pm 0.2$ and $\Omega_{\rm m}=0.27 \pm 0.01$
(at $2\sigma$ level).
When the subsample of 11 LEC clusters, that are expected to be more relaxed
and with a well-formed central cooling core, is considered, we measure
$\gamma = 0.55 \pm 0.05$, $\Gamma=0.40\pm0.02$
$\sigma_8 = 0.85 \pm 0.16$ and $\Omega_{\rm m}=0.27 \pm 0.01$
(at $2\sigma$ level).

We confirm that, assumed correct the ones measured with E01, 
NFW tends to overestimate the predicted concentrations and, therefore,
requires lower normalization $\sigma_8$ of the power spectrum, whereas
B01+M08 compensate with larger values of $\sigma_8$
the underestimate of $c_{200}$ with respect to E01.

We assess the systematics affecting our results by comparing
the cosmological constraints obtained by assuming
(i) different algorithms to relate the cosmological models to the derived
$c-M$ relation, 
(ii) biases both in the concentration parameter ($b_c = 0.9$), 
from the evidence in numerical simulations that relaxed halos have 
an higher concentration by about 10 per cent (e.g. Duffy et al. 2008), 
and in the dark matter ($b_M = 1.1$) measurements from the evidence
provided from hydrodynamial simulations that the hydrostatic equilibrium
might underestimate the true mass by 5-20 per cent (e.g. see recent work
in Meneghetti et al. 2010).
As expected, lower concentrations and higher masses push the best-fit
values to lower normalizations of the power spectrum at fixed $\Omega_{\rm m}$, 
with an offset of about 10 per cent with $b_c = 0.9$ and of few
per cent $b_M = 1.1$ and $M_{\rm tot}$.

\section{Summary and Conclusions}

We present the reconstruction of the dark and gas mass from the \xmm\ 
observations of 44 massive X-ray luminous galaxy clusters in the redshift range $0.1-0.3$. 
We estimate a dark ($M_{\rm tot} - M_{\rm gas}$) mass within $R_{200}$ 
in the range (1st and 3rd quartile) $4-10 \times 10^{14} M_{\odot}$, 
with a concentration $c_{200}$ between 2.7 and 5.3, and a gas mass fraction
within $R_{500}$ between 0.11 and 0.16.

By applying the equation of the hydrostatic equilibrium 
to the spatially resolved estimates of the spectral temperature and normalization,
we recover the underlying gravitational potential of the dark matter halo, 
assumed to be well described from a NFW functional form, with two independent techniques.

Our dataset is able to resolve the temperature profiles up to about $0.6-0.8 R_{500}$
and the gas density profile, obtained from the geometrical deprojection
of the PSF--deconvolved surface brightness, up to a median radius of $0.9 R_{500}$.
Beyond this radial end, our estimates are the results of an 
extrapolation obtained by imposing a NFW profile for the total mass
and different functional forms for $M_{\rm gas}$.
 
We estimate, with a relative statistical uncertainty of $15-25\%$, 
the concentration $c_{200}$ and the mass $M_{200}$ of the dark matter
(i.e. total$-$gas mass) halo.
We constrain the $c_{200}-M_{200}$ relation to have a normalization
$c_{15} = c_{200} \times (1+z) \times \left( M_{200}/10^{15} M_{\odot} \right)^{-B}$
of about $2.9-4.2$ and a slope $B$ between $-0.3$ and $-0.7$ (depending on the
methods used to recover the cluster parameters and to fit the linear
correlation in the logarithmic space), 
with a relative error of about 5\% and 15\%, respectively.
Once the slope is fixed to the expected value of $B=-0.1$, the normalization,
with estimates of $c_{15}$ in the range $3.8-4.6$, agrees with
results of previous observations and simulations for calculations
done assuming a low density Universe.
We conclude thus that the slope of the $c_{200}-M_{200}$ relation 
cannot be reliably determined from the fitting over a narrow mass range
as the one considered in the present work, altough the steeper values
measured are not significantly in tension with the results
for simulated halos when the subsamples of the most robust 
estimates are considered (see Sect.~5 and Appendix~B).
We measure a total scatter in the logarithmic space of about 0.15
at fixed mass. This value decreases to 0.08 when the subsample
of LEC clusters is considered, where a slightly lower normalization and
flatter distribution is measured. 
This is consistent in a scenario where disturbed systems have an 
estimated concentration through the hydrostatic equilibrium equation 
that is biased higher (and with larger scatter)
than in relaxed objects up to a factor of 2 due to the action of the ICM motions
(see e.g. Lau et al. 2009).

We put constraints on the cosmological parameters ($\sigma_8, \Omega_{\rm m}$)
by using the measurements of $c_{200}$ and $M_{200}$ and by comparing 
the estimated values with the predictions tuned from numerical simulations of CDM universes.
In doing that, we propagate the statistical errors (with a relative value
of about $15-25$\% at $1 \sigma$ level) and consider the systematic uncertainties 
present both in the simulated datasets ($\sim 20$\%) and in our measurements
($\sim 10$\%; see Table~\ref{tab:syst}).
To represent the observed degeneracy, we constrain the parameters
of the power-law fit $\sigma_8 \, \Omega_{\rm m}^{\gamma} = \Gamma$
and obtain $\gamma = 0.60 \pm 0.03$ and $\Gamma=0.45\pm0.02$
(at $2\sigma$ level) when the E01 formalism is adopted.
Different formalisms (like the ones in B01, revised after M08, and NFW) 
induce variations in the best-fit parameters in the order of 20 per cent.
A further variation of about 10 per cent occurs if a bias of the order of 10 per cent
is considered on the estimates of $c_{200}$ and $M_{200}$.

We break the degeneracy of the best-fit values in the ($\sigma_8, \Omega_{\rm m}$) plane
by assuming that the cluster baryon fraction represents the cosmic value well.
We obtain that $\sigma_8 = 1.0 \pm 0.2$ and $\Omega_{\rm m}=0.26 \pm 0.01$ 
(at $2\sigma$ level; statistical only).

When the subsample of 11 LEC clusters, that are expected to be more relaxed
and with a well-formed central cooling core, is considered, we measure
$\gamma = 0.56 \pm 0.04$, $\Gamma=0.39\pm0.02$
$\sigma_8 = 0.83 \pm 0.1$ and $\Omega_{\rm m}=0.26 \pm 0.02$
(at $2\sigma$ level).

All these estimates agree well with similar constraints
obtained for an assumed low-density Universe in Buote et al. (2007;
$0.76 < \sigma_8 < 1.07$ at 99\% confidence for a $\Lambda$CDM model
with $\Omega_{\rm m}=0.3$) and with the results obtained 
by analysing the mass function of rich galaxy clusters
[see, e.g., Wen, Han \& Liu (2010) that summarizes
recent results obtained by this cosmological tool], showing that the study
of the distribution of the measurements in the $c-M_{\rm DM}-f_{\rm gas}$
plane provides a valid technique already mature and competitive 
in the present era of precision cosmology.

However, we highlight the net dependence of our results on the models
adopted to relate the properties of a DM halo to the background cosmology.
In this context, we urge the $N-$body community to generate
cosmological simulations over a large box to properly predict the expected concentration 
associated to the massive ($>10^{14} M_{\odot}$) DM halos as function of 
$\sigma_8$, $\Omega_{\rm m}$ and redshift. 
The detailed analysis of the outputs of these datasets will provide
the needed calibration to make this technique more reliable and robust.

\section*{ACKNOWLEDGEMENTS}
The anonymous referee is thanked for suggestions that have improved the
presentation of the work.
We acknowledge the financial contribution from contracts ASI-INAF
I/023/05/0 and I/088/06/0.
This research has made use of the X-Rays Clusters Database (BAX)
which is operated by the Laboratoire d'Astrophysique de Tarbes-Toulouse (LATT),
under contract with the Centre National d'Etudes Spatiales (CNES).

\begin{appendix}

\section{Conversion between different overdensity and $c-M$ relations}

The total mass within a given overdensity $\Delta$ is defined 
in the present work as 
\begin{equation}
M_{\Delta} = \frac{4}{3}\pi R_{\Delta}^3 \, \Delta \rho_{c, z}, 
\label{eqa:mass}
\end{equation}
where $\rho_{c, z} = 3 H_z^2 / (8 \pi G)$ is the critical density of the Universe
at the cluster's redshift $z$, $R_{\Delta} = c_{\Delta} r_{\rm s}$
is the radius within which the mean cluster overdensity is $\Delta$ times 
$\rho_{c, z}$ and the relation with the concentration $c_{\Delta}$ and
the scale radius $r_{\rm s}$ holds by definition of the NFW mass profile. 
We assume $\Delta = 200$. Hereafter, we refer to $\Delta$ as any other 
assumed overdensity. In case it is referred to the background density of the
Universe, $\rho_b = \Omega_{\rm m, z} \rho_{c, z}$, it is straightforward
to correct $\Delta$ by $\Omega_{\rm m, z} = \Omega_{\rm m} (1+z)^3 / H_z^2$
to recover the definition in equation~\ref{eqa:mass}.

To convert the tabulated values to our definition of the $c-M$ relation,
\begin{equation}
c_{200} = \frac{c_{15}}{1+z} \left( \frac{M_{200}}{M_{15}} \right)^B,
\end{equation}
where $M_{15} = 10^{15} h_{70}^{-1} M_{\odot}$, 
we proceed as follows:
\begin{enumerate}

\item by definition, $M_{\Delta} / (R_{\Delta}^3 \Delta)$ is constant 
and $R_{\Delta}/c_{\Delta}$ is fixed from the measurement of the scale radius.
Therefore, we can write
\begin{equation}
\frac{M_{\Delta}}{c_{\Delta}^3} = \frac{M_{200}}{c_{200}^3} \frac{\Delta}{200} 
\end{equation}

\item $c_{\Delta}$ and $c_{200}$ are related through the assumed NFW mass density profile
\begin{equation}
\left( \frac{c_{200}}{c_{\Delta}} \right)^3 
\frac{\ln (1+c_{\Delta}) - c_{\Delta}/(1+c_{\Delta}) }
{\ln (1+c_{200})-c_{200}/(1+c_{200})} = \frac{\Delta}{200}.
\label{eqa:rdelta}
\end{equation}
This function is monotonic and easily to resolve numerically to estimate
$C = c_{\Delta} / c_{200}$, that is a quantity that depends mostly
on $\Delta$ and only marginally on the guessed $c_{200}$,
 as shown in Fig.~\ref{fig:cdelta}.
For instance, for $\Delta = 178 \Omega_{\rm m, z}^{0.45}$, which
estimate the virial overdensity predicted from the spherical collapse model
in a flat Universe with a contribution from dark energy (Eke et al. 2001),
$C = 1.34$ and $1.22$ at $z=0$ and $z=0.3$, respectively, for $\Omega_{\rm m}=0.3$,
with deviations within 2\% in the range $c_{200} = 3-6$.

\begin{figure}
 \epsfig{figure=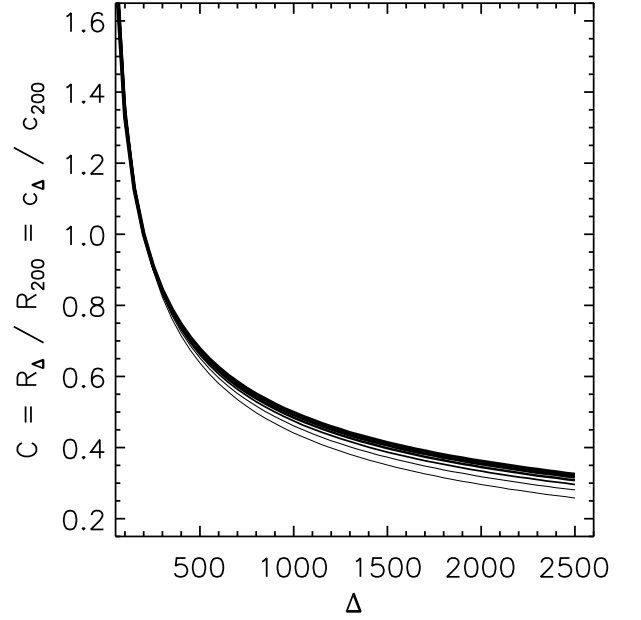,width=0.5\textwidth}
\caption{Numerical solution to equation~\ref{eqa:rdelta} for an assumed 
$c_{200}$ in the range $3-8$ (from the thinnest to the thickest line).
} \label{fig:cdelta}
\end{figure}

\item for a given relation $c_{\Delta} = c_0 (1+z)^{-1} \left(M_{\Delta}/M_{\ast} \right)^B$,
we substitute the above relations to obtain after simple algebrical operations:
\begin{equation}
c_{200} = \frac{c_0 C^{3B-1}}{1+z} \left( \frac{\Delta}{200} 
\frac{M_{15}}{M_{\ast}} \frac{M_{200}}{M_{15}} \right)^B,
\end{equation} 
or
\begin{equation}
c_{15} = c_0 C^{3B-1} \left( \frac{\Delta}{200} \frac{M_{15}}{M_{\ast}} \right)^B.
\end{equation}

\end{enumerate}

\section{Monte-Carlo realizations of the $c_{200}-M_{200}$ relation}

\begin{figure}
\epsfig{figure=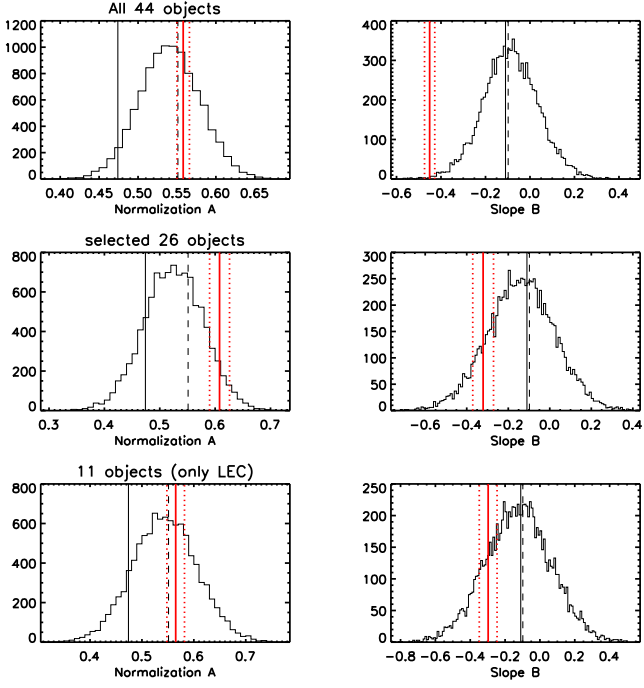,width=0.5\textwidth}
\caption{Distribution of the best-fit values of the normalization $A$ and slope $B$
after 10,000 MC realizations. The input values (in these plots from M08 / WMAP-5)
are indicated with vertical solid (for all the simulated objects) and dashed
(for the relaxed ones) lines. The red solid line represents the central value for
the corresponding sample as quoted in Table~\ref{tab:cm}. 
The red dotted lines show the $1 \sigma$ uncertainties.
} \label{fig:hist_cm}
\end{figure}

\begin{table*}
\caption{Results of the 10,000 MC runs of the $c-M$ relation fitted
using the expression in equation~\ref{eq:cm}.
}
\centering
\begin{tabular}{l c c c c c c}
\hline \\
Model & N obj & mean (rms) $A$ & mean (rms) $B$ & $B_{\rm obs} \pm \sigma$ & $P_{1\sigma}(B_{\rm obs})$ & $P_{3\sigma}(B_{\rm obs})$ \\
 & \\
\hline \\
N07 & 44 & $0.585 (0.034)$ & $-0.140 (0.106)$ & $-0.451 \pm 0.023$ & 0.5 & 1.5 \\
N07 & 26 & $0.583 (0.047)$ & $-0.130 (0.136)$ & $-0.321 \pm 0.050$ & 14.5 & 38.2 \\
N07 & 11 & $0.570 (0.051)$ & $-0.134 (0.153)$ & $-0.297 \pm 0.051$ & 23.4 & 47.4 \\
M08 / WMAP-1 & 44 & $0.610 (0.041)$ & $-0.095 (0.135)$ & $-0.451 \pm 0.023$ & 0.8 & 1.8 \\
M08 / WMAP-1 & 26 & $0.591 (0.057)$ & $-0.154 (0.175)$ & $-0.321 \pm 0.050$ & 24.6 & 46.1 \\
M08 / WMAP-1 & 11 & $0.611 (0.067)$ & $-0.140 (0.206)$ & $-0.297 \pm 0.051$ & 29.9 & 49.3 \\
M08 / WMAP-3 & 44 & $0.524 (0.040)$ & $-0.073 (0.130)$ & $-0.451 \pm 0.023$ & 0.4 & 0.9 \\
M08 / WMAP-3 & 26 & $0.510 (0.056)$ & $-0.118 (0.170)$ & $-0.321 \pm 0.050$ & 18.0 & 37.7 \\
M08 / WMAP-3 & 11 & $0.526 (0.065)$ & $-0.107 (0.200)$ & $-0.297 \pm 0.051$ & 23.6 & 42.6 \\
M08 / WMAP-5 & 44 & $0.541 (0.039)$ & $-0.087 (0.125)$ & $-0.451 \pm 0.023$ & 0.4 & 1.1 \\
M08 / WMAP-5 & 26 & $0.525 (0.054)$ & $-0.139 (0.162)$ & $-0.321 \pm 0.050$ & 20.5 & 41.9 \\
M08 / WMAP-5 & 11 & $0.544 (0.062)$ & $-0.125 (0.189)$ & $-0.297 \pm 0.051$ & 26.0 & 45.9 \\
\hline \\
\end{tabular}

\label{tab:hist_cm}
\tablefoot{
$B_{\rm obs}$ is the best-fit result quoted in Table~\ref{tab:cm}.
$P_{1\sigma}(B_{\rm obs})$ and $P_{3\sigma}(B_{\rm obs})$ indicate 
the percentage of MC runs that provides an estimate of $B$ lower
than $B_{\rm obs} +1\sigma$ and $B_{\rm obs} +3\sigma$, respectively.
} 
\end{table*}

We have run Monte-Carlo (MC) simulations to test the robustness of the observed
deviations in the $c_{200}-M_{200}$ relation described in Section~5.
We have used as input values the best-fit results (defined in the following
analysis as $\bar{c}_{15}$ and $\bar{B}$) obtained in the numerical
simulations from Neto et al. (2007; see their equations~4 and 5) 
and Macci\`o et al. (2008; see Table~A1 and A2) and listed in Tab.~\ref{tab:cm}.
We have considered the results for both the complete sample and the relaxed
objects only. 
To each cluster in our sample with measured mass $M_{200, i}$ and 
redshift $z_i$, we assign the concentration $c_{200, i}$ defined as 
\begin{eqnarray}
c_{200, i} & = & 10^{l_i} \\
l_i & = & \log_{10} \left[ \bar{c}_{15} \times \left(M_{200, i}/10^{15}\right)^{\bar{B}}
/ (1+z_i) \right] +R \times \epsilon_{\log c} \nonumber,
\label{eq:mc}
\end{eqnarray}
where $R$ is a random value extracted from a Gaussian distribution and 
$\epsilon_{\log c}$ is the scatter in the log-Normal distribution
measured in the numerical simulations ($\sim 0.13$ for samples including all 
the simulated objects and $\sim 0.1$ for the sample of the relaxed ones; the 
actual values are quoted in N07 and in Table~A1 and A2 of M08).
We assume that (1) our LEC objects follow the distribution obtained
for relaxed simulated clusters; (2) all the remaining clusters
follow the distribution estimated for the complete simulated halo sample;
(3) the 3 samples considered in our analysis with 44, 26 and 11 clusters, 
respectively, are built considering whether each object is
a LEC and/or has an upper limit at $1 \sigma$ on the scale radius lower 
than $R_{\rm sp}$, as discussed in Section~5.
To be conservative in our approach, we fit equation~\ref{eq:cm} 
to the distribution in the $c_{200}-M_{200}$ with the {\sl FITEXY}
technique that is the one that provides the most significant deviations
from the results obtained in numerical simulations.
We repeat this process 10,000 times and obtain the plots shown in 
Fig.~\ref{fig:hist_cm} for each of the case investigated.
In Table~\ref{tab:hist_cm}, we summarize our finding from which
we conclude that, while the best-fit values estimated for the samples 
of 26 and 11 clusters are within the overall distribution expected 
in numerical simulations, the sample of 44 clusters provides results 
on the slope $B$ that lie on the lower end of the distribution,
as probable consequence of the uncertainties present both on the estimates
of $c_{200}$ and $M_{200}$ for the 18 clusters that are indeed not selected 
for the further analysis and on the residual bias affecting the measurements
of $c_{200}$ under the hypothesis of the hydrostatic equilibrium 
(see Section~5.1 and, e.g., Lau et al. 2009). 

\end{appendix}

\end{document}